\documentclass[12pt]{iopart}
\usepackage[dvips]{graphics}
\usepackage{graphicx}
\usepackage{subfigure}

\usepackage{graphicx}
\def\rootfig{./}

\begin{document}


\title[Structure and stability of 2D BECs under harmonic and lattice confinement]{
Structure and stability of two-dimensional Bose-Einstein condensates 
under both harmonic and lattice confinement}
\author{K.J.H. Law$^1$ and P.G. Kevrekidis$^1$, B.P. Anderson$^2$, R. Carretero-Gonz{\'a}lez$^3$, and D.J. Frantzeskakis$^4$}

\address{$^1$
Department of Mathematics and
Statistics, University of Massachusetts, Amherst, MA 01003-4515 USA
\eads{\mailto{law@math.umass.edu}}}
\address{$^2$
College of Optical Sciences and
Department of Physics, University of Arizona, Tucson, AZ 85721 USA
}
\address{$^3$
Nonlinear Dynamical Systems Group,
Department of Mathematics and Statistics,
San Diego State University, San Diego, California 92182-7720 USA
\footnote{URL: {\tt http://nlds.sdsu.edu/}}}
\address{$^4$
Department of Physics,
University of Athens, Panepistimiopolis,
Zografos, Athens 15784, Greece
}

\begin{abstract}
In this work, we
study
two-dimensional Bose-Einstein
condensates confined by both a 
cylindrically symmetric harmonic
potential
and an optical lattice with equal periodicity
in two orthogonal directions.
We first identify
the spectrum of the underlying two-dimensional linear
problem through multiple-scale techniques. Then, we
use the results obtained in the linear limit
as a starting point for the 
existence and stability analysis of the lowest energy states,
emanating from the linear ones, in the nonlinear problem. Two-parameter continuations
of these states are performed for increasing nonlinearity and optical
lattice strengths, and
their instabilities
and temporal evolution are investigated.
It is found that the ground state as well as some of the excited states
may be stable or 
weakly unstable for both 
attractive and repulsive interatomic interactions.
Higher excited states are typically found to be increasingly
more unstable.
\end{abstract}

\submitto{\JPB}
\maketitle

\vspace{6mm}


\section{Introduction}

The last decade has witnessed a tremendous amount
of research efforts in the physics
of atomic Bose-Einstein
condensates (BECs) \cite{book1,book2}. The
study of BECs has
yielded a wide array of interesting phenomena,
not only because of very precise experimental control
that exists over the relevant experimental procedures \cite{rmp,ourbook}, but also
because of the intimate connections of the
description of
dilute-gas BECs with other areas of physics, such as superfluidity, superconductivity, lasers and coherent optics, nonlinear optics, and
nonlinear wave theory.
Of particular emphasis in much experimental and theoretical work is the setting of a BEC trapped 
in periodic potentials, usually combined with an additional harmonic trapping potential. 
From the standpoint of nonlinear interactions, mathematical descriptions of BECs held in purely harmonic traps are now well known.
Nevertheless, apart from studies focusing on the transition between superfluidity and an 
insulator state \cite{bloch:02}, relatively little attention has been given to an understanding of the varieties of 
many-body states that may possibly exist with intermediate lattice strengths, where phase coherence is maintained across the sample. 
A more complete understanding of BEC behavior in such lattice potentials is relevant and important to current work with BECs, 
and to an even broader array of topics, in particular discrete nonlinear optics and nonlinear wave theories. 
Such regimes of BEC physics are experimentally and theoretically accessible, and comparisons between theoretical 
and experimental results are certainly possible.  Here, we present a theoretical stability examination of 
BECs with either attractive or repulsive interatomic interactions in a combined harmonic and periodic potential.

Many of the common elements between BECs and other areas of physics,
and in particular optics, originate in the existence of macroscopic coherence in the many-body state of the system. 
Mathematically, BEC dynamics are therefore often accurately described by a mean-field model, namely a partial 
differential equation of the nonlinear Schr{\"o}dinger (NLS) type, the so-called Gross-Pitaevskii (GP) equation \cite{book1,book2,rmp}. 
The GP equation is particularly successful in drawing connections between BEC physics, nonlinear optics and nonlinear wave 
theories, with vortices and solitary waves examples of common elements between these areas.
The GP equation is a classical nonlinear evolution equation (with the nonlinearity
originating from the interatomic interactions) and, as such, it permits the study of a variety
of interesting nonlinear phenomena. These phenomena have primarily been studied by treating
the condensate as a purely nonlinear coherent matter-wave, i.e., from the viewpoint of the nonlinear dynamics
of solitary waves. Relevant studies have already been summarized in various books (see, e.g., Ref.~\cite{ourbook})
and reviews (see, e.g., Refs.~\cite{abdu} for bright matter-wave solitons, \cite{fetter,pgk1} for
vortices in BECs, \cite{pgk2} for dynamical instabilities in BECs, \cite{konotop,morsch} for
nonlinear dynamics of BECs in optical lattices).

On the other hand, many static and dynamic properties of BECs confined in various types of external potentials
can be studied by starting from the non-interacting limit, where the nonlinearity is considered
to be negligible. The basic idea of such an approach is that in the absence of interactions
the GP equation is reduced to a linear Schr{\"o}dinger equation for a confined single-particle state;
in this limit, and in the case of, e.g., a harmonic external potential, the linear problem becomes the equation
for the quantum harmonic oscillator characterized by discrete energies and corresponding eigenmodes \cite{landau}.
Exploiting this simple physical picture, one may then use analytical and/or numerical techniques for
the {\it continuation} of these linear eigenmodes supported by the particular type of the external trapping
potential into nonlinear states as the
interactions become stronger. This idea has been
explored at the level of one-dimensional (1D) \cite{yuri1} and higher-dimensional states \cite{kivshar}
in the case of a harmonic trapping potential, where nonlinear stationary modes were found
from a continuation of the (linear) states of the quantum harmonic oscillator. The same problem has
been studied in the framework of the so-called Feshbach resonance management technique in Ref.~\cite{konotop1},
where a linear temporal variation of the nonlinearity was considered.
The continuation of the
linear states to their nonlinear counterparts has also been explored from the point of
view of bifurcation and stability theory \cite{mtol1}. Finally, in the same spirit but in the two-dimensional (2D) setting,
radially symmetric nonlinear states of harmonically trapped ``pancake-shaped'' condensates were recently investigated
in Ref.~\cite{gregh}.

Importantly, all of the above studies provide a clear physical picture of how
genuinely nonlinear states of harmonically
confined BECs (such as dark and bright matter-wave solitons in 1D or ring solitons and vortices in 2D), are connected to and emanate
from the eigenmodes of the quantum harmonic oscillator. Similar considerations also hold for BECs confined in optical lattices.
In this case, pertinent nonlinear stationary states (such as spatially extended nonlinear Bloch waves, truncated nonlinear
Bloch waves, matter-wave gap solitons in 1D, and gap vortices in 2D and 3D), can be understood by the structure of the
band-gap spectrum of the linear Bloch waves supported in the non-interacting limit (see, e.g., Ref.~\cite{yuri2} and references therein).
However, there are only few studies for condensates confined in {\it both} harmonic and
optical lattice potentials, and these are basically devoted to the dynamics of particular nonlinear structures
(such as dark \cite{gt} and bright \cite{bs} solitons in 1D, and vortices in 2D \cite{vortex}). Thus, the
structure of condensates confined in such superpositions of harmonic and periodic potentials remains, to the best of our knowledge, largely unexplored.

Nevertheless, such a study is particularly relevant to current work with BECs, and even suggests new avenues for exploration. 
In particular, one might ask whether the addition of a weak optical lattice might increase the stability of excited states 
that are known to be unstable in harmonic traps. Stability, if it is found, may add new realistic options for the engineering 
of new quantum states of BECs. It may also be interesting to investigate the Mott insulator transition by starting from a stable 
excited state of a weak optical lattice. Also, the transport of excited states (which is not discussed in this paper) through a 
lattice structure may have application in future BEC interferometry experiments.
Finally, the advances of far-off-resonant optical trapping techniques
allow for the creation of strongly pancake-shaped condensates that may be confined by harmonic and spatially periodic components, 
and we expect that the theoretical considerations described here may be directly explored using current experimental techniques.

Our aim in the present work is to contribute to this direction and study the structure and the stability
of a pancake-shaped condensate confined by the combination of a harmonic trap and a periodic potential, with periodicity in two orthogonal directions.
We will adopt the above mentioned approach of continuation of linear states to
nonlinear ones, thus
providing a host of interesting solutions that have not been explored previously and yet should be tractable within
presently available experimental settings.  In particular, our analysis starts by first considering the non-interacting limit. 
In this regime, we employ a multiscale perturbation method (which uses the harmonic trap strength as a formal small parameter) to find
the discrete energies and the corresponding eigenmodes of the pertinent single-particle Schr\"{o}dinger equation with
the combined harmonic and periodic potential.
We then use this linear limit as a starting point for initializing a 2D
``nonlinear solver'' that identifies the relevant stationary nonlinear eigenstates as a
function of the chemical potential (i.e., the nonlinearity strength) and of the optical lattice depth.

Once the basic structure of the condensate is found, we subsequently perform
a linear stability analysis
of the nonlinear modes that can be initiated
by the non-interacting ground and first few excited states. When
nonlinear states are found to be unstable, we use direct numerical simulations to study their dynamics and
monitor the evolution of the relevant instability. Essential results that will be presented below are
the following: for a fixed harmonic trap strength, there exist certain regions in the parameter plane
defined by the chemical potential and the optical lattice depth, where not only the ground state, but also
excited states are stable or only weakly unstable. Particularly, an excited state with a shape resembling an
out-of-phase matter-wave soliton pair (for attractive interactions) is found to persist for long times, being stable (weakly unstable) for
attractive (repulsive) interatomic interactions. Thus, the ground state and 
the aforementioned excited state
have a good chance to be observed in a real experiment with either 
attractive or repulsive pancake BECs. Similar conclusions can be drawn 
for more complex states, such as a quadrupolar one which may also be 
stable in the attractive case, however, higher excited states are typically 
more prone to instabilities, as is shown in our detailed numerics below.

The paper is organized as follows: In Section II we present the model and study analytically the non-interacting regime.
The continuation of the linear states to the nonlinear ones, as well as the stability properties of the nonlinear states
are presented in Section III. Finally, in Section IV, we summarize our findings and present our conclusions.

\section{
The model and its analytical consideration}


At sufficiently low temperatures, and in the framework of the
mean-field approach, the condensate dynamics can be described by the order parameter
$\Psi(\mathbf{r},t)$. We assume that the condensate is kept in a highly
anisotropic trap, with the transverse ($x, y$) and longitudinal ($z$) trapping frequencies
chosen so that $\omega_{x} = \omega_{y} \equiv \omega_{\perp}\ll \omega_{z}$ . In such a case, the condensate
has a nearly planar, so-called ``pancake'' shape (see, e.g., Refs.~\cite{2dbec} for relevant experimental realizations),
which allows us to assume a separable wave function, $\Psi = \Phi(z)\psi(x,y)$, where $\Phi(z)$ is the ground state
of the respective quantum harmonic oscillator.
Then, averaging of the underlying
three-dimensional (3D) GP equation along the longitudinal direction $z$ \cite{gpe1d}
leads to the following 2D GP equation for the transverse component of the wave function
(see also Refs.~\cite{pgk1,pgk2,ourbook}):
\begin{eqnarray}
i \hbar \partial_{t} \psi = -\frac{\hbar^2}{2m} \Delta \psi + g_{2D} |\psi|^2 \psi + V_{\rm ext}(x,y) \psi.
\label{gpe}
\end{eqnarray}
Here, $\Delta \equiv \partial_{x}^{2}+\partial_{y}^{2}$ is the 2D Laplacian, $m$ is the atomic mass, and
$g_{2D}=g_{3D}/\sqrt{2}  \pi a_z$ is an effective 2D coupling constant, where $g_{3D}=4\pi \hbar^2 a/m$
($a$ being the scattering length) and $a_z = \sqrt{\hbar/m\omega_z}$ is the longitudinal
harmonic oscillator length. Finally, the potential $V_{\rm ext}(x,y)$ in the GP Eq.~(\ref{gpe}) is
assumed to consist of a harmonic component and a square 2D optical lattice (OL) created by two pairs of interfering
laser beams of wavelength $\lambda$:
\begin{eqnarray}
V_{\rm ext}(x,y)
&=& \frac{1}{2}m \omega_{\perp}^2 r^2
+ V_0 [\cos^{2}(kx) + \cos^{2}(ky)].
\nonumber \\[1.0ex]
&\equiv& V_{\rm H}(r) + V_{\rm OL}(x,y)
\label{extpot}
\end{eqnarray}
In the above expression, $r^2 \equiv x^2+y^2$, while the optical lattice is characterized by two parameters, namely its depth $V_0$
and its periodicity $d=\pi/k =(\lambda/2)/\sin(\theta/2)$, where $\theta$ is the angle between
the two beams that create the $x$-direction lattice, and between the two beams that create the $y$-direction lattice.

Measuring length in units of $a_{L}=d/\pi$, time in units of $\omega_{L}^{-1}=\hbar/E_{L}$,
and energy in units of $E_{L}=2E_{\rm rec}=\hbar^2/m a_{L}^{2}$
(where $E_{\rm rec}$ is the lattice recoil energy),
the GP Eq.~(\ref{gpe}) can be put into the following dimensionless form:
\begin{eqnarray}
i \partial_t \psi = - \frac{1}{2} \Delta \psi + s |\psi|^2 \psi + V(x,y) \psi.
\label{eqn1}
\end{eqnarray}
%
%
In the normalized GP Eq.~(\ref{eqn1}),
the wavefunction is rescaled as $\psi \rightarrow \sqrt{|g_{2D}|/E_{L}}\psi \exp\left[i(V_{0}/E_{L})t\right]$,
the parameter $s$ is given by
$s={\rm sign}(g_{2D})=\pm 1$ (with $s=+1$ or $s=-1$ corresponding, respectively, to repulsive or attractive interatomic interactions),
%
%
while the normalized trapping potential $V(x,y)$
is now given by:
\begin{eqnarray}
V(x,y)=
\frac{1}{2} \Omega^2 r^2
+ V_0 (\cos(2 x) + \cos(2 y)).
\label{eqn2}
\end{eqnarray}
In the above equation, the normalized lattice depth $V_0$ is measured in units of $4E_{\rm rec}$,
while the normalized harmonic trap strength is given by
%
\begin{eqnarray}
\Omega = \frac{a_{L}^2}{a_{\perp}^2} = \frac{\omega_{\perp}}{\omega_{L}},
\label{hts}
\end{eqnarray}
%
%
where $a_{\perp}=\sqrt{\hbar/m \omega_{\perp}}$ is the transverse harmonic oscillator length. Note that
using realistic parameter values (see, e.g., Ref.~\cite{becol}),
namely, a lattice periodicity 0.3 $\mu$m, a
recoil energy $E_{\rm rec}/h \sim 6$KHz (assuming an atomic mass corresponding to $^{87}$Rb), 
and using a transverse trap frequency $\omega_{\perp}=2\pi\times 5$Hz, the parameter $\Omega$ is of order of 
$10^{-4}$; thus, it is a natural small parameter of the problem.

Our analysis starts by considering the non-interacting limit $s \rightarrow 0$, in which
the GP equation becomes a linear Schr\"{o}dinger equation. Then, seeking stationary localized solutions
of the form $\psi(x,y,t)=\exp(-i E_{m,n} t) u_{m,n} (x,y)$ [where $E_{m,n}$ are discrete energies
and $u_{m,n} (x,y)$ are the corresponding linear eigenmodes],
%
and rescaling spatial variables by $\sqrt{\Omega}$,
we obtain the following equation:
\begin{eqnarray}
\frac{E_{m,n}}{\Omega} u_{m,n}&=&
-\frac{1}{2} \Delta u_{m,n} + \frac{1}{2} (x^2+y^2) u_{m,n} 
\\[1.0ex]
\nonumber
&& + \frac{V_0}{\Omega} \left[\cos\left(\frac{2x}{\sqrt{\Omega}}\right) + \cos\left(\frac{2y}{\sqrt{\Omega}}\right)\right] u_{m,n}.
\label{eqn3}
\end{eqnarray}
%
The next step is to separate variables through $u(x,y)=u_m(x) u_n(y)$ and split
the energy into $E_{m,n}=E_m+E_n$,
to obtain two 1D eigenvalue problems of the same type as above, namely,
%
\begin{eqnarray}
-\frac{1}{2} \frac{d^2 u_m}{dx^2} + \frac{1}{2} x^2 u_m
+ \frac{V_0}{\Omega}  \cos\left(\frac{2x}{\sqrt{\Omega}}\right)
u_m = \frac{E_m}{\Omega} u_m,
\label{eqn4}
\end{eqnarray}
and a similar one for $y$ (with $x$
replaced by $y$ and the subscript $m$
replaced by $n$).

\begin{figure}[t]
\begin{center}
\includegraphics[width= 5.0cm,height=3.5cm]{\rootfig 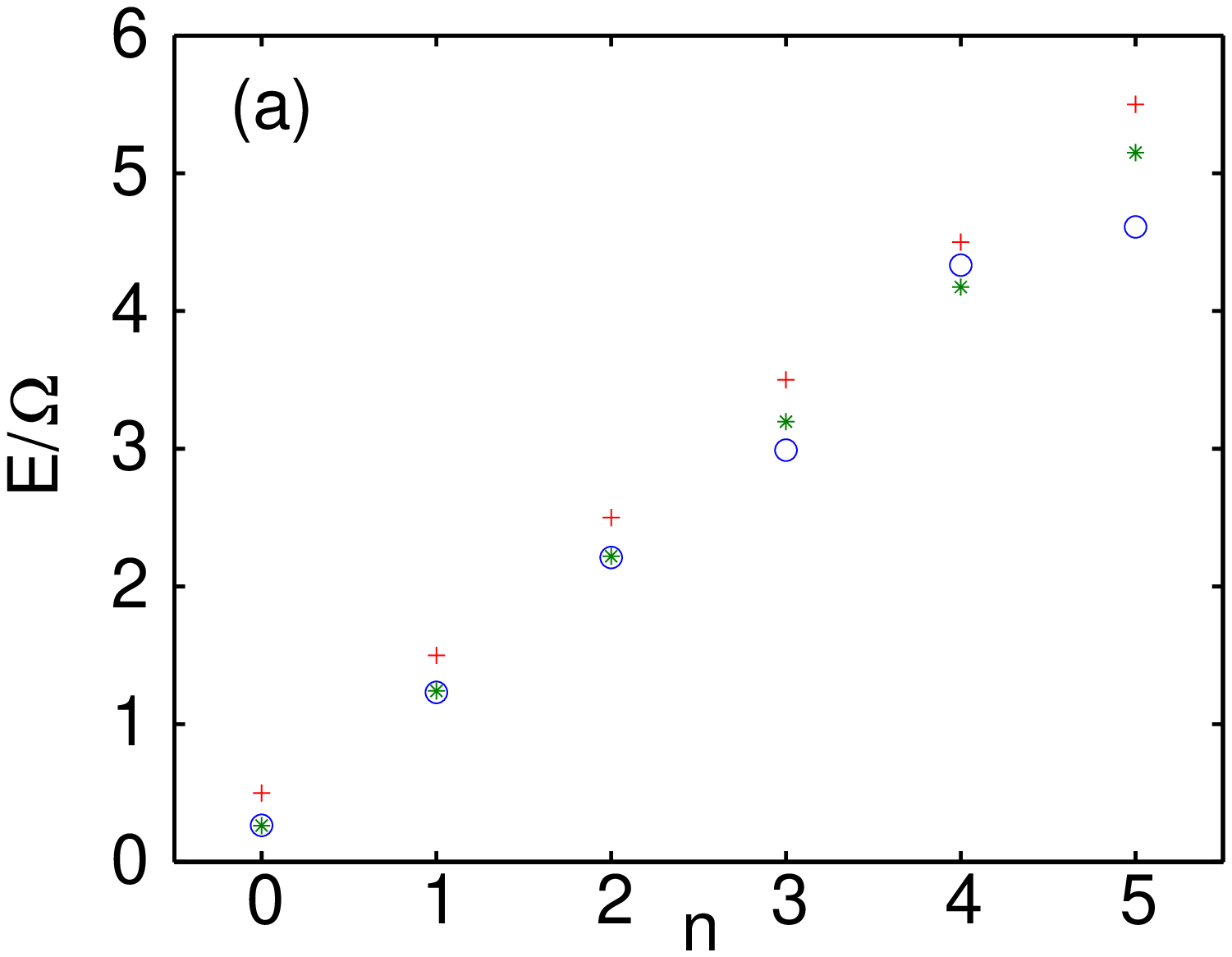} ~~
\includegraphics[width= 5.0cm,height=3.5cm]{\rootfig 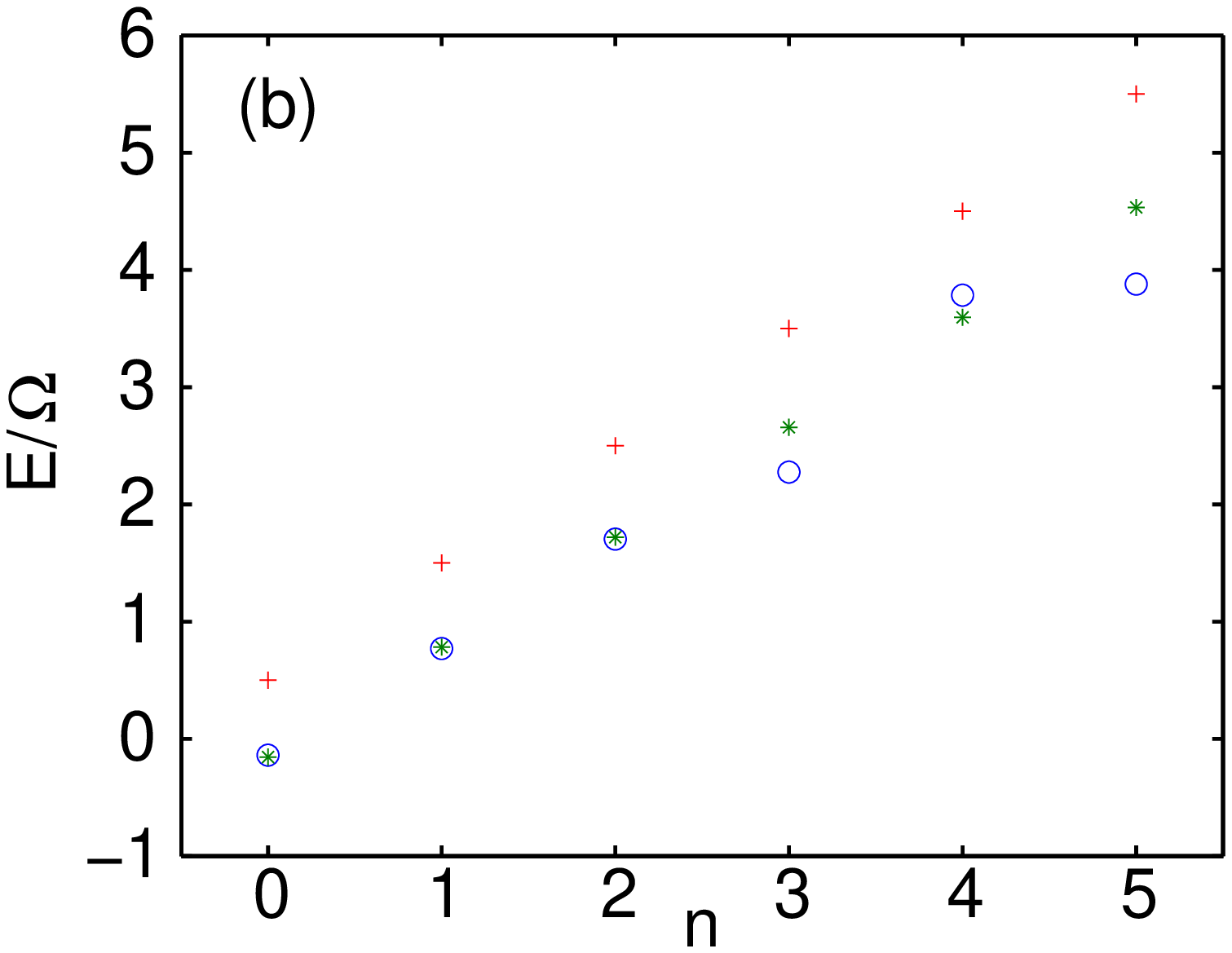}\\[2.0ex]
\includegraphics[width=16.0cm]{\rootfig 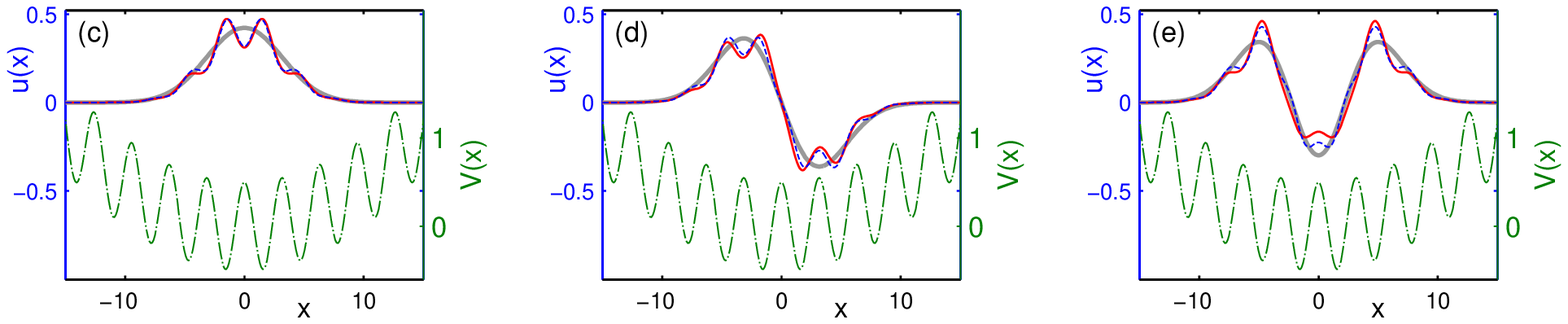} 
\begin{center}
\caption{ \label{vfig0} 
(Color Online) 
Panel (a) shows the energy spectra corresponding to a purely
parabolic potential (pluses), a parabolic and lattice one found numerically
(circles) and parabolic and
lattice potential found analytically
(stars).
Shown are only the first few eigenvalues for $\Omega=0.1$ and $V_0=0.3$.
A similar result is demonstrated
in panel (b) but for a larger lattice depth, $V_0=0.5$ (and the same value of $\Omega$).
For the latter case ($V_0=0.5$), panels (c), (d) and (e) show the first few eigenmodes
for the parabolic potential (thick solid), parabolic and
lattice potential computed numerically (thin solid), and the same ones given by Eq.~(\ref{eqn11})
(dashed). The potential (rescaled for visibility) is shown by the
dash-dotted line.}\end{center}
\end{center}
\end{figure}

We now restrict ourselves to the physically relevant regime of $0 < \Omega \ll 1$ as discussed above.
In this case, we may use $\nu \equiv \sqrt{\Omega}$
as a formal small parameter and develop methods of multiple scales and
homogenization techniques \cite{mtol1} in order to obtain analytical
predictions for the linear spectrum.
In particular, introducing the
fast (i.e., rapidly varying) variable $X=x/\nu$ and
rescaling the energy as $\varepsilon_m=E_m/\Omega$, the eigenvalue problem of Eq.~(\ref{eqn4}) is expressed as follows:
\begin{eqnarray}
\left(\nu^2 {\cal L}_{\rm H} -\nu \frac{\partial^2}{\partial x \partial X} +
{\cal L}_{\rm OL}\right) u_m= \nu^2 \varepsilon_m u_m,
\label{eqn5}
\end{eqnarray}
where
\begin{eqnarray}
{\cal L}_{\rm H} &=& -\frac{1}{2} \frac{\partial^2}{\partial x^2} +\frac{1}{2} x^2,
\label{eqn6}
\\
{\cal L}_{\rm OL} &=& -\frac{1}{2} \frac{\partial^2}{\partial X^2}
+ V_0 \cos(2 X)
\label{eqn7}
\end{eqnarray}
(note that $V_0/\Omega$ was treated as an ${\cal O}(1)$ parameter). Additionally, we consider
a formal series expansion (in $\nu$) for $u_m$ and $\varepsilon_m$
\cite{mtol1}, namely,
\begin{eqnarray}
u_m &=& u_0+ \nu u_1 + \nu^2 u_2 + \dots,
\label{eqn8}
\\
\varepsilon_m &=& \varepsilon_{-2} {\nu^{-2}} + \varepsilon_{-1} {\nu^{-1}} + \varepsilon_0
+ \varepsilon_1 \nu + \dots.
\label{eqn9}
\end{eqnarray}
To this end, substitution of this expansion in the eigenvalue problem of Eq.~(\ref{eqn5})
and use of the solvability conditions for the first three orders of the expansion
[i.e., ${\cal O}(1)$, ${\cal O}(\nu)$ and ${\cal O}(\nu^2)$] yields the following results for
the eigenvalue problem of the original operator. The
energy of the $m$-th mode
can be approximated by:
\begin{eqnarray}
E_{m} =-\frac{1}{4} V_0^2 + \left(1-\frac{1}{4} V_0^2 \right)
\Omega \left(m+\frac{1}{2}\right),
\label{eqn10}
\end{eqnarray}
%
while the corresponding eigenfunction
is given by:
\begin{eqnarray}
u_m(x)&=&c_m H_m \left(\frac{x}{\sqrt{1-\frac{V_0^2}{4}}}\right)
             \exp\left(-\frac{x^2}{2-\frac{V_0^2}{2}}\right)
\nonumber\\[1.0ex]
&\times& \frac{1}{\sqrt{\pi}} \left[1-\frac{V_0}{2} \cos \left(\frac{2 x}{\sqrt{\Omega}}
\right) \right],
\label{eqn11}
\end{eqnarray}
where $c_m=(2^m m! \sqrt{\pi})^{-(1/2)}$ is the normalization factor and
$H_m(x)=e^{x^2} (-1)^m (d^m/dx^m) e^{-x^2}$ are the Hermite polynomials.

Combining the results in the two orthogonal directions $x$ and $y$
yields a total energy eigenvalue
%
\begin{eqnarray}
E_{m,n}=-\frac{1}{2} V_0^2 + \left(1-\frac{1}{4} V_0^2 \right)
\Omega \left(n+ m + 1\right),
\label{eqn12}
\end{eqnarray}
and a corresponding eigenfunction which up to normalization factors
can be written as:
\begin{eqnarray}
u_{m,n}(x,y) &\propto &
\nonumber
 H_m \left(\frac{x}{\sqrt{1-\frac{V_0^2}{4}}}\right)
 \left[1-\frac{V_0}{2} \cos \left(\frac{2 x}{\Omega^{1/2}}\right) \right]\\[1.0ex]
&\times& H_n \left(\frac{y}{\sqrt{1-\frac{V_0^2}{4}}}\right)
\nonumber
 \left[1-\frac{V_0}{2} \cos \left(\frac{2 y}{\Omega^{1/2}}\right) \right]\\[1.0ex]
&\times& \exp \left(-\frac{r^2}{2-\frac{V_0^2}{2}}\right)
.
\label{eqn13}
\end{eqnarray}
The
particularly appealing feature of this expression is that it
allows us to combine various (ground or excited) states
in the $x$ direction with different ones along the $y$ direction.
In this work we will focus on the
simplest possible combinations of $m,n \in \{0,1,2\}$ and
examine the various states generated by the combinations of
these, which we hereafter denote $|m,n\rangle$.
In particular, below we will focus on
the ground state $|0,0\rangle$
and the excited states $|1,0\rangle$, $|1,0\rangle + |0,1\rangle$, $|1,1\rangle$, and
$|2,0\rangle$. Our aim is to investigate which of these states persist in the nonlinear regime
in both cases of attractive and repulsive interatomic interactions,
and study the stability of these states in detail. Notice that
we will only illustrate (by an appropriate
curve in the numerical results that follow) the linear
limit, $E_{m,n}$ of Eq.~(\ref{eqn12}) above, for the case of attractive interactions.


\section{Numerical
Results
}

\subsection{The non-interacting limit
}

We start by examining the validity of the above analytical predictions
concerning the linear limit of the problem, namely, Eqs.~(\ref{eqn12}) and (\ref{eqn13}).
The results are summarized
in Fig.~\ref{vfig0}.
Panel (a) shows the
1D harmonic oscillator
energy spectrum (for $\Omega=0.1$) and compares it with the
energy spectrum obtained
from numerical and approximate theoretical [see Eq.~(\ref{eqn10})] solutions
of the combined harmonic and optical lattice potential for $V_0=0.3$.
Panel (b) offers a similar comparison but for a larger lattice depth, namely
$V_0=0.5$. One can clearly see that the theoretical calculation
approximates very accurately the numerical results for the first
few states (i.e., $n=0,1,2$), while deviations become more significant
for higher-order excited states. Panels (c), (d) and (e) show the zeroth,
first and second eigenfunction of the purely harmonic potential
(thick solid line), as well as of the harmonic trap and optical lattice
potential as found numerically (thin solid line) and analytically
[given by
Eq.~(\ref{eqn11})]. The green dash-dotted
line represents the form of the combined potential.
We once again note the good agreement of our analytical results above
in comparison with the full numerical computation.


\subsection{The approach for the interacting case
}
%
%
We now consider the full nonlinear
problem of Eq.~(\ref{eqn1}).
In the following, we will monitor the two-dimensional, $V_0 \times \mu$, parameter space
(where $\mu$ is the chemical potential of the relevant modes) 
for nonlinear excitations that stem from the linear spectrum of the
problem. 
We perform the relevant analysis first in the case of attractive interatomic interactions
and then in the case of repulsive ones. Notice that the parameter $\Omega$ will be fixed 
to a relatively large value, namely $\Omega =0.1$; this is done for convenience in our numerical simulations 
(such ``large'' values of $\Omega$ correspond to smaller condensates that can be analyzed numerically 
with relatively coarser spatial grids), but we have checked that our results remain qualitatively similar 
for smaller values of $\Omega$ (results not shown here). Nevertheless, it should be noted that even such a value 
of $\Omega$, together with the considered range of values of the other normalized parameters (chemical potential, 
number of atoms, lattice depth, etc -- see below) is still physically relevant. For example, our choice 
may realistically correspond to a $^{87}$Rb condensate containing $\sim 15,000$ atoms, confined by a harmonic 
potential with frequencies $\omega_z = 20\omega_{\perp} = 2\pi\times 240$Hz (so that $\omega_{\perp}=2\pi\times 12$ Hz) 
and an optical lattice potential with a periodicity $d \approx 3\mu$m. The recoil energy in this case is 
$E_{\rm rec}/h = 60$ Hz (so that $\omega_L = 2\pi\times 120$ Hz, giving $\Omega = 0.1$), and a lattice depth of 
$V_0 = 0.3$ corresponds to $\sim$1.2 $E_{\rm rec}$. To further set the scale for the simulations described below, 
such a BEC with repulsive interactions in the purely harmonic trap (where the optical lattice is not applied) would have 
a chemical potential of $\mu \sim 0.5$ in our dimensionless units. 

In the following sections, stationary solutions of the full nonlinear problem are sought in the form
\begin{equation}
\psi(x,y,t)= \exp(-i \mu_{m,n} t) u_{m,n}(x,y), 
\label{nlans}
\end{equation}
where $\mu_{m,n}$
(which is the nonlinear analog of the energy $E_{m,n}$ found in the non-interacting limit) represents the chemical potential.
Note that we will henceforth avoid using subscripts $m$ and $n$ when the meaning is clear, in the interest of
avoiding notational clutter.

\subsection{Attractive interatomic interactions}

\subsubsection{Existence and Stability}

\begin{figure}
\includegraphics[width=8cm]{\rootfig 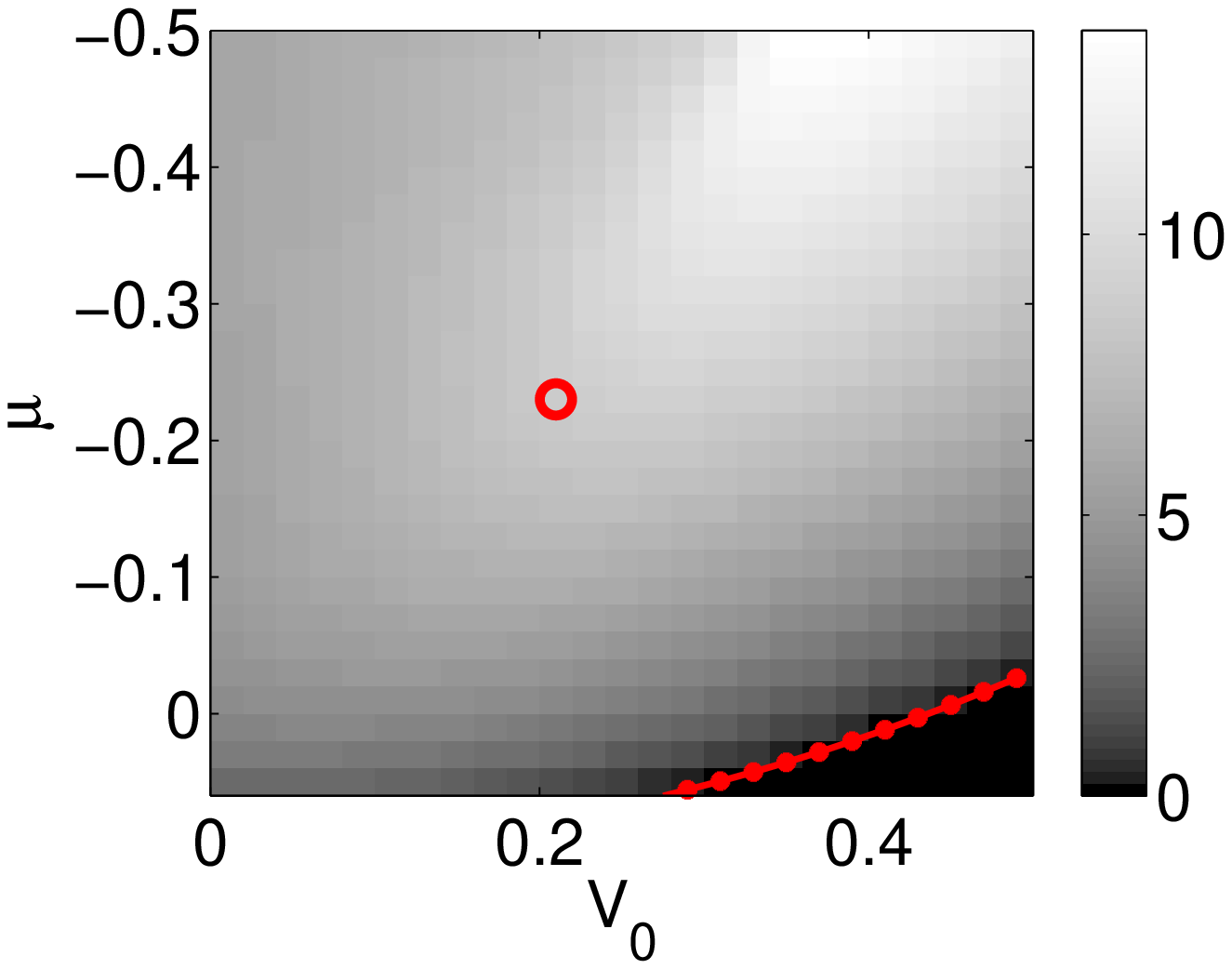}
\includegraphics[width=8cm]{\rootfig 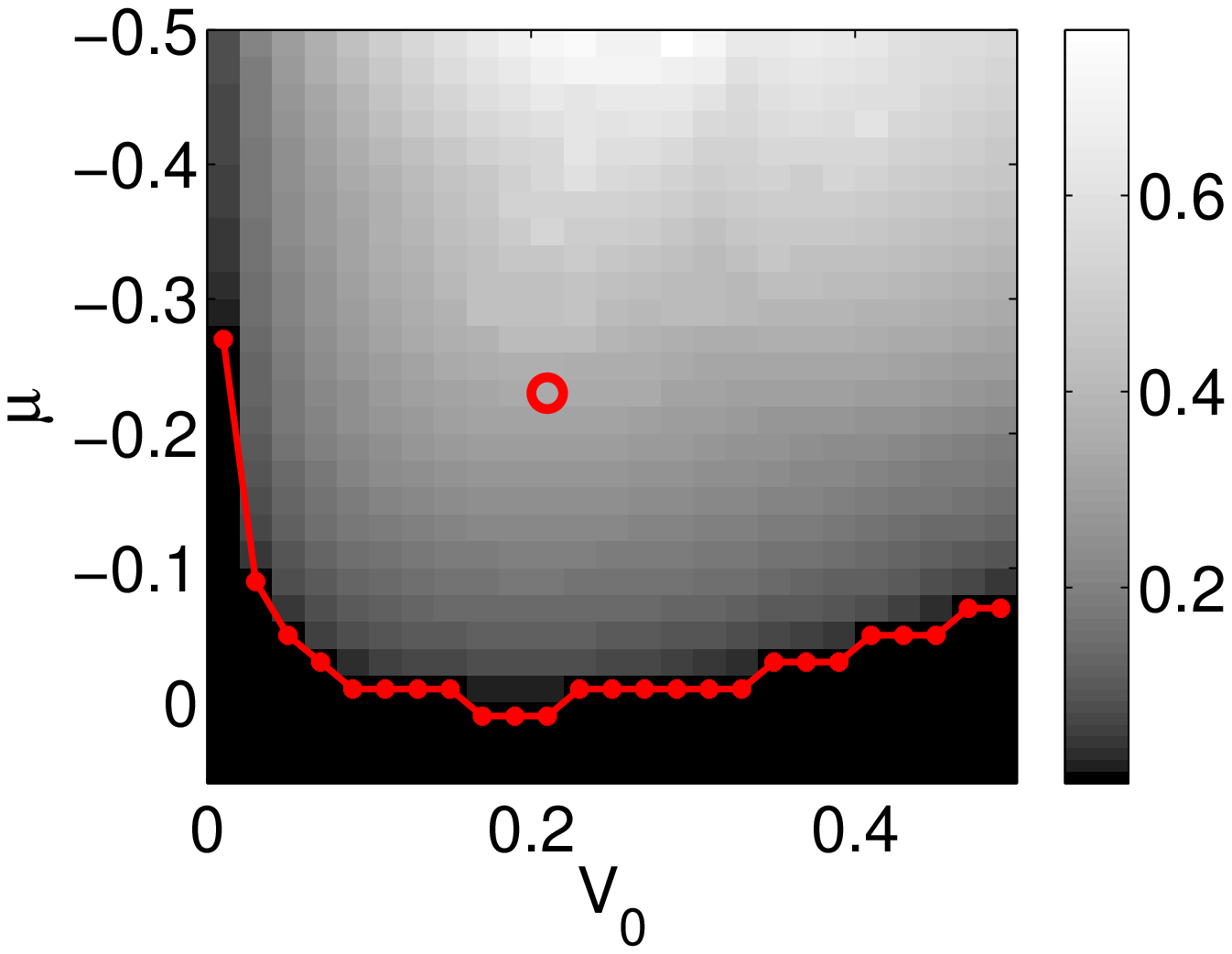}
\\
\includegraphics[width=8cm]{\rootfig 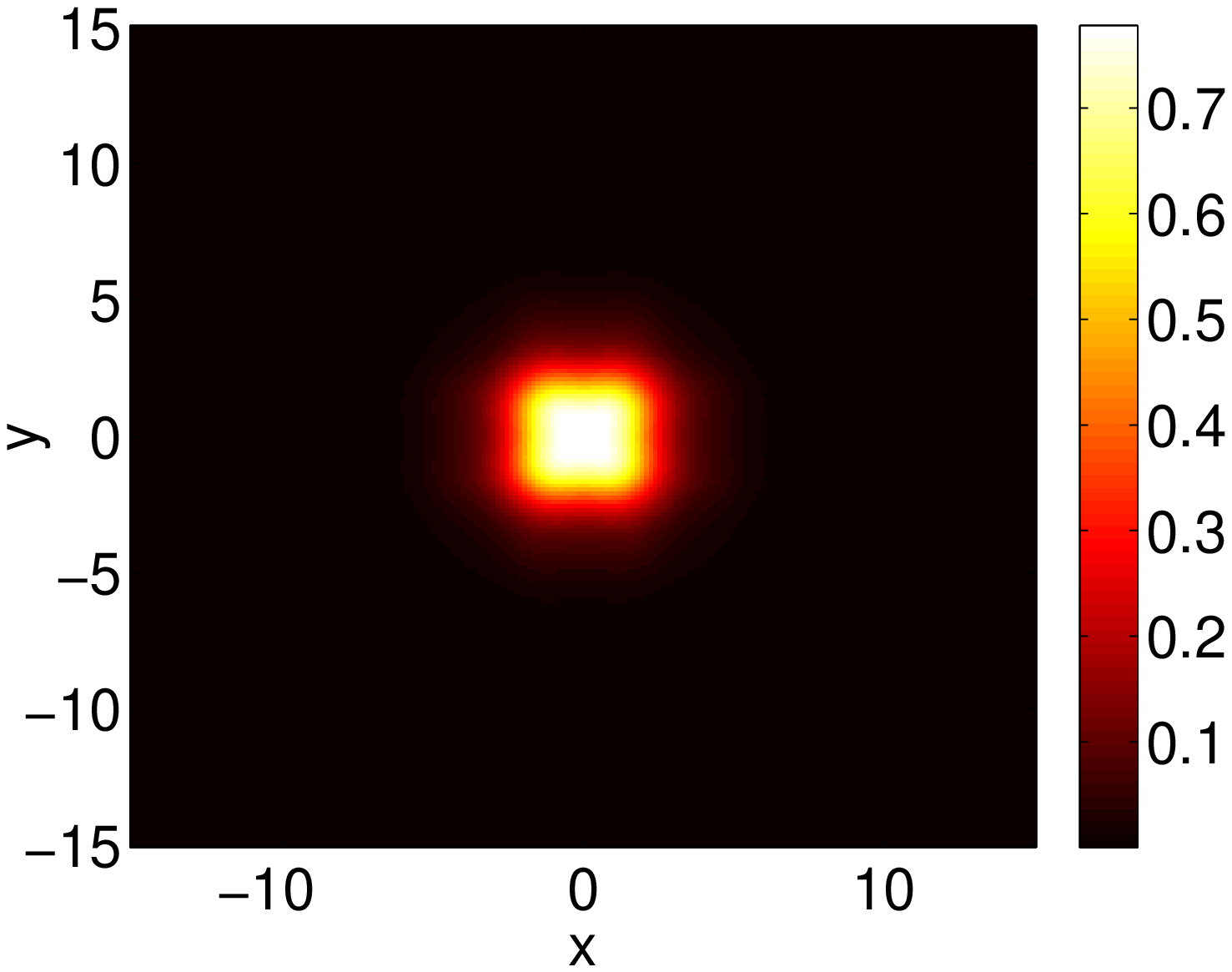}
\includegraphics[width=7.65cm,height=6.75cm]{\rootfig 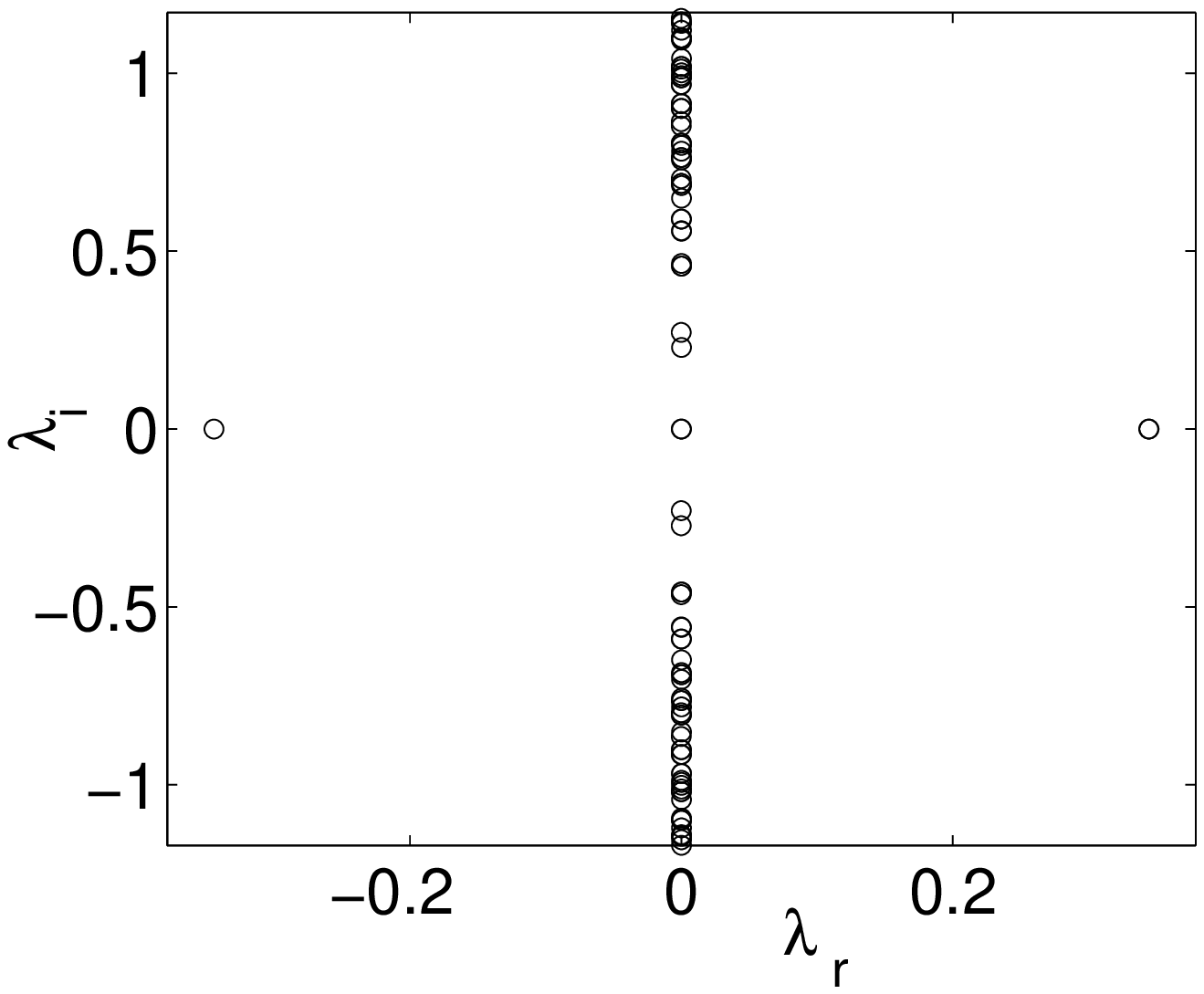}
\\
\caption{(Color Online) 
The ground state in the case of attractive interatomic interactions. The top left panel shows the
rescaled number of particles $N(V_0,\mu)=\int |u|^2 dx dy$ as a function of the amplitude of the optical lattice $V_0$ and the
chemical potential $\mu$;
the red line represents
the approximation of the energy eigenvalue $E(V_0)$ of the linear
problem given by Eq.~(\ref{eqn12}). For each $V_0$, the number
of atoms, $N_{V_0}(\mu)$ approaches zero in the limit
$\mu \rightarrow E(V_0)$. The top right panel shows the stability domain
$S(V_0,\mu)={\rm max}(\lambda_r)$;
the red line here corresponds to the stability window $S<10^{-4}$.
It is clear that for each $V_0$, there is a window of values of $\mu$
for which $S_{V_0}(\mu)<10^{-4}$. This is expected, since the
attractive nature of the interatomic interactions
leads to blowup above a critical value of $N$.
The bottom left and right panels depict, respectively,
a contour plot of an unstable solution $u$
in the $(x,y)$ plane
and its corresponding spectral plane $(\lambda_r,\lambda_i)$
[for $(V_0,\mu)=(0.21,-0.23)$ corresponding to the parameter
value depicted by the red circle in the panels of the top row]. 
The bottom-left colorbar represents atomic density.  In the bottom-right plot, the presence of
real eigenvalue pairs denotes instability (its growth rate is given by the
magnitude of the real part), while the imaginary pairs indicate
the frequencies of oscillatory modes.}
\label{vfig6}
\end{figure}

We begin by looking at the case of attractive interatomic interactions.
The most fundamental solutions are those belonging to the $|0,0\rangle$ branch,
which represents the ground state of the system and
is shown in Fig.~\ref{vfig6}.
The top left panel of this
figure shows the diagnostic that we will typically use to follow the
$V_0 \times \mu$ surface, namely the rescaled number of particles
$N(V_0,\mu)=\int |u_{m,n}|^2 dx dy$
as a function of the chemical potential $\mu$ introduced above,
and the
optical lattice depth $V_0$. Essentially, the grayscale values in this plot correspond to 
the number of atoms needed to obtain a particular chemical potential with a particular lattice depth; 
lighter values correspond to more atoms.
As $N$ becomes smaller, through the appropriate variation
of $\mu$, we approach the linear limit so one expects the
solution to degenerate to the corresponding linear eigenmode
(for $\mu$ tending to the corresponding eigenvalue of the linear problem).
Figure \ref{vfig6} shows our observations for this fundamental branch, which
seems to disappear for
$\mu_{m,n}(V_0)\approx E_{m,n}(V_0)=-\frac{1}{2} V_0^2 + \left(1-\frac{1}{4} V_0^2 \right)
\Omega \left(n+ m + 1\right)$.
%
Naturally the surface degenerates to its linear
limit for $\mu_{m,n}(V_0) \rightarrow E_{m,n}(V_0)$ and
the number of particles is a decreasing function of $\mu$,
contrary to what is the case in the repulsive
nonlinearity (see below).
This is a well-known difference between the two cases
that has been documented elsewhere (see, e.g., Refs.~\cite{konotop1,mtol1}).
It is relevant to indicate here that although there is a direct
correspondence between the atom number $N$ and the chemical potential
$\mu$, we opt to illustrate our results as a function of $\mu$ (and $V_0$),
since the latter is the relevant parameter entering the mathematical
setup of the problem and it is the one for which we developed an 
analytical prediction in the linear limit. Nevertheless, we
also give $N(V_0,\mu)$, so as to associate in each case the relevant
chemical potential (and lattice strength) for a given configuration with the
corresponding physical quantity, i.e., the atom number.

\begin{figure}
\includegraphics[width=8cm]{\rootfig 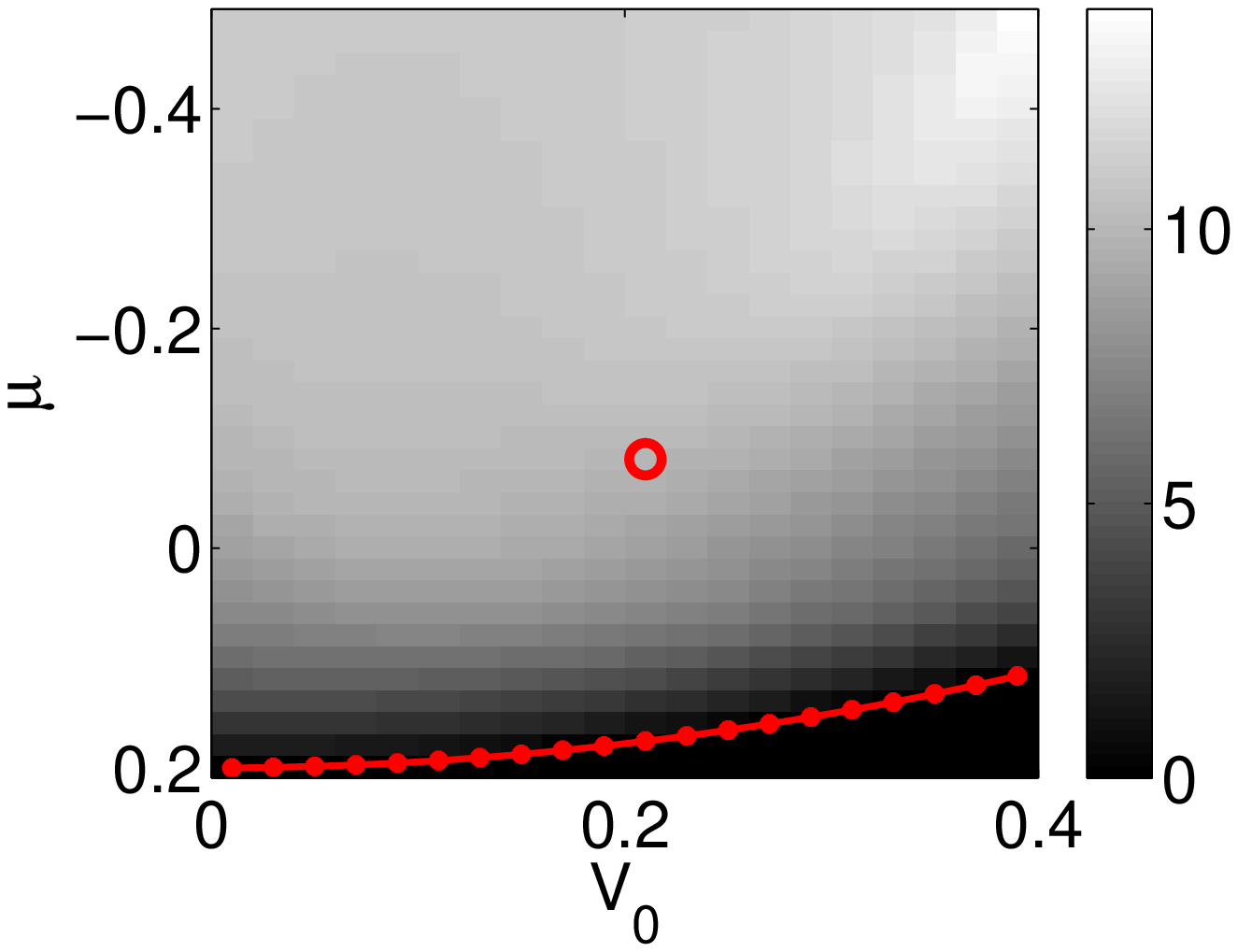}
\includegraphics[width=8cm]{\rootfig 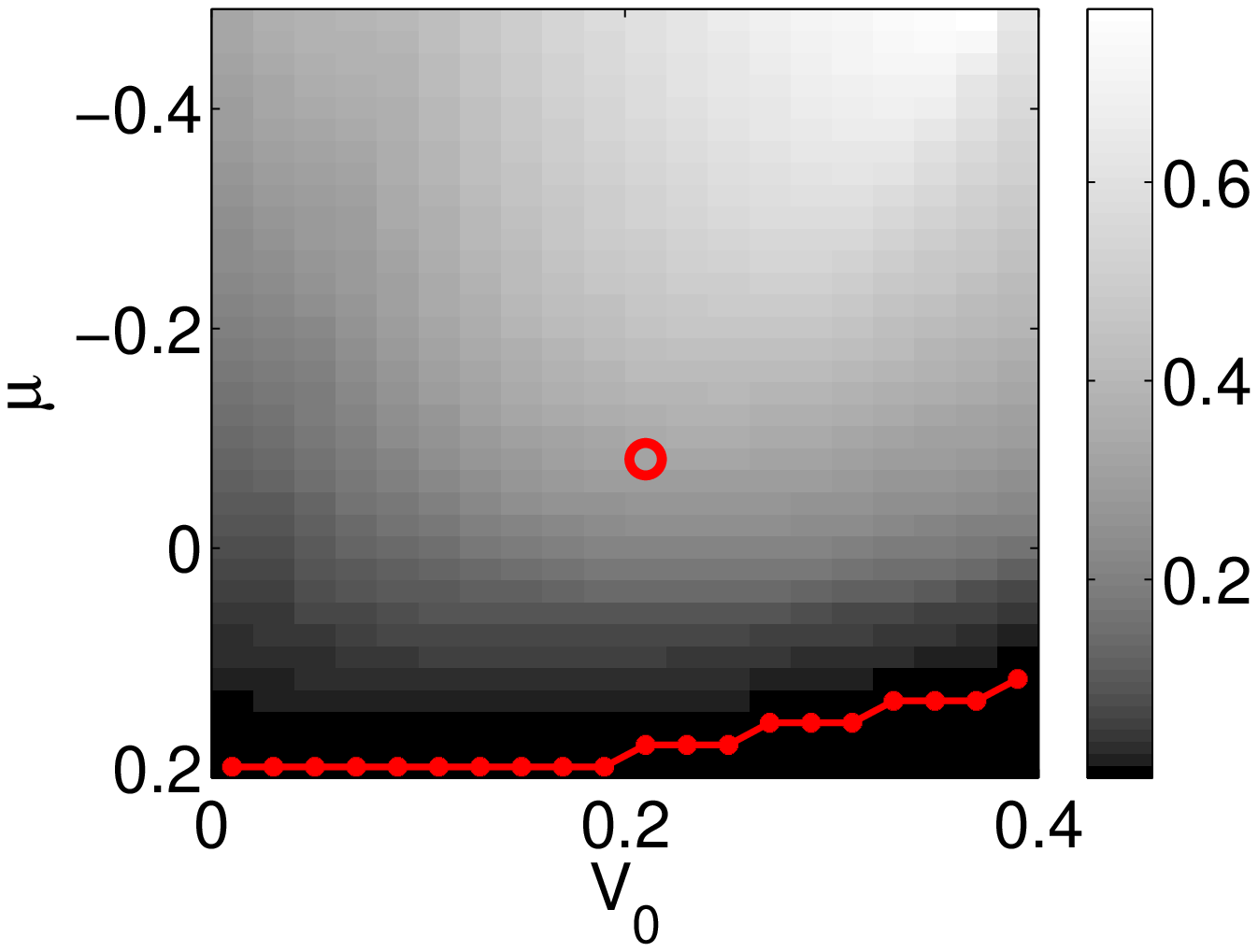}
\\
\includegraphics[width=8cm]{\rootfig 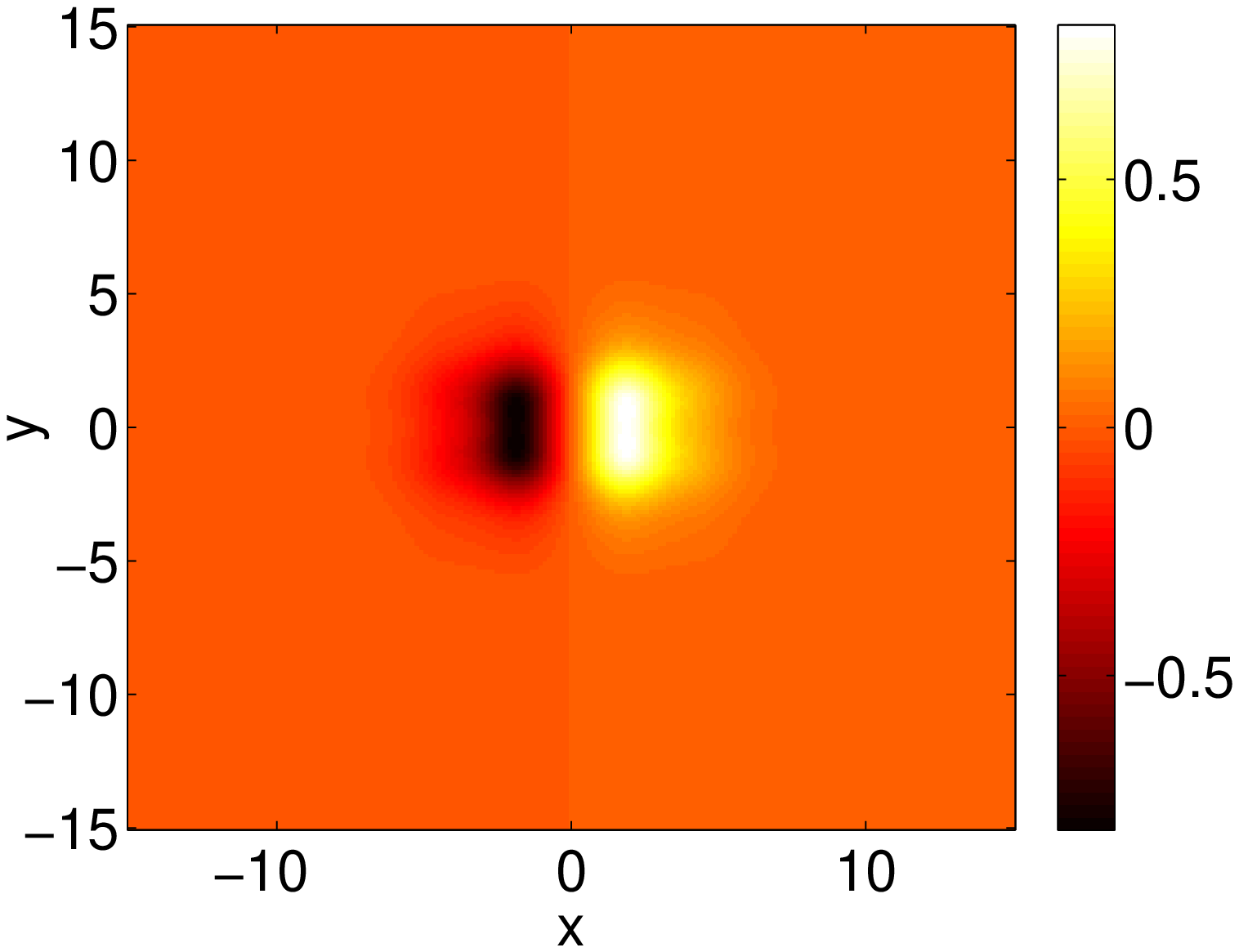}
~~\includegraphics[width=7.65cm,height=6.65cm]{\rootfig 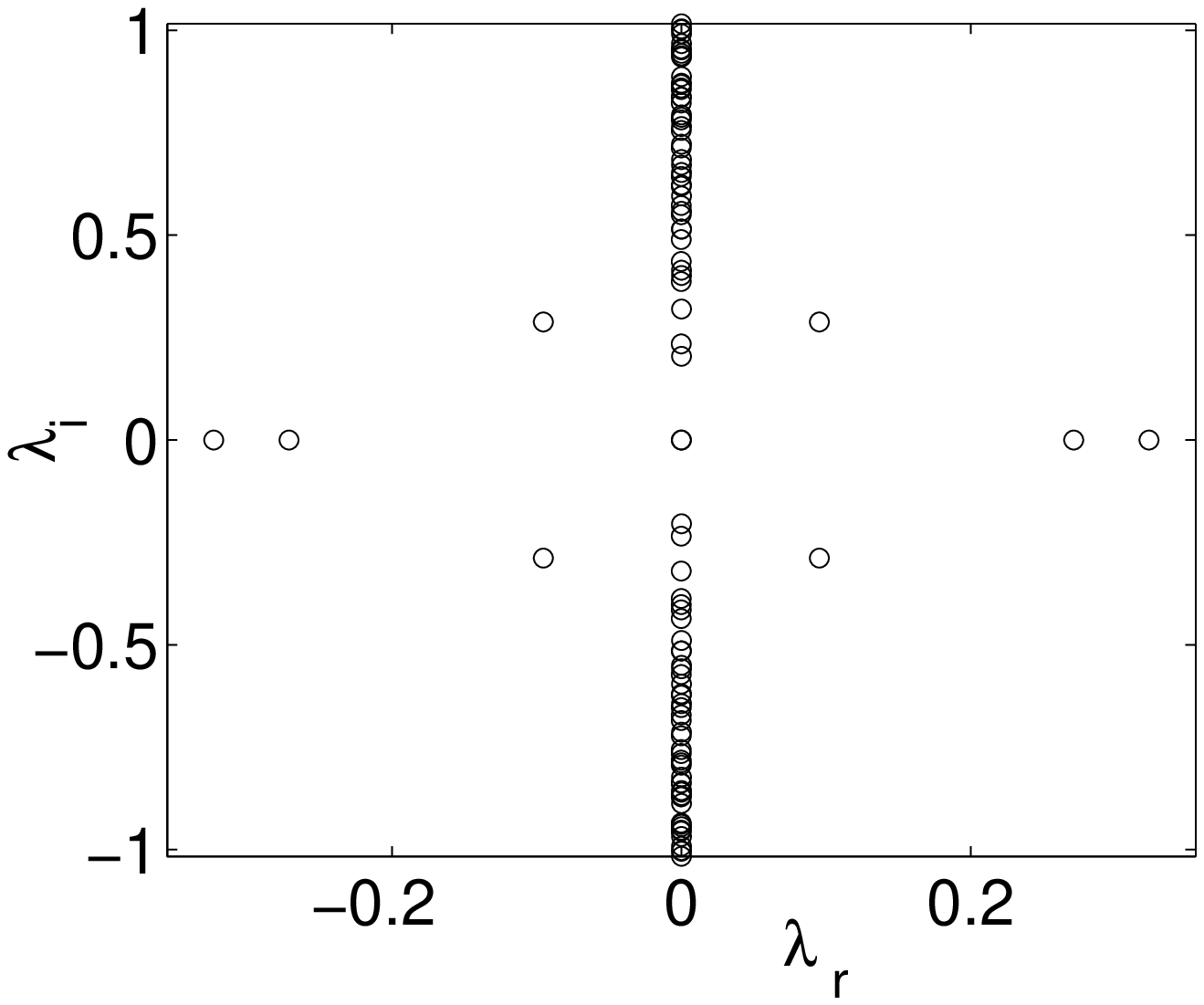}~~~

\caption{(Color Online) Similar to Fig.~\ref{vfig6} but for the
case of the $|1,0\rangle$ state for attractive interactions.
The results shown in the bottom row correspond to parameter values 
$(V_0,\mu)=(0.21,-0.081)$ (see red circle in top panels).}
\label{vfig7}
\end{figure}

\begin{figure}[tbp]
\includegraphics[width=8cm]{\rootfig 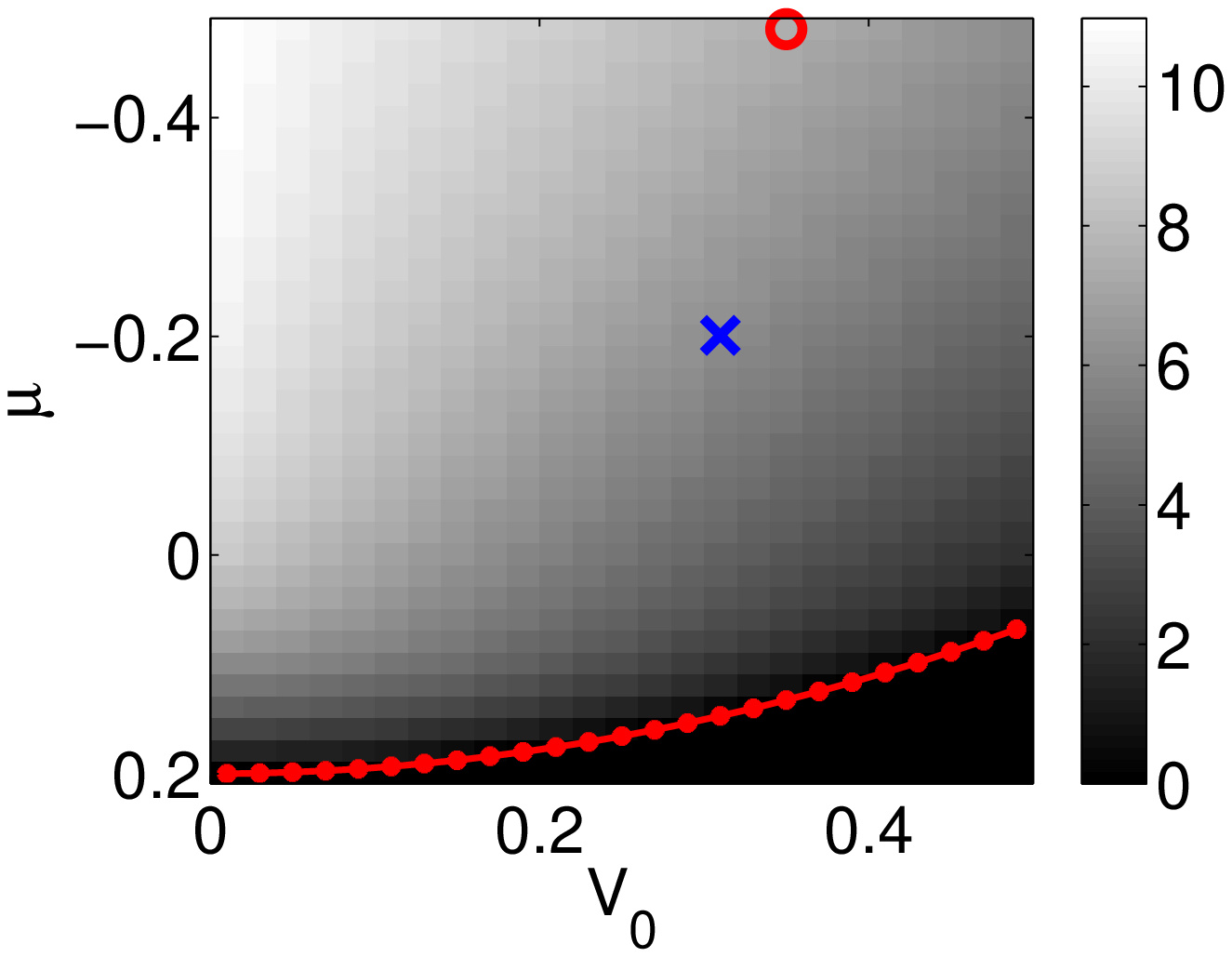}
\includegraphics[width=8cm]{\rootfig 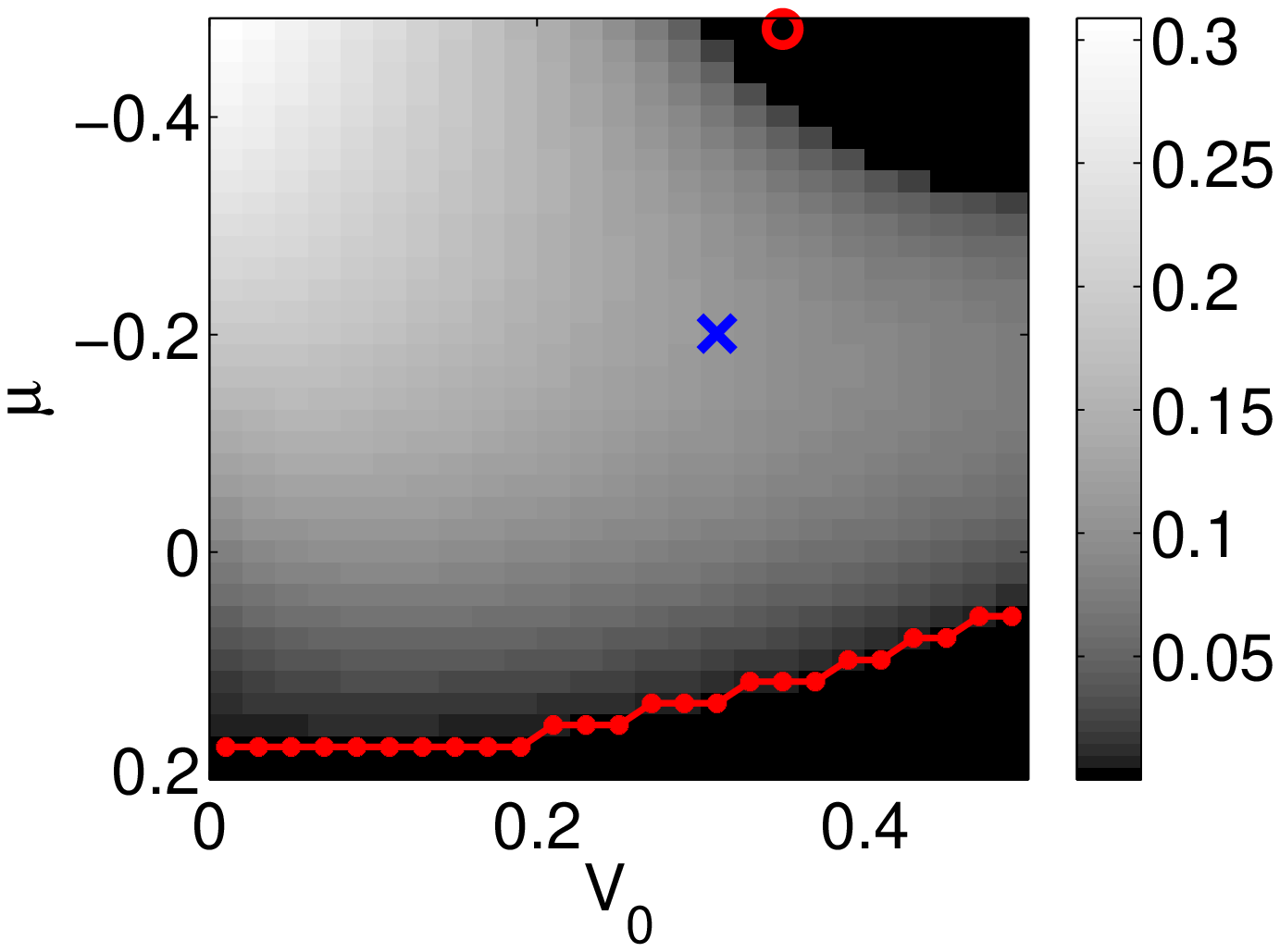}
\\
\includegraphics[width=8cm]{\rootfig 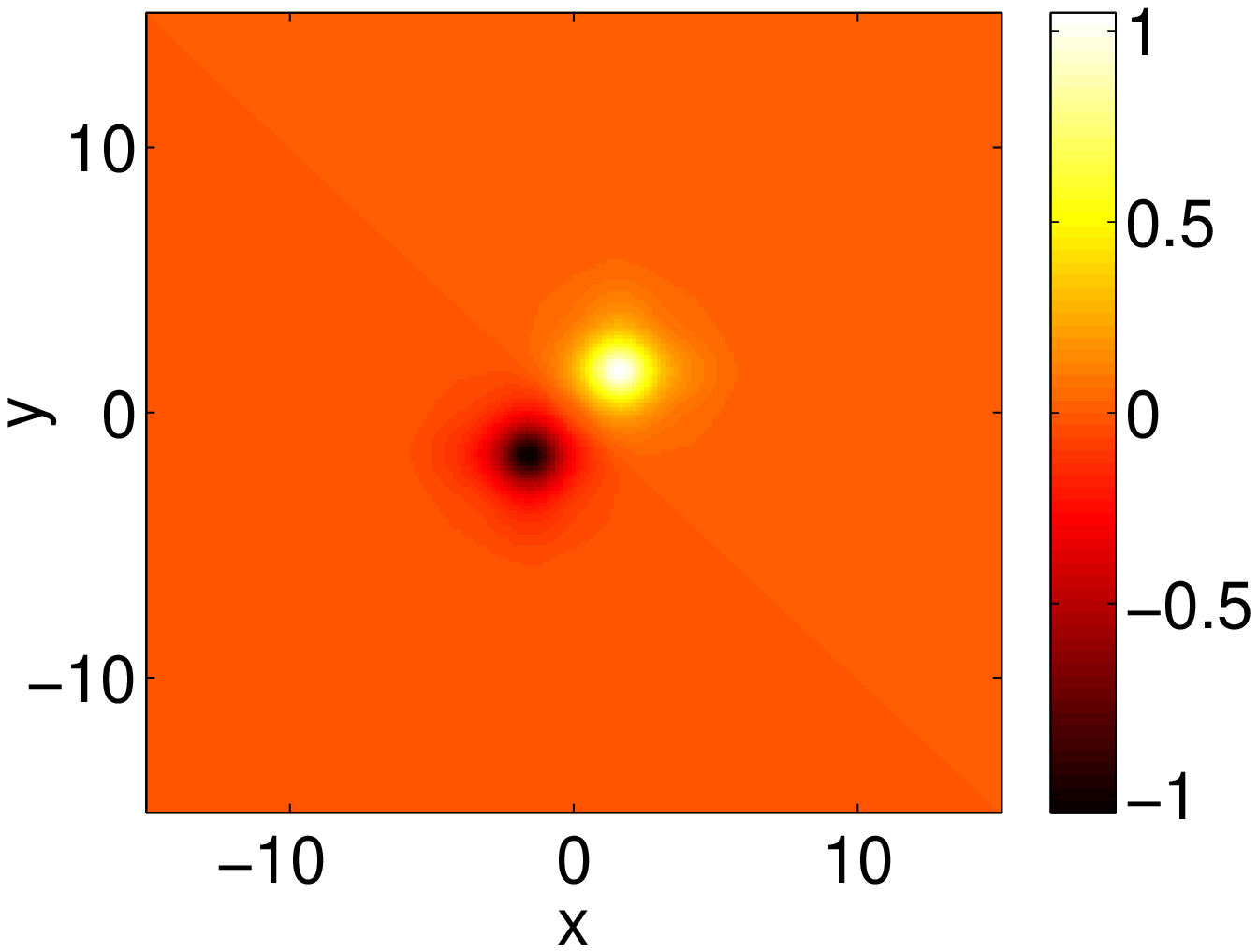}
~~\includegraphics[width=7.65cm,height=6.75cm]{\rootfig 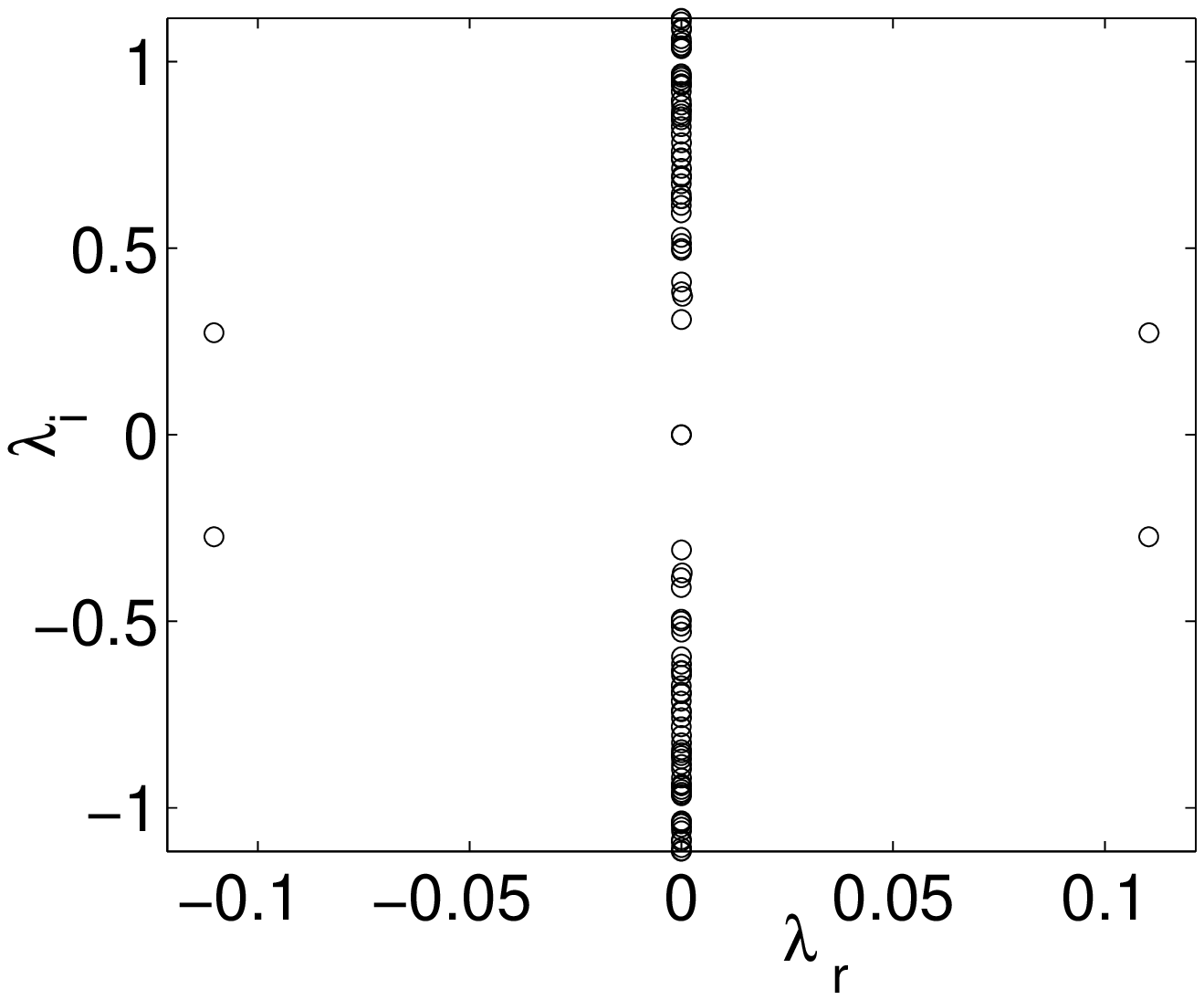}~~~
\\
\includegraphics[width=8cm]{\rootfig 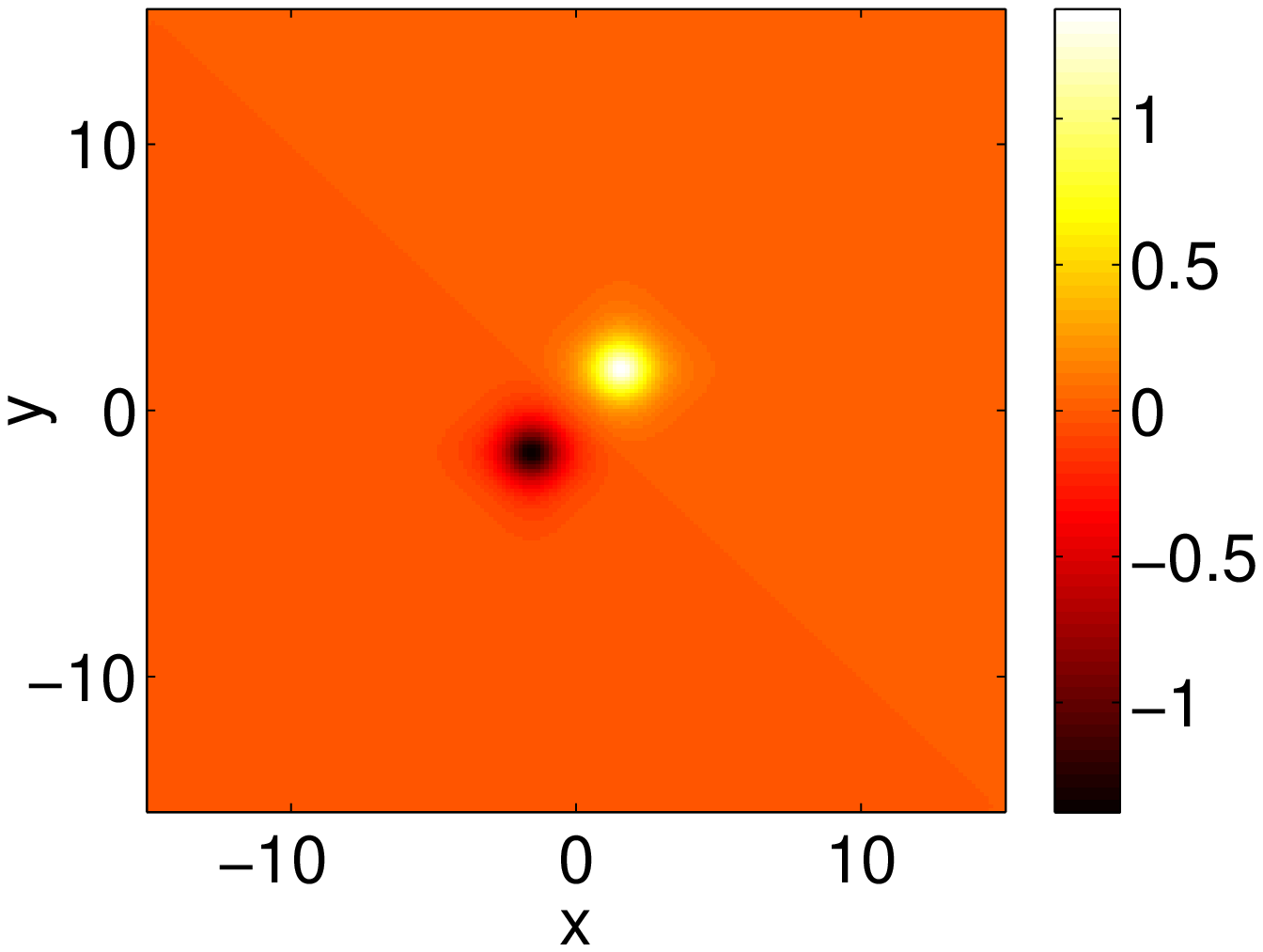}
~~\includegraphics[width=7.65cm,height=6.75cm]{\rootfig 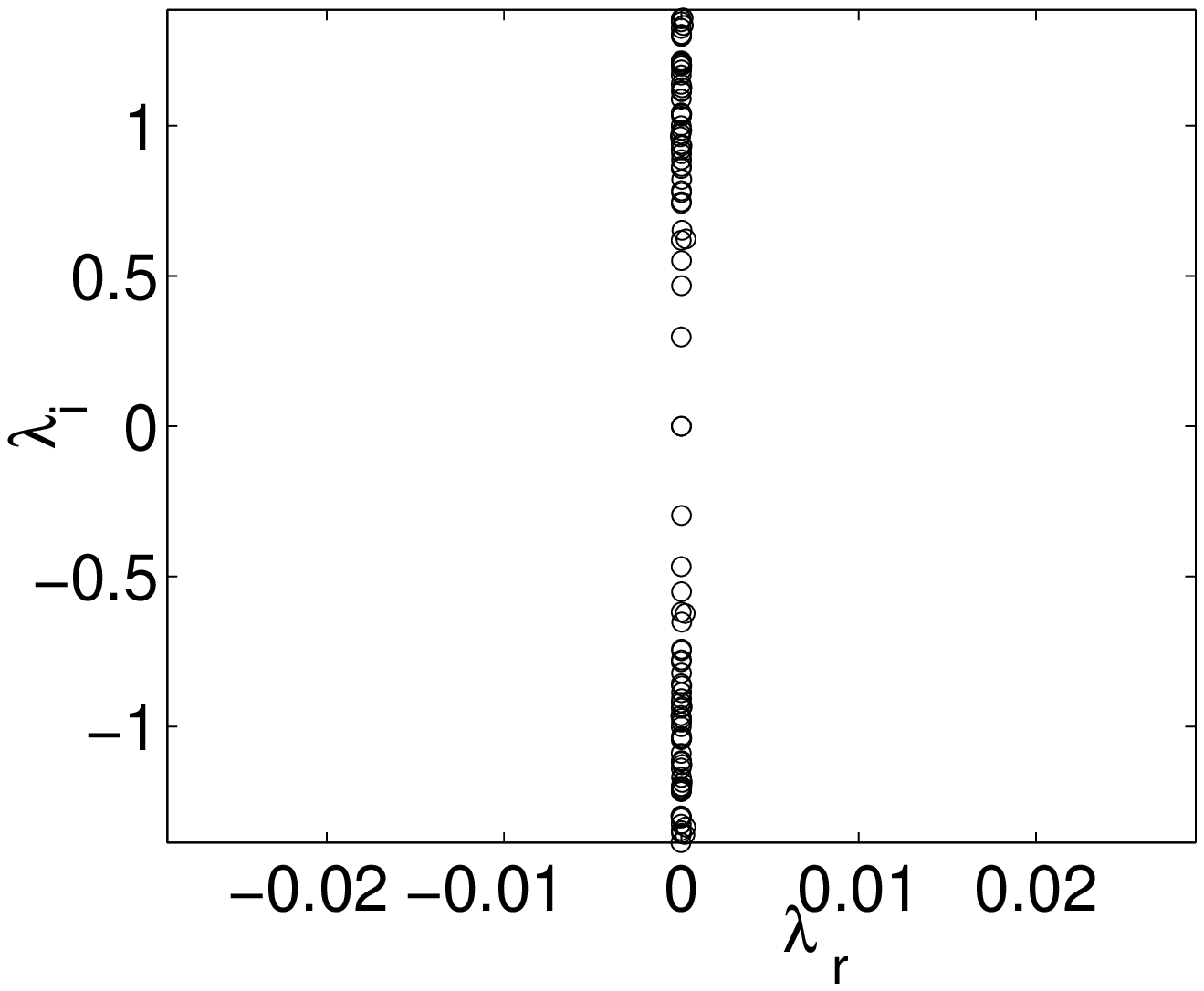}~~~
\vspace{-0.2cm}
\caption{(Color Online) The state
$|1,0\rangle+|0,1\rangle$ for attractive interatomic interactions.
The top row is similar to Fig.~\ref{vfig0}, 
the middle row is for parameter values $(V_0,\mu)=(0.31,-0.201)$
(see blue cross in top panels), and the bottom one
is for $(0.35,-0.481)$ (see red circle in top panels).}
\label{vfig10}
\end{figure}

It is important to highlight here that that the numerical
computations have been performed in a domain of size $201 \times 201$,
with $\Delta x=\Delta y=0.15$.
The size of the grid weakly affects
the value of the respective eigenvalues. In particular, the energy
eigenvalues in the non-interacting limit of the one-dimensional
decoupled eigenvalue problem (with $V_0=0.5$) for the
$n=0$ and $n=1$ mode that we report below
are $E_0\approx-0.01453$ and $E_1\approx 0.07639$ 
respectively
for this coarser domain, while 
for the a domain of size $3001$ with $\Delta x=0.01$ 
they become $ -0.01408$ and $0.07692$ respectively (computations show the 
discrepancy is uniformly smaller for smaller values of $V_0$).
This feature will weakly affect the quantitative aspects of the
results that follow
(in essence, providing an error bar in the
estimates below of $\Delta \mu \approx 5 \times 10^{-4}$ and similar for the
eigenvalues $\lambda$ introduced below), but is essentially
necessary, given the limitations of standard eigenvalue solvers
for such big Jacobian eigenvalue problems.


The linear stability of the solutions is analyzed by using the following standard ansatz for the perturbation,
\begin{eqnarray}
\psi
=e^{-i \mu t} \left[u(x,y) + \left(a(x,y) e^{\lambda t}
+ b^{\star}(x,y) e^{\lambda^{\star} t} \right) \right], 
\label{eqn14}
\end{eqnarray}
where $\lambda=\lambda_r + i \lambda_i$ is the generally complex eigenvalue 
(subscripts {\it r} and {\it i} denote, respectively, the real and imaginary parts of $\lambda$) 
of the ensuing 
Bogoliubov-de Gennes equations \cite{book1,book2,rmp,ourbook},  
and $(a,b)^T$ is the corresponding eigenvector. 
The real part $\lambda_r$ of the eigenvalue 
then determines the growth rate of the potential
instability of the solution, since for
positive real values the perturbation will grow exponentially
in time. The imaginary part $\lambda_i$ denotes the oscillatory part (frequency) of the relevant eigenmode.
The top right of Fig.~\ref{vfig6} depicts the stability domain, denoted 
by $S(V_0,\mu)={\rm max}_{\lambda}(\lambda_r)$, in terms
of the maximum real part of all eigenvalues as a function of the
lattice depth $V_0$ and the chemical potential $\mu$.
This quantity $S$ is a particularly important one from a physical point of
view since if a perturbation to the system has initially a projection
$p(0)$ onto the most unstable eigenmode of the linearization, then
this perturbation will grow in time according to 
$||p(t)||=||p(0)|| \exp(S t)$ while the solution is sufficiently close 
in space to the original profile. Hence, given the initial condition profile
[which determines $p(0)$] and $S$, we can determine for an unstable
configuration the time $t$ required for the instability to manifest 
itself, i.e., for $p(t)$ to become of the order of the solution 
amplitude.

\begin{figure}[tbp]
\includegraphics[width=8cm]{\rootfig 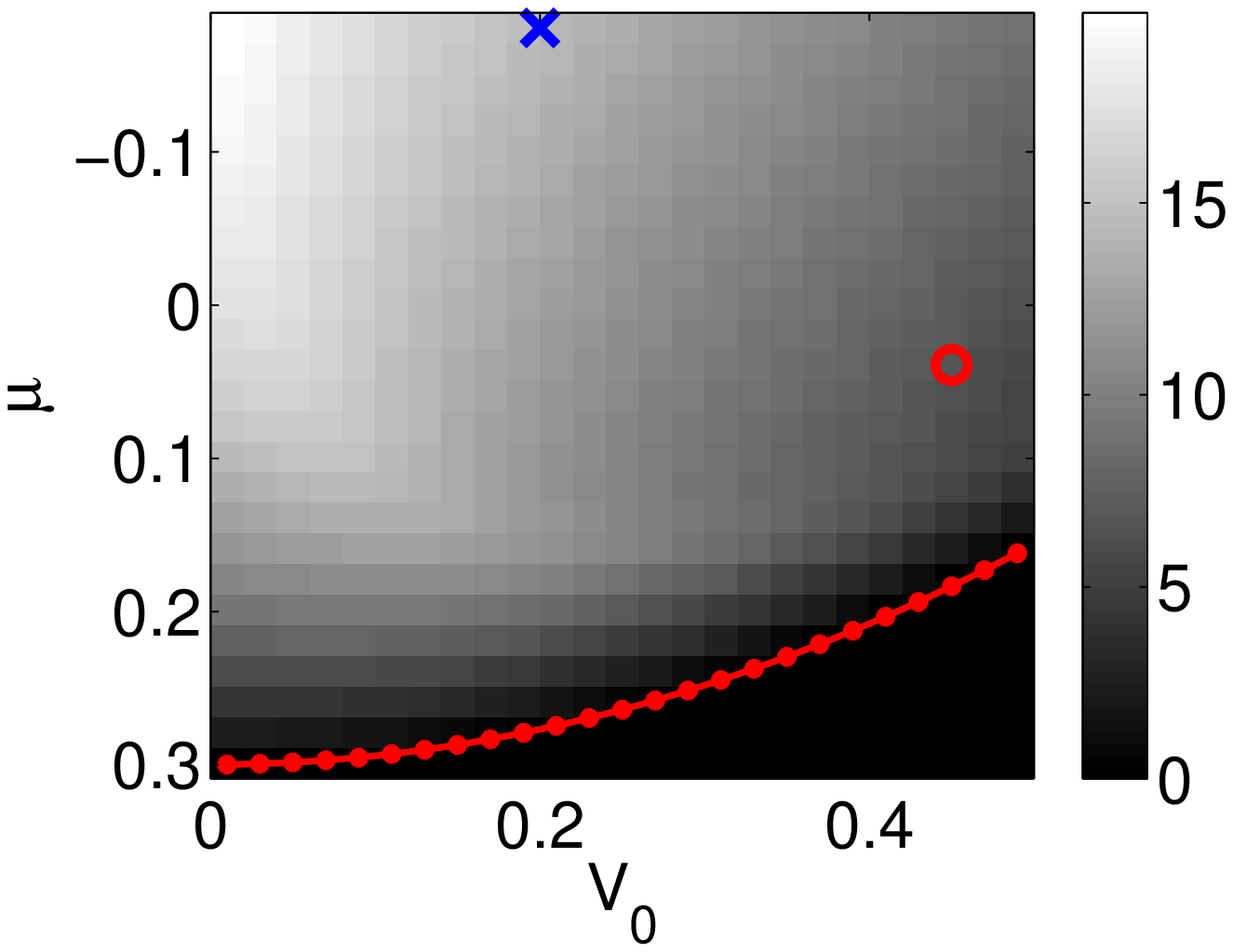}
\includegraphics[width=8cm]{\rootfig 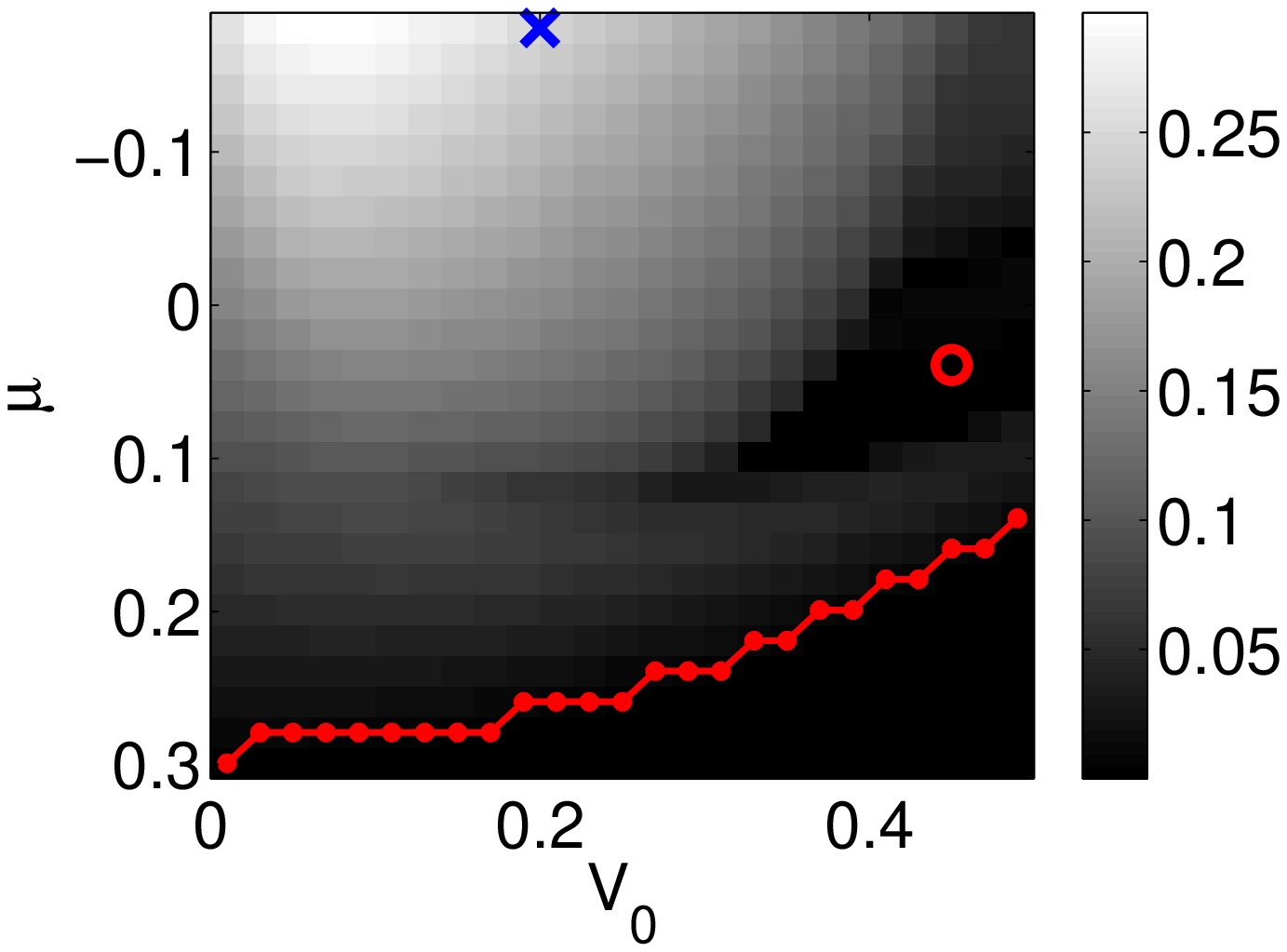}
\\
\includegraphics[width=8cm]{\rootfig 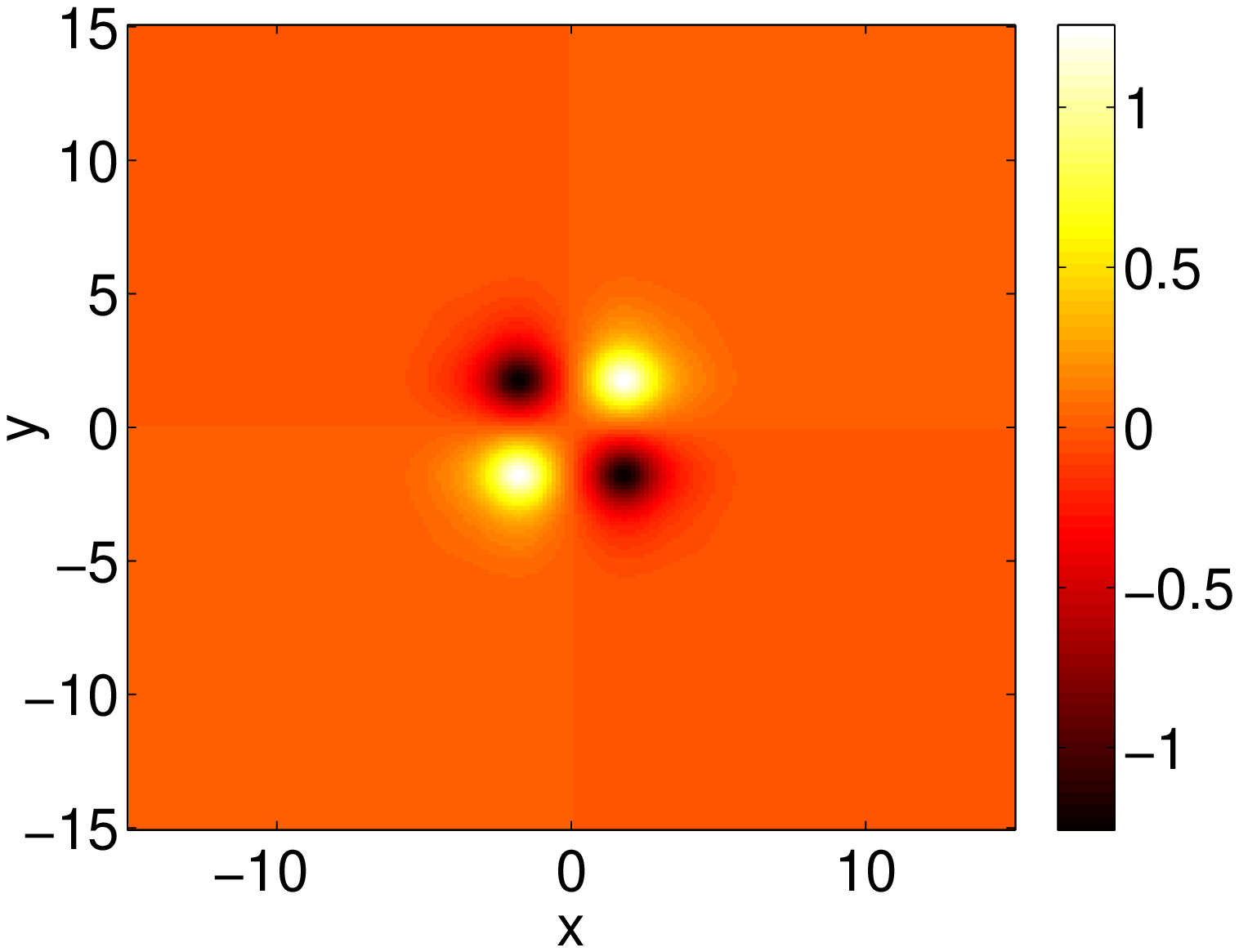}
~~\includegraphics[width=7.65cm,height=6.75cm]{\rootfig 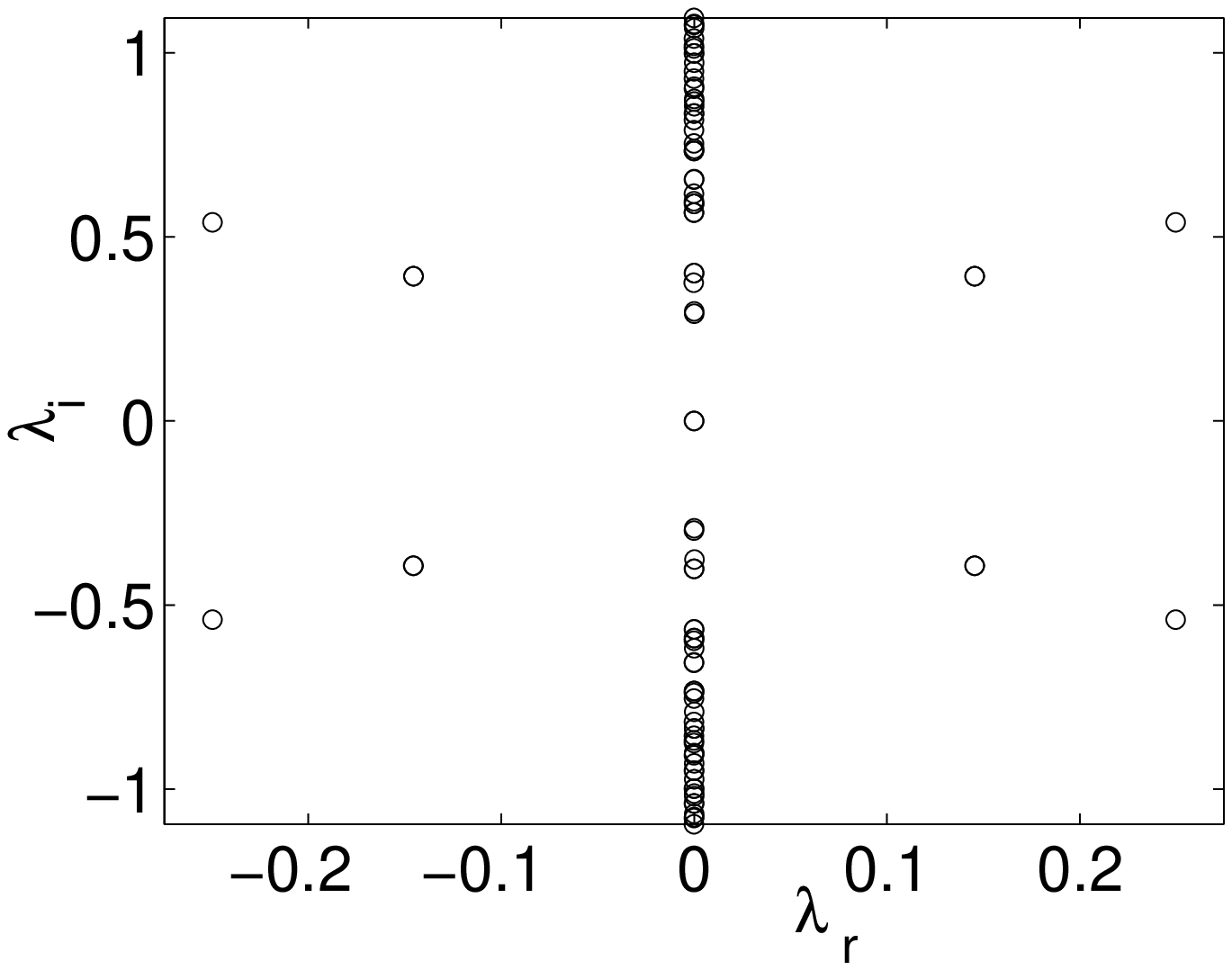}~~~
\\
\includegraphics[width=8cm]{\rootfig 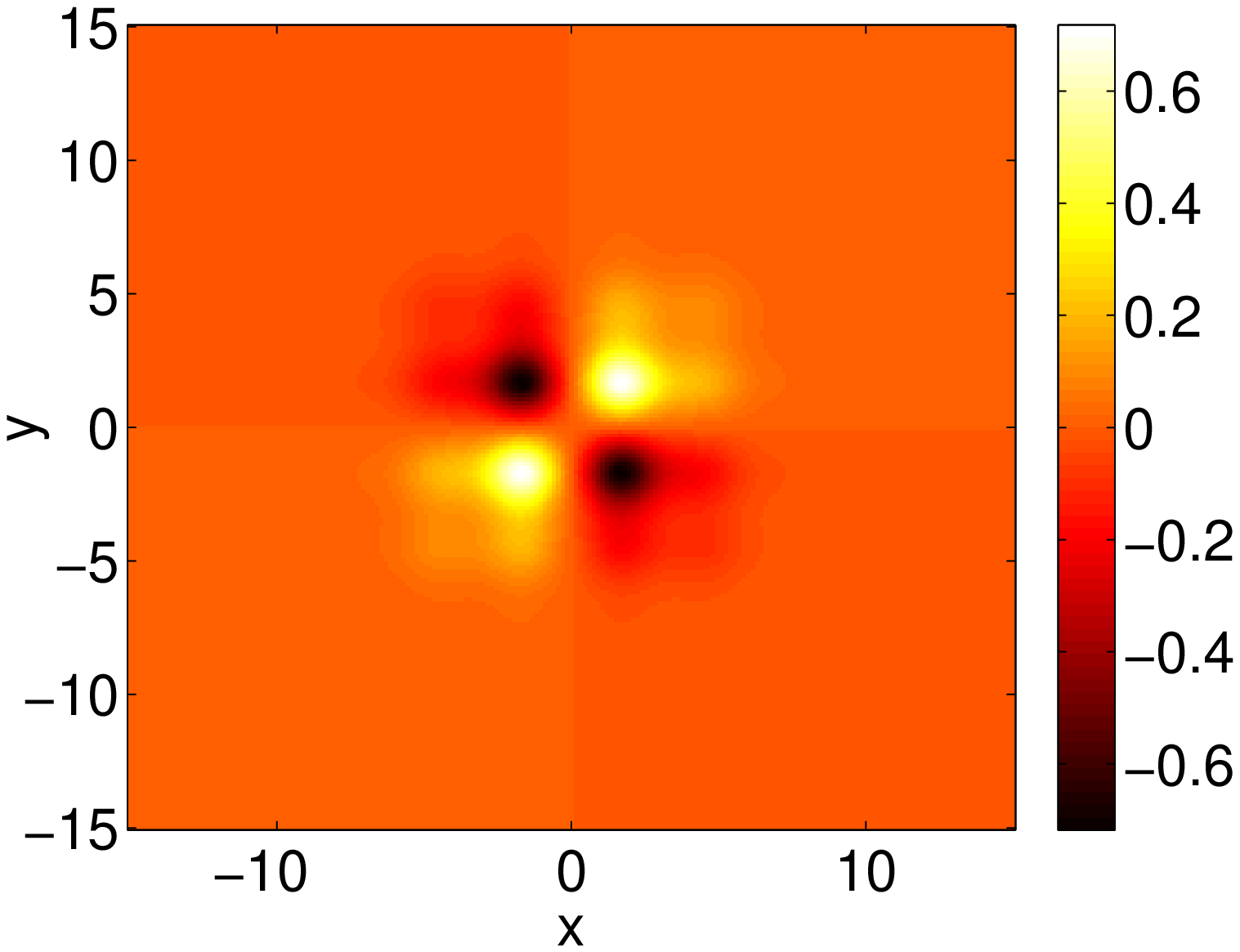}
~~\includegraphics[width=7.65cm,height=6.75cm]{\rootfig 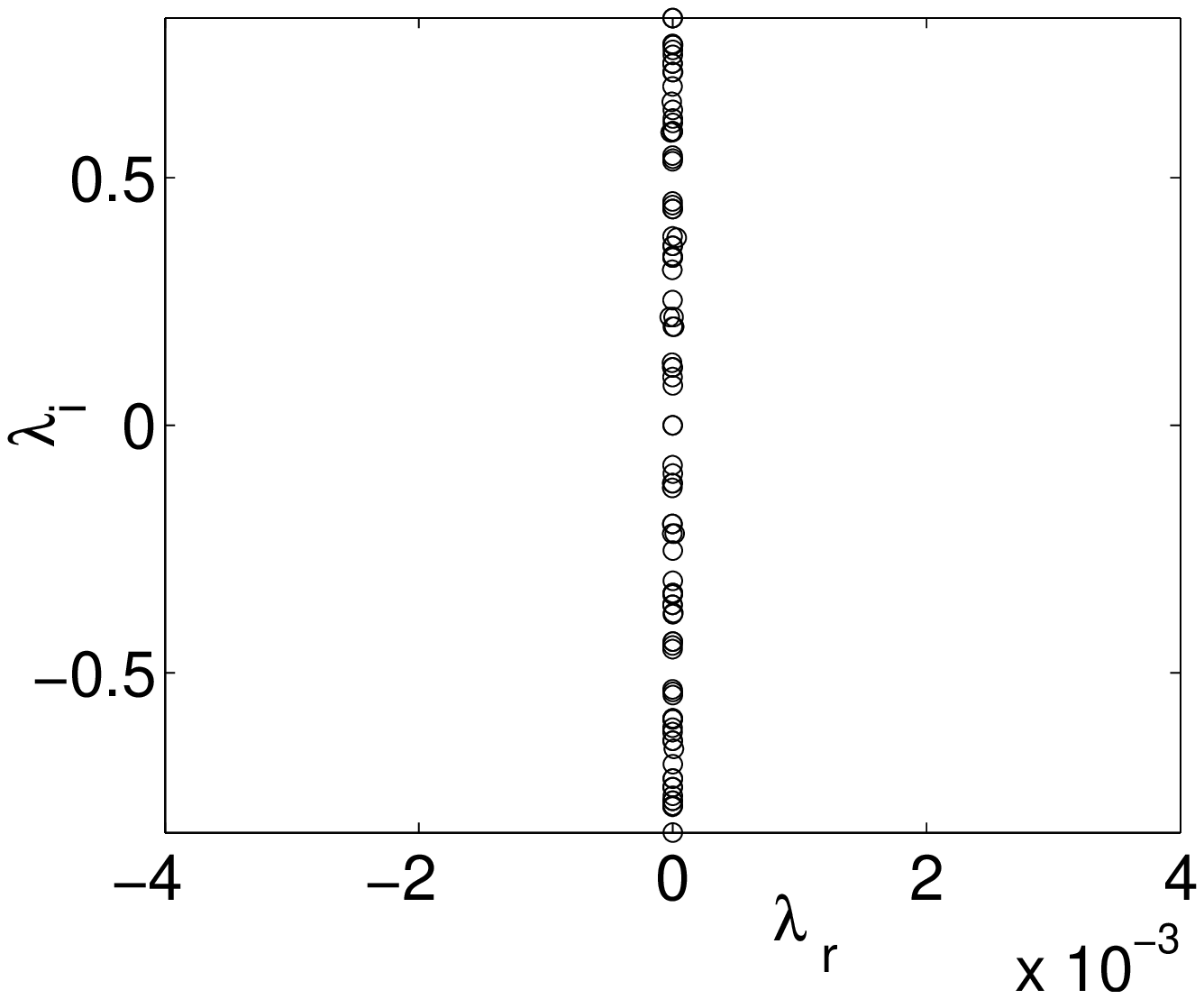}~~~
\vspace{-0.2cm}
\caption{(Color Online) The state
$|1,1\rangle$ for attractive interatomic interactions.
The layout of the figure is similar to the one used in the previous figures. The parameters for the solution
depicted in the middle and bottom rows are $(V_0,\mu)=(0.2,-0.181)$ (see blue cross in top panels)
and $(0.45,0.039)$ respectively (see red circle in top panels).}
\label{vfig8}
\end{figure}

It is important to note, in connection to our numerical linear stability
results, that the
$|0,0\rangle$ branch can become unstable for $\mu<\mu_{\rm cr}(V_0)$
(see top right panel of Fig.~\ref{vfig6}) due to
the appearance of a real
pair of eigenvalues.
This instability for large $N$ is
something that may be expected in the case of attractive interactions under consideration, as the
corresponding
2D GP equation for an
a homogeneous BEC
(i.e., without any external potential) is well-known to be subject to
collapse \cite{sulem}. However, it should also be expected that
very
close to the linear limit 
the growth rate
of the instability is essentially zero (cf. with the top left
panel of Fig.~2 of Ref.~\cite{gregh} for $V_0=0$ which is not shown here). 
Essentially, the potential
appears to {\it stabilize} the solitary wave against dispersion in
this regime (i.e., close to the linear limit),
but cannot stabilize it against the catastrophic collapse-type instability.
Furthermore, in the presence of the optical lattice we can observe
that there is always a range of chemical potentials for which the condensate
is stabilized, in accordance with what was originally suggested in Ref.~\cite{baizakov}.
Furthermore, even in the 3D case it is in principle possible to arrest 
collapse by appropriate choices of the parameters \cite{Mihalache05}.

\begin{figure}[tbp]
\includegraphics[width=8cm]{\rootfig 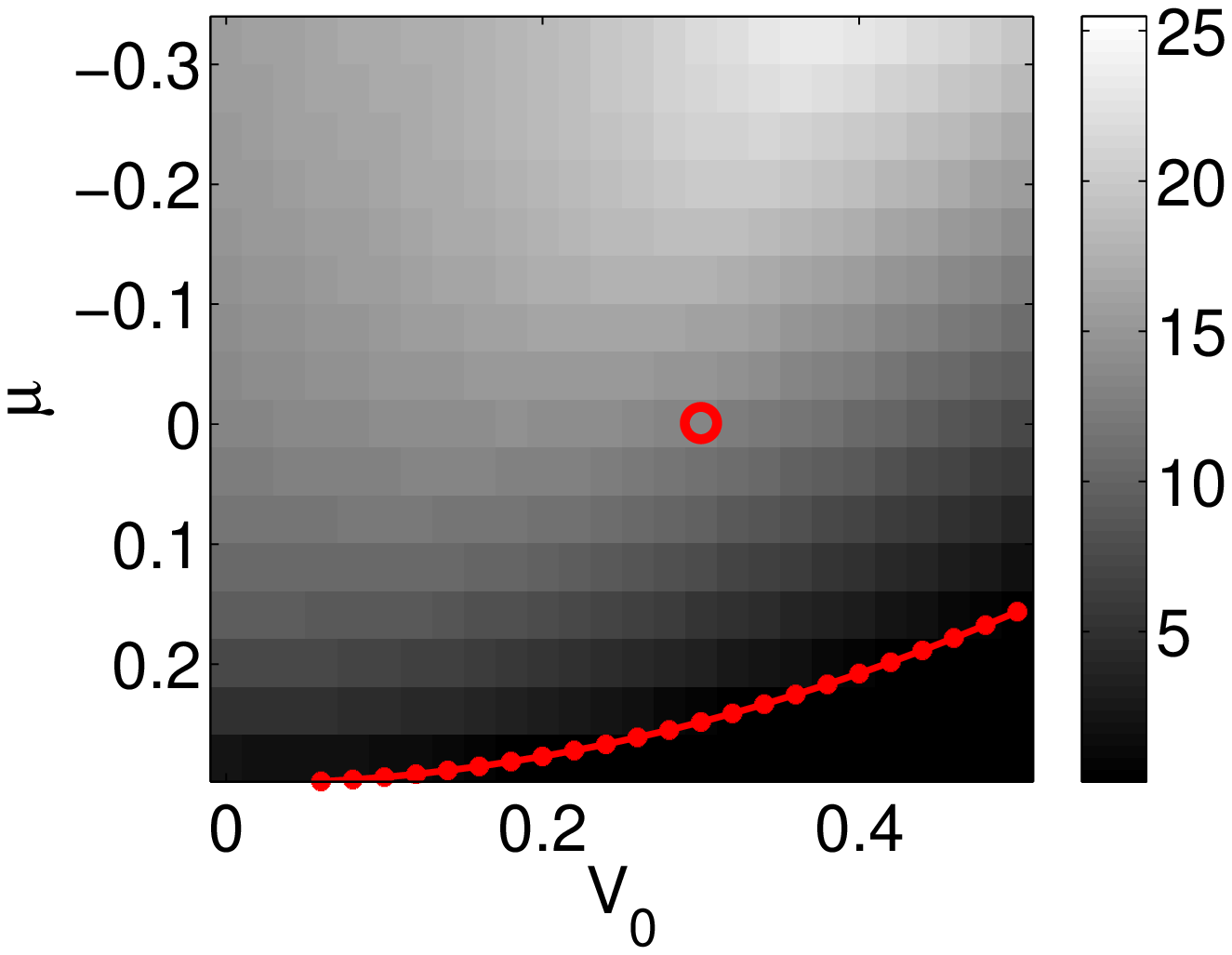}
\includegraphics[width=8cm]{\rootfig 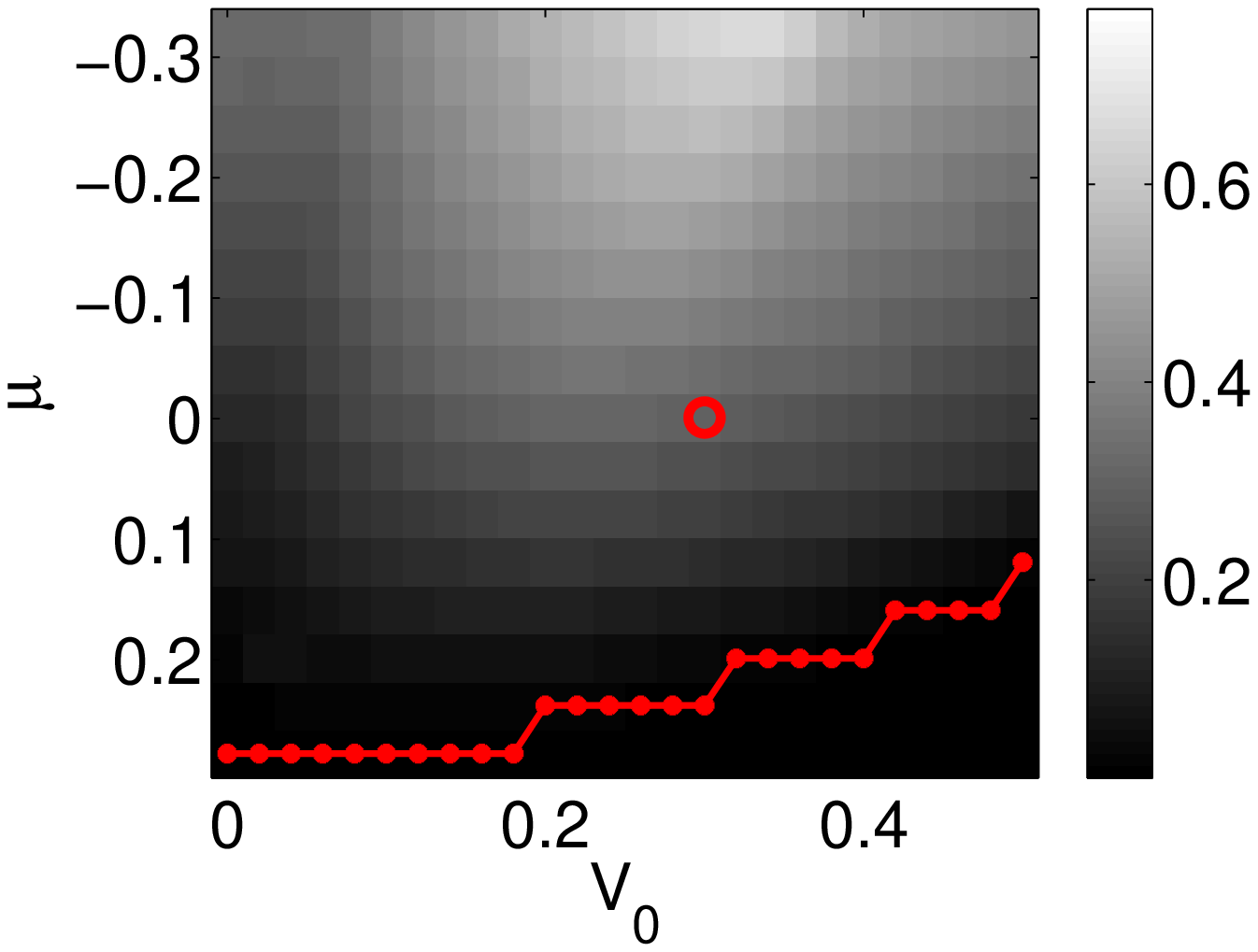}
\\
\includegraphics[width=8cm]{\rootfig 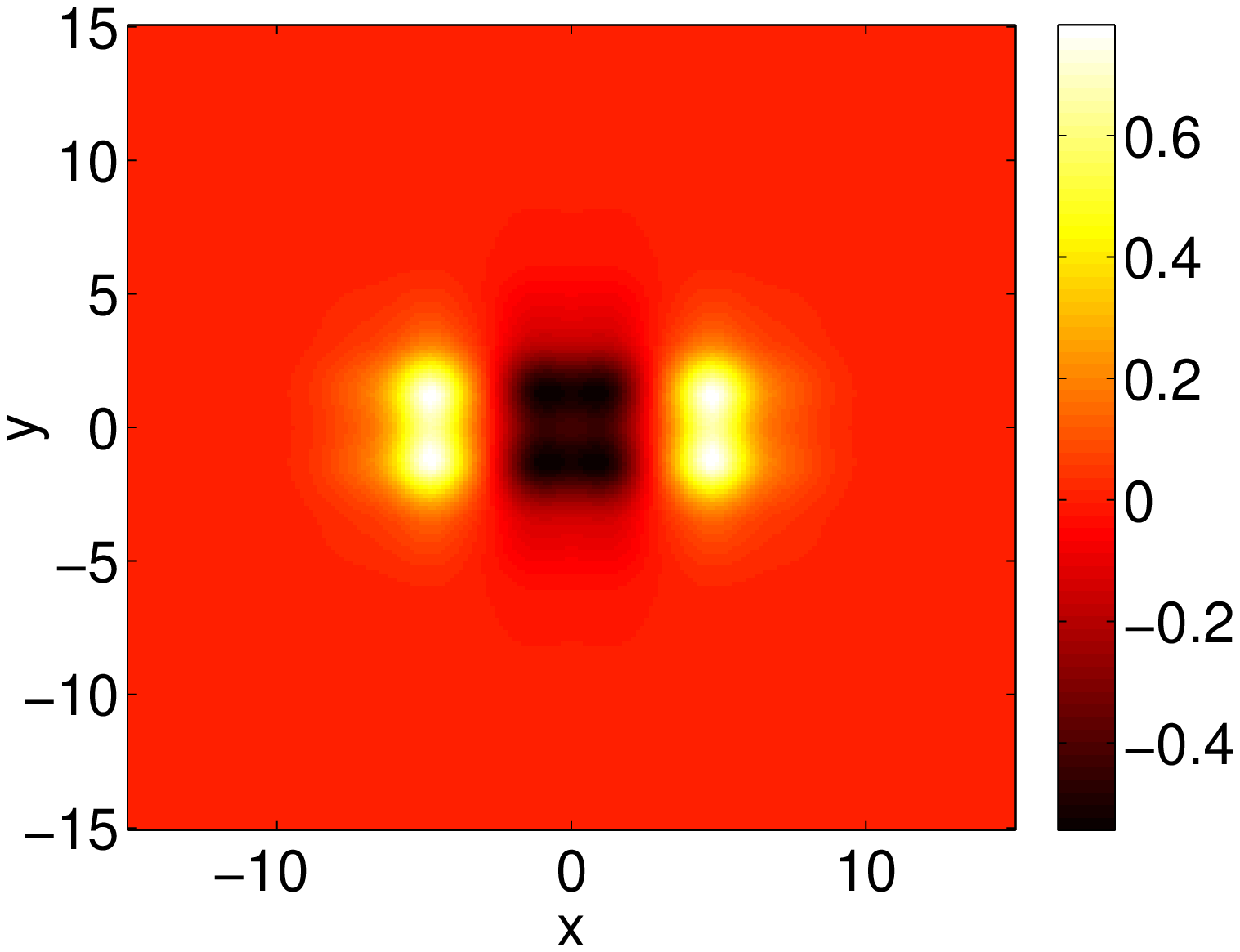}
~~\includegraphics[width=7.65cm,height=6.75cm]{\rootfig 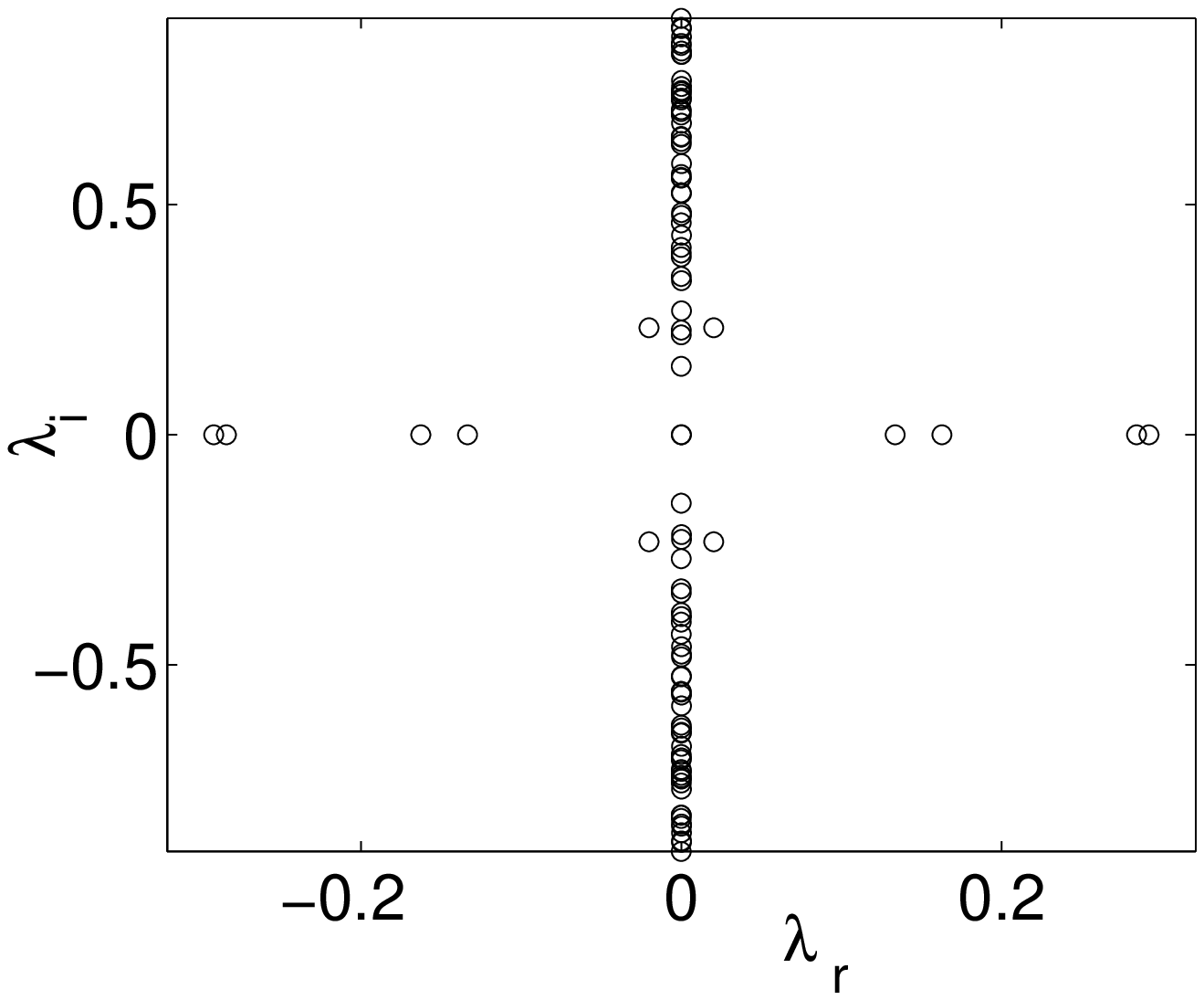}~~~
\vspace{-0.2cm}
\caption{(Color Online) The state
$|2,0\rangle$ for attractive interatomic interactions.
The layout of the figure is similar to the one used in the previous figures. 
The parameters for the solution depicted in the bottom row 
are $(V_0,\mu)=(0.3,-0.001)$ (see red circle in top panels).}
\label{vfig20}
\end{figure}

\begin{figure}[t]
\begin{center}
\begin{tabular}{lllll}
&~~~~~~~~~(a)&
~~~~~~~~~(b)\\
&\includegraphics[height=5.2cm]{\rootfig 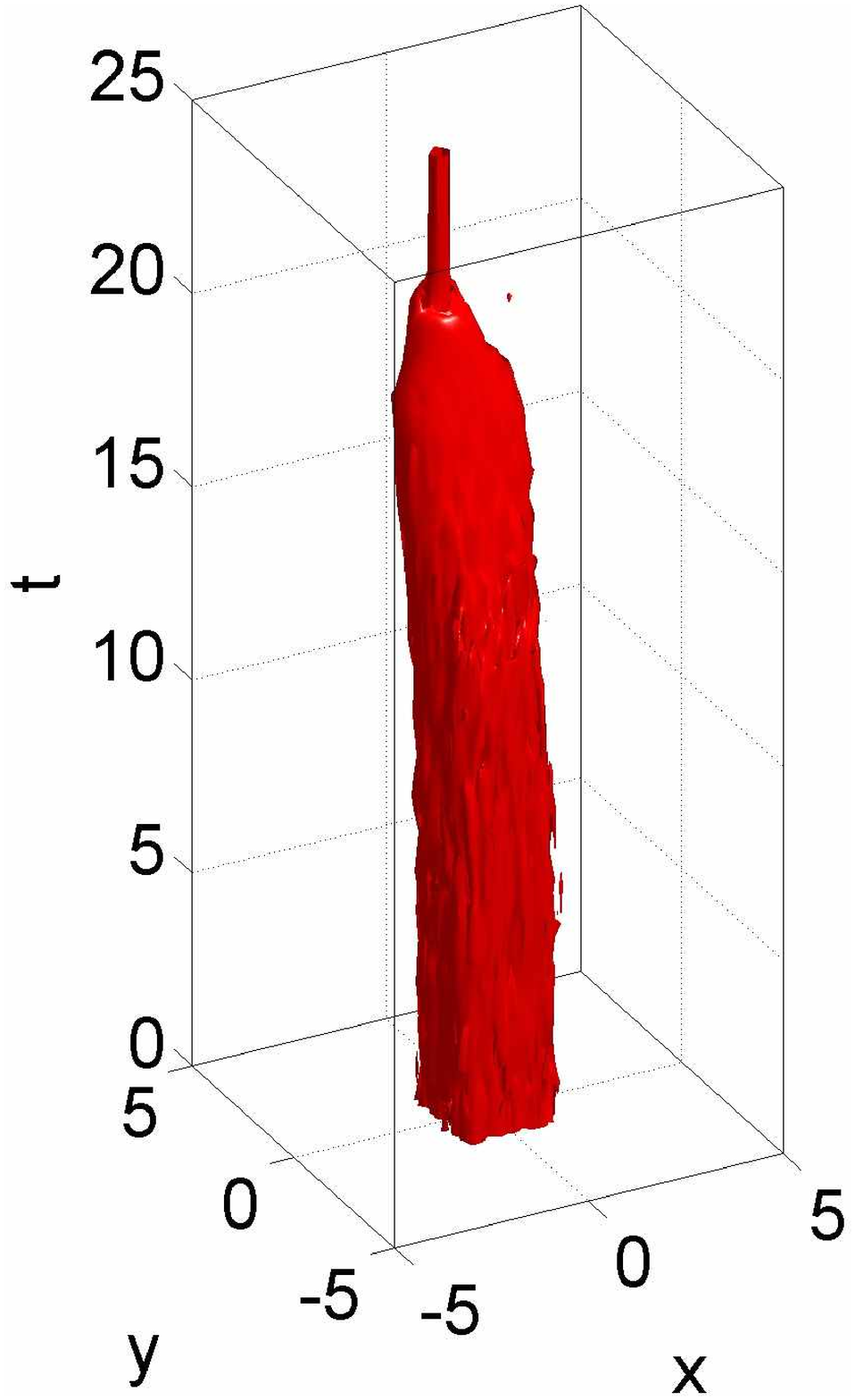}&
\includegraphics[height=5.2cm]{\rootfig 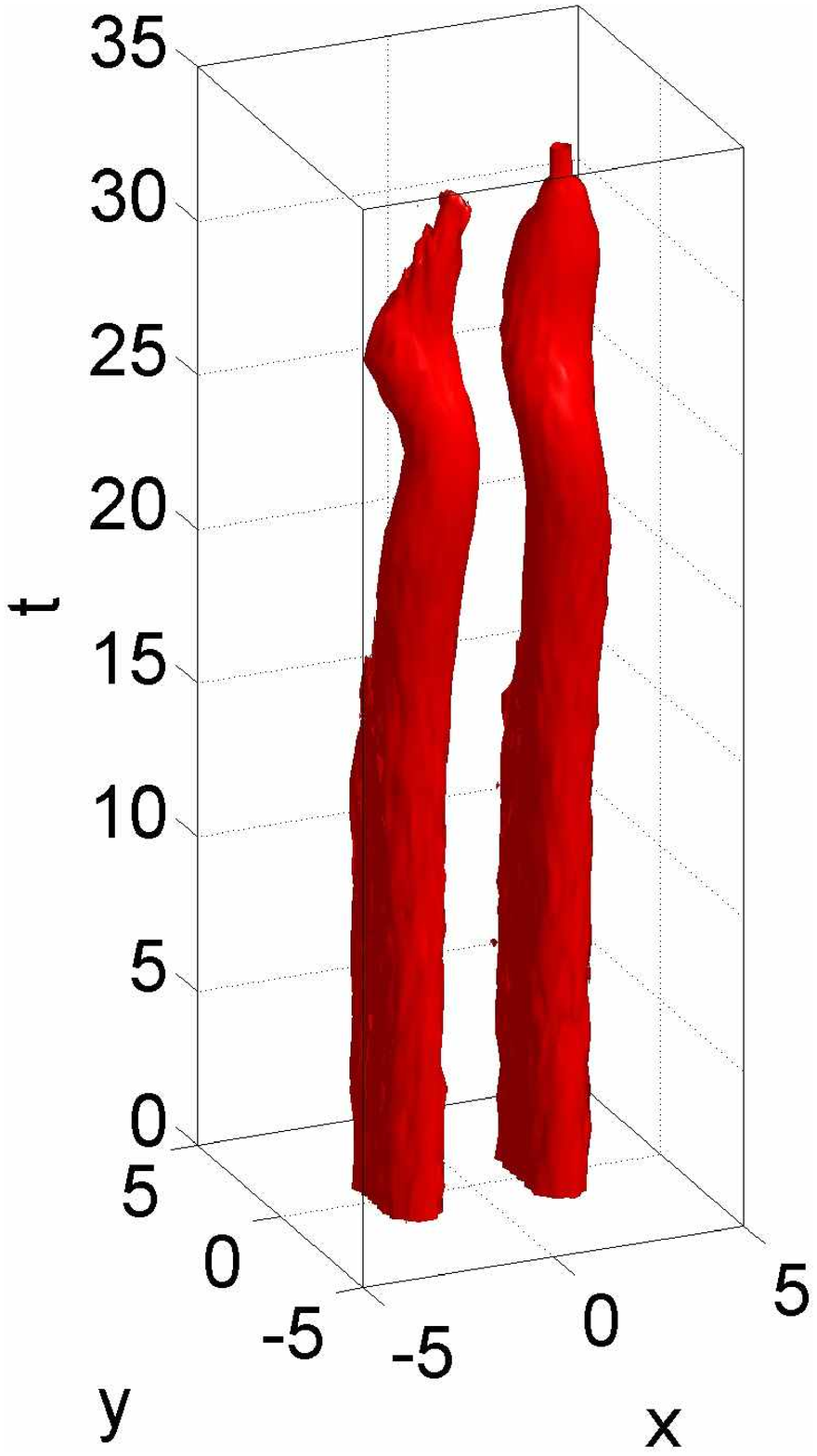}\\
~~~~~~~~~(c)&
~~~~~~~~~(d)&
~~~~~~~~~(e)\\
\includegraphics[height=5.2cm]{\rootfig 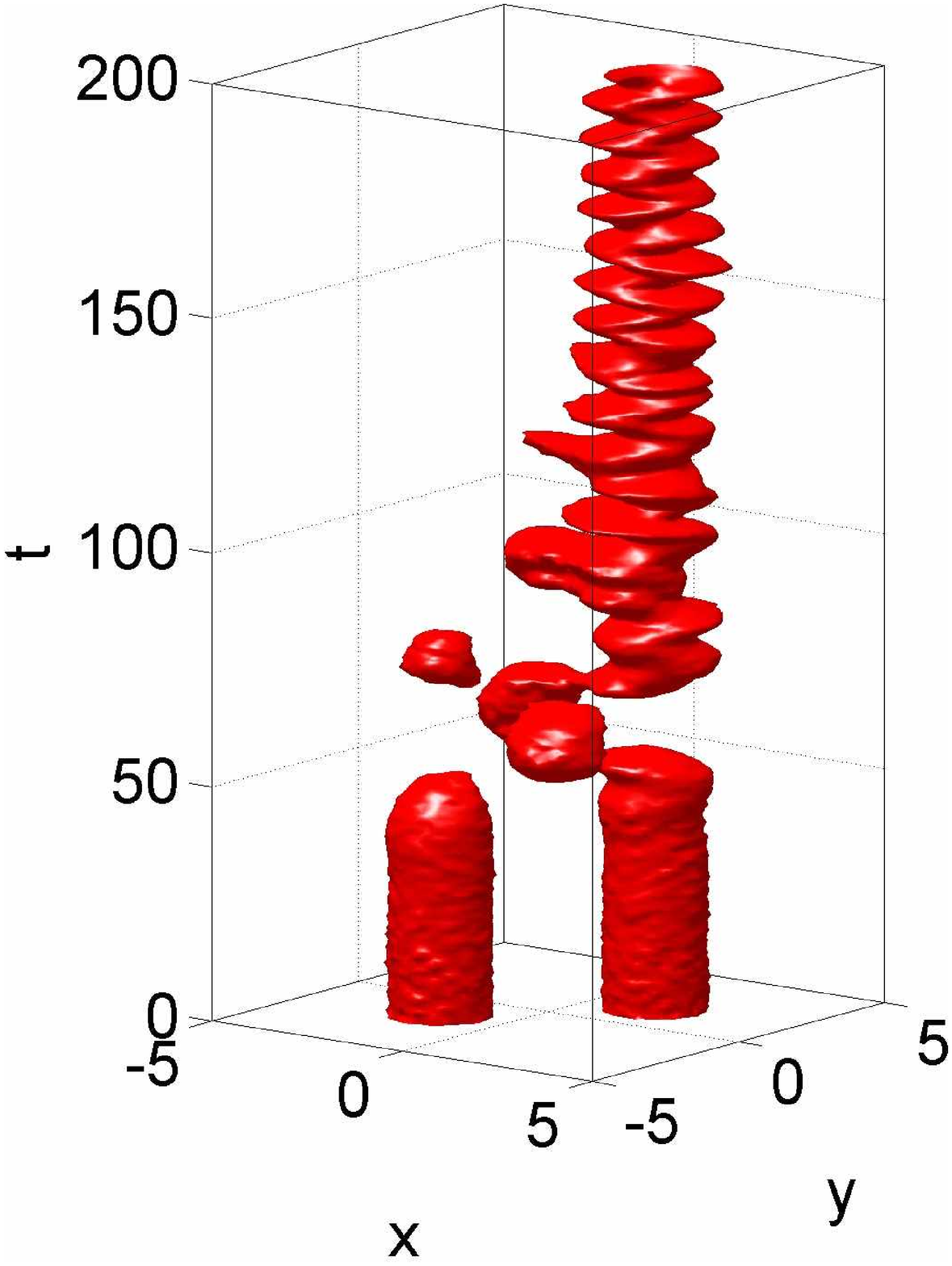}&
\includegraphics[height=5.2cm]{\rootfig 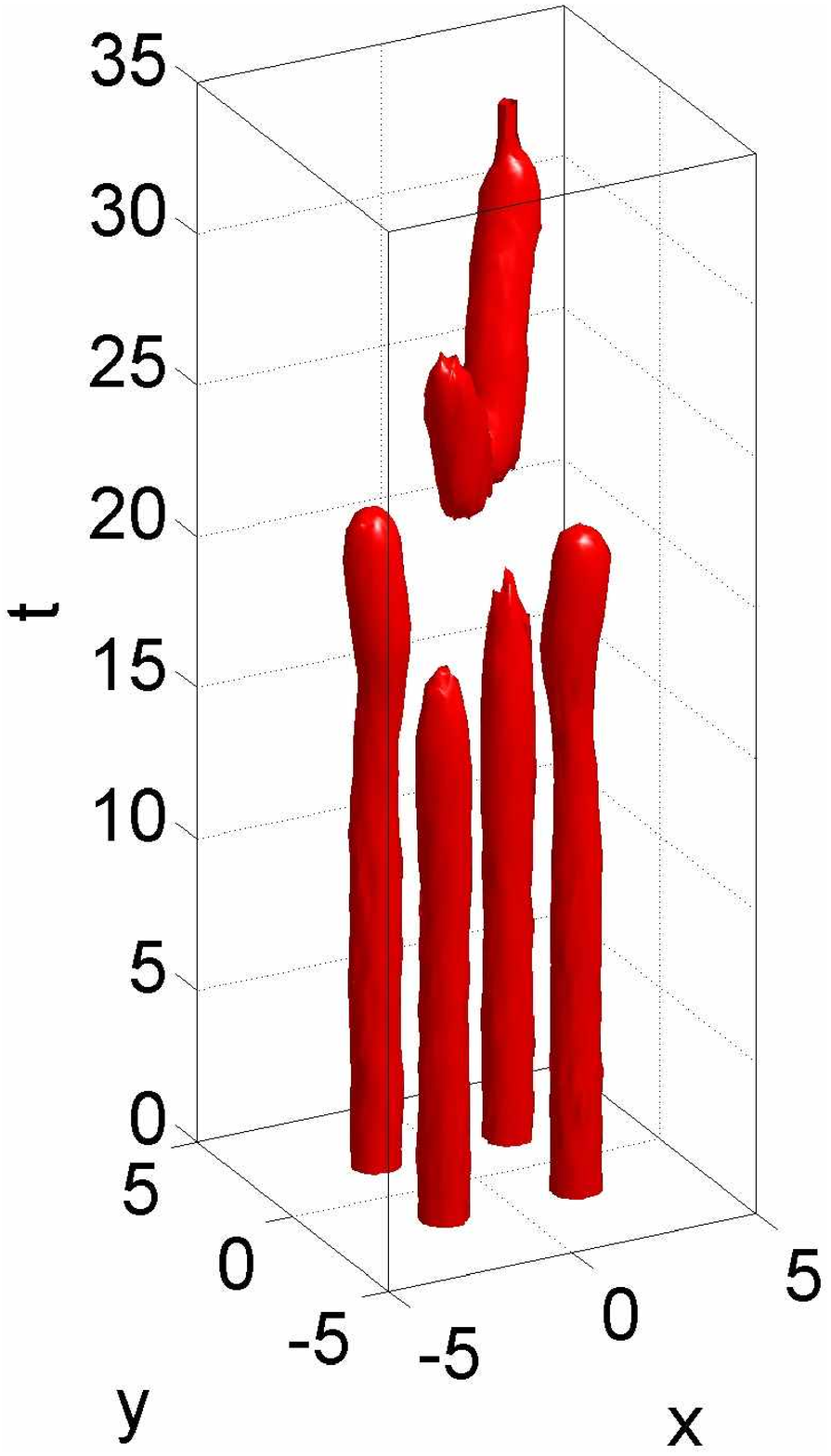}&
\includegraphics[height=5.2cm]{\rootfig 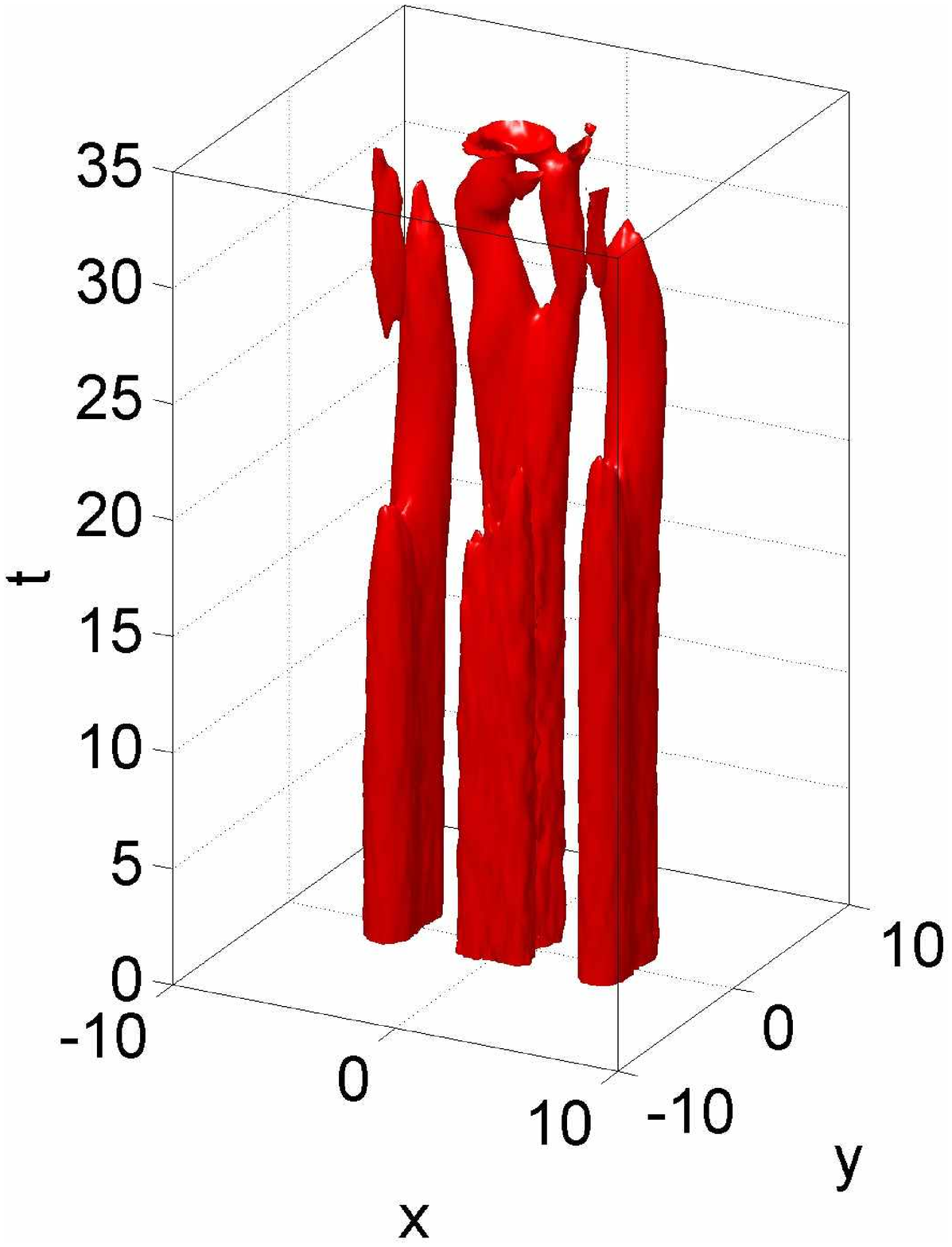}
\end{tabular}
\caption{(Color Online) 
Dynamics of the unstable states (in the case of attractive interatomic interactions)
that were shown in the previous figures.
Shown are space-time evolution plots given by a characteristic density isosurface
$D_k=\{x,y,t \mid |u(x,y,t)|^2=k \}$, where $k=a ({\rm max}_{\{x,y\}}\{|u(x,y,0|^2\})$ and  
[a=0.7 for (a), 0.5 for (b) and (d), and 0.3 for (c) and (e)].
(a) Ground state $|0,0\rangle$, which collapses very quickly.
(b) First excited state $|1,0\rangle$, which collapses shortly after
the ground state.
(c) Degeneration of a $|1,0\rangle+|0,1\rangle$ state
into an eventual single-pulse structure that survives
for a long time after the merger. 
The unstable $|1,1\rangle$ (d) and $|2,0\rangle$ (e) states
deform, for very short times as expected from the strong instabilities 
identified in their spectra, and subsequently collapse.}
\label{f_dyno}
\end{center}
\end{figure}

Next, we consider real-valued solutions with $m+n=1$.
The $|1,0\rangle$ state
(again in the case of attractive interactions) is shown in 
Fig.~\ref{vfig7}. This branch is always unstable, due to up to two
real eigenvalue pairs and one complex quartet. A typical example
of the branch in the bottom panels of the figure reveals this
instability.

The $|1,0 \rangle+|0,1 \rangle$ configuration for the attractive interactions case is
shown in Fig.~\ref{vfig10}. This configuration turns out to
be unstable in a large fraction of the regime of parameters considered
due to a quartet of complex eigenvalues.
However, remarkably, as $V_0$ and $\mu$ are increased 
and decreased respectively, it is possible
to actually trap this state in a linearly stable form (eliminating the
relevant oscillatory instability). This indicates that it would be of
particular interest to try to identify such a state
(which resembles an out-of-phase soliton pair) in
a real experiment.
Also, as expected, the solution degenerates to its
linear counterpart as $\mu \rightarrow E$.
Images of a typical unstable solution and its complex quartet are
shown along with a stable solution from the top right-hand region
of the two-parameter diagram.

We should also note in passing that states in the form of
$|1,0\rangle+ i |0,1\rangle$  would produce a vortex waveform;
however since such states have been
studied in some detail earlier in Ref.~\cite{vortex} in a similar setting (i.e., in the presence
of an external potential containing both harmonic
and
lattice components), we do not examine them in more detail here.

We now turn to solutions featuring $m+n=2$.
First, we consider the $|1,1 \rangle$ branch for the attractive case
in Fig.~\ref{vfig8}. In this case the solution may
possess between one and three complex
eigenvalue quartets in its linearization (the middle panel of
the figure shows a particular unstable case where there are two
such quartets).
However, once again, there exists a region
in the 
right side of the relevant parameter space [i.e., for
appropriate $(V_0,\mu)$] where the solution is found to
be linearly stable and all potential oscillatory instabilities
are suppressed.
The bottom panel of Fig.~\ref{vfig8} shows such a linearly stable case of the
quadrupolar configuration, which, again, should be experimentally
accessible.

Finally, we consider the state $|2,0\rangle$, as depicted in 
Fig.~\ref{vfig20}. This configuration is highly unstable throughout
our parameter space, with up to four real pairs and one
complex quartet of eigenvalues. A typical example of the
unstable configuration and its spectral plane of eigenvalues
is shown in the bottom panel of Fig.~\ref{vfig20}.

\subsubsection{Dynamics}

Now we corroborate our existence and stability results (for the attractive interactions case)
with an investigation of the actual dynamics
of typical unstable solutions selected from the above families.
For each case, the particular solution
presented in the corresponding figures is perturbed
with random noise distributed uniformly between $-0.05$ and $0.05$ and
integrated over time. It is important to note that although a
random perturbation is used here to ``emulate'' the experimental noise,
the system is deterministic and the sole relevant feature of
any (generic) random perturbation is its projection onto the 
most unstable eigendirection(s) of the perturbed solution profile.
These projections, as indicated above, will grow (determinstically)
according to the corresponding growth rate.
For the time propagation, we implement a standard
4th-order Runge-Kutta integrator scheme where we have
numerical consistency and stability for the conservative
time step of $\Delta t=10^{-3}$. The results are compiled
in Fig.~\ref{f_dyno}.

Panel (a) in Fig.~\ref{f_dyno} depicts the
catastrophic instability of the ground state, $|0,0\rangle$, which
is subject to collapse, occurring at $t \approx 25$.
Panel (b) shows similar behavior for the $|1,0\rangle$ state,
in which the two lobes appear to self-focus independently, although
one eventually prevails and collapses for $t \approx 35$.
It is very interesting
to note that while the $|1,0\rangle$ state 
collapses,
its more stable superposition with the $|0,1\rangle$ state survives for
longer times, as expected, and also
eventually merges into a ground-state-like (single pulse)
configuration, which was found to have a number of atoms
just on the unstable side of the boundary of stability for
such a structure.
The resulting state actually survives
for \textit{very} long times, oscillating
within one of the wells where it originally collected itself
[see the panel (c)],
apparently stabilized by the ensuing
oscillations. 
Panels (d) and (e) show, respectively, the
relatively rapid break up and subsequent collapse of the
$|1,1 \rangle$ and $|2,0\rangle$ states.



\subsection{Repulsive interatomic interactions
}

\subsubsection{Existence and Stability}

\begin{figure}[htbp]
\includegraphics[width=8cm]{\rootfig 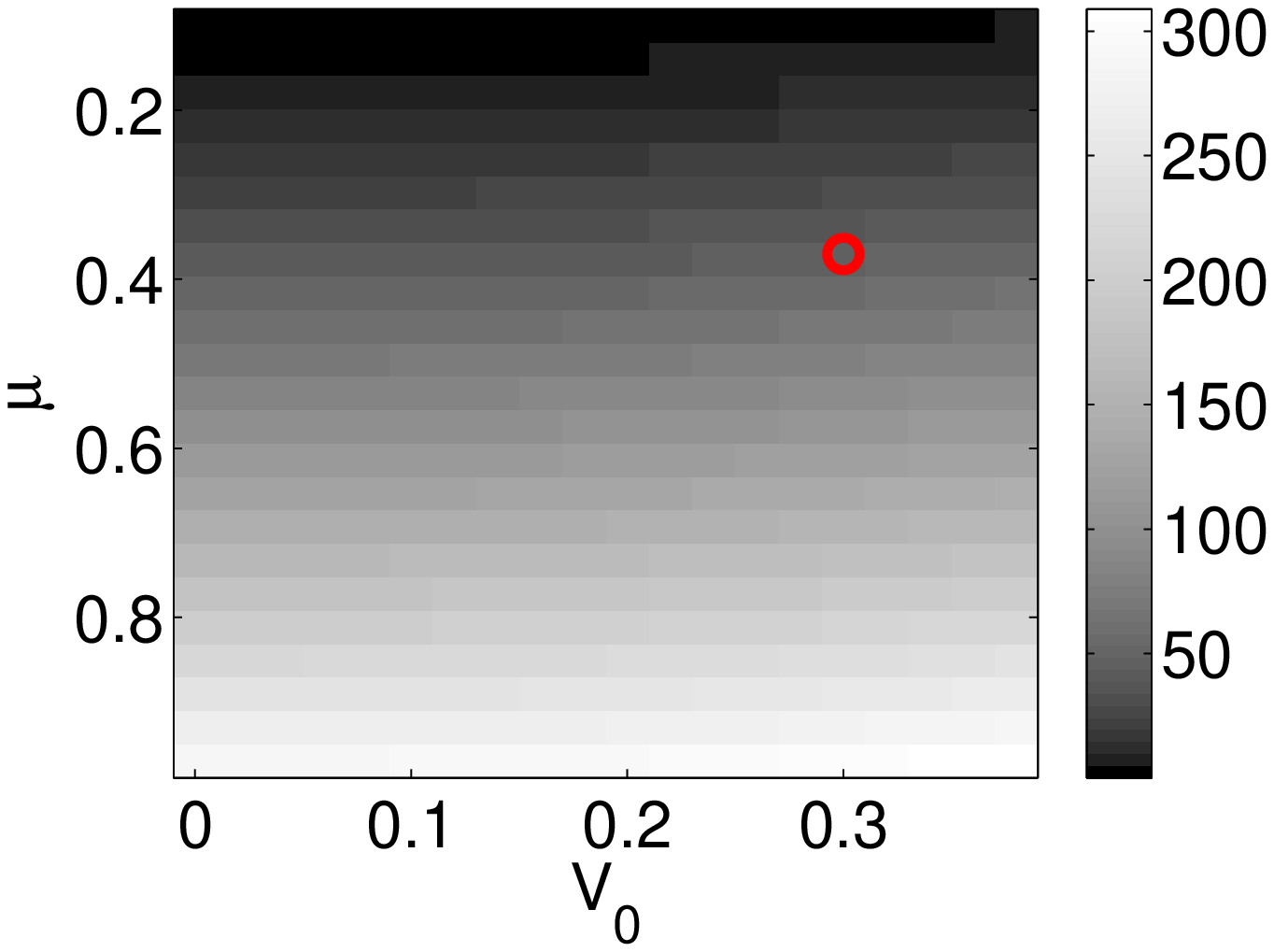}
\\
\includegraphics[width=8cm]{\rootfig 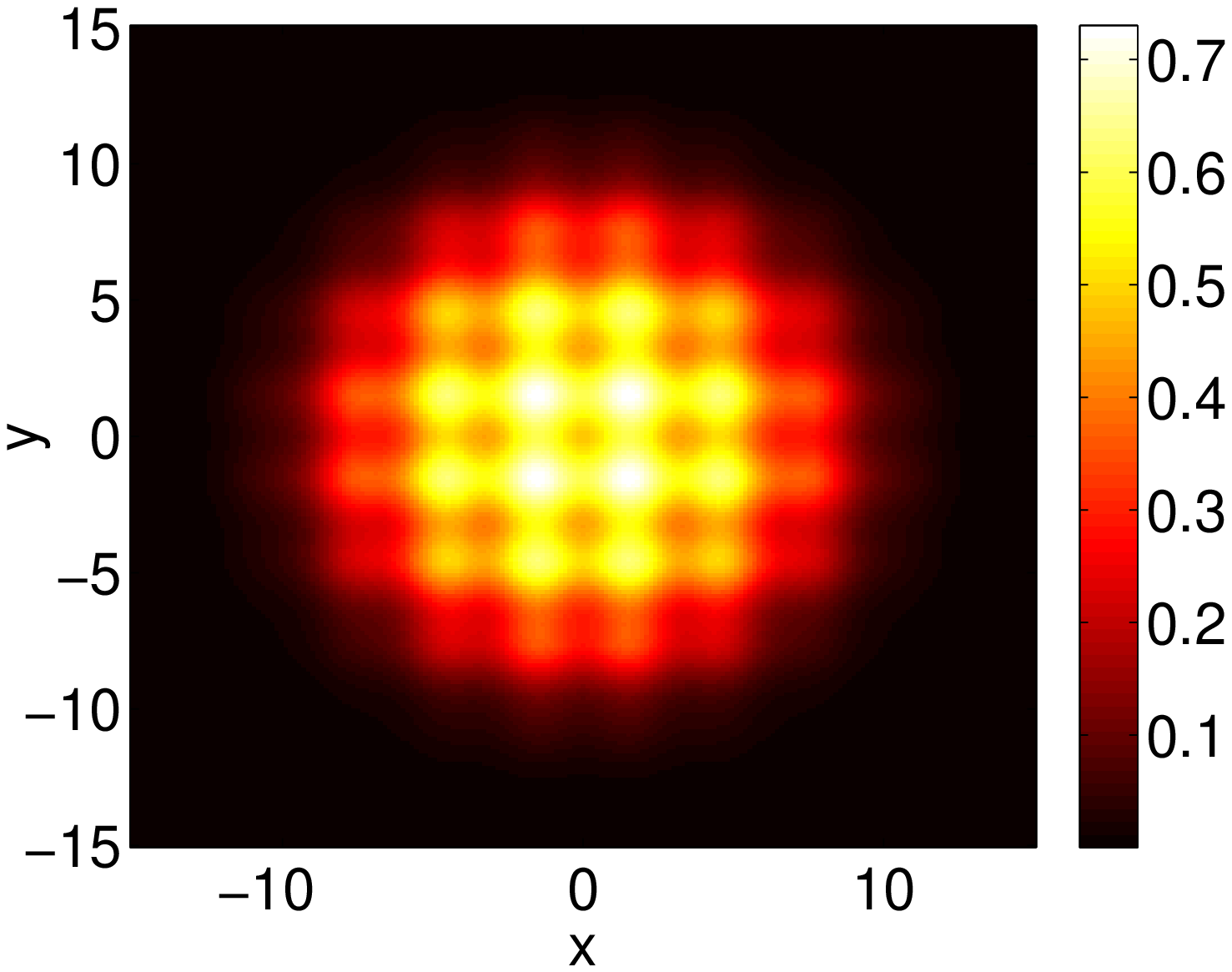}
~~\includegraphics[width=7.65cm,height=6.75cm]{\rootfig 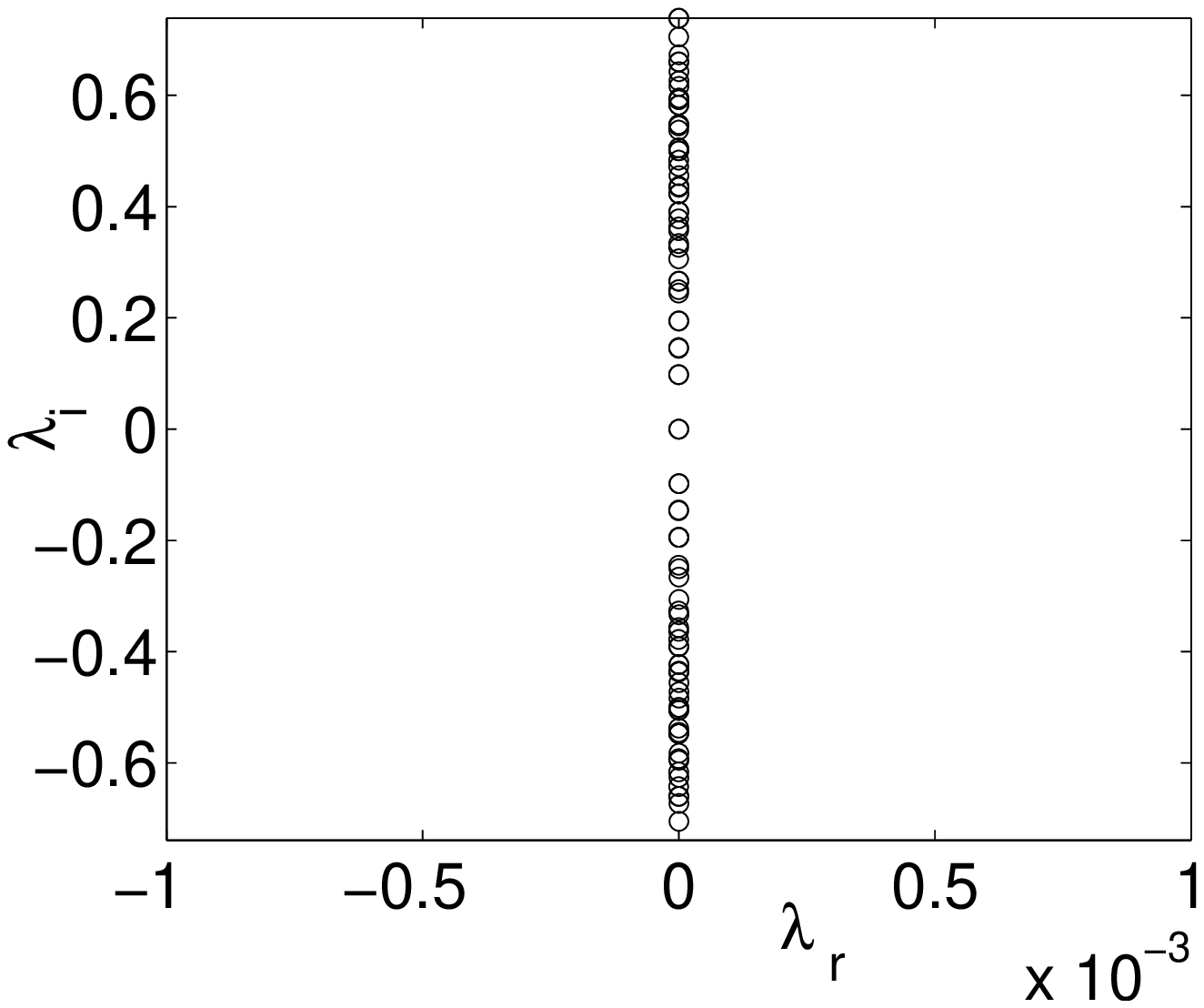}~~~
\caption{(Color Online)
The ground state state $|0,0\rangle$ solution in the case of repulsive interatomic interactions.
The bottom panels correspond to $(V_0,\mu)=(0.3,0.37)$.
The stability of the ground state 
persists over the entire parameter space, and hence the stability
surface is omitted from this set.}
\label{vfig1}
\end{figure}

Now we will investigate the results
pertaining to repulsive interatomic interactions
for the same linear states examined above.
We once again start with the ground state $|0,0\rangle$ branch shown in Fig.~\ref{vfig1}.
The top left panel of Fig.~\ref{vfig1} shows the number of atoms $N(V_0,\mu)$
as in the previous section. However,
the linear stability $S(V_0,\mu)$ for this case is omitted
because, as may be expected, this branch
is stable throughout the parameter space,
in contrast to its attractive counterpart (which is subject to collapse).

\begin{figure}[htbp]
\includegraphics[width=8cm]{\rootfig 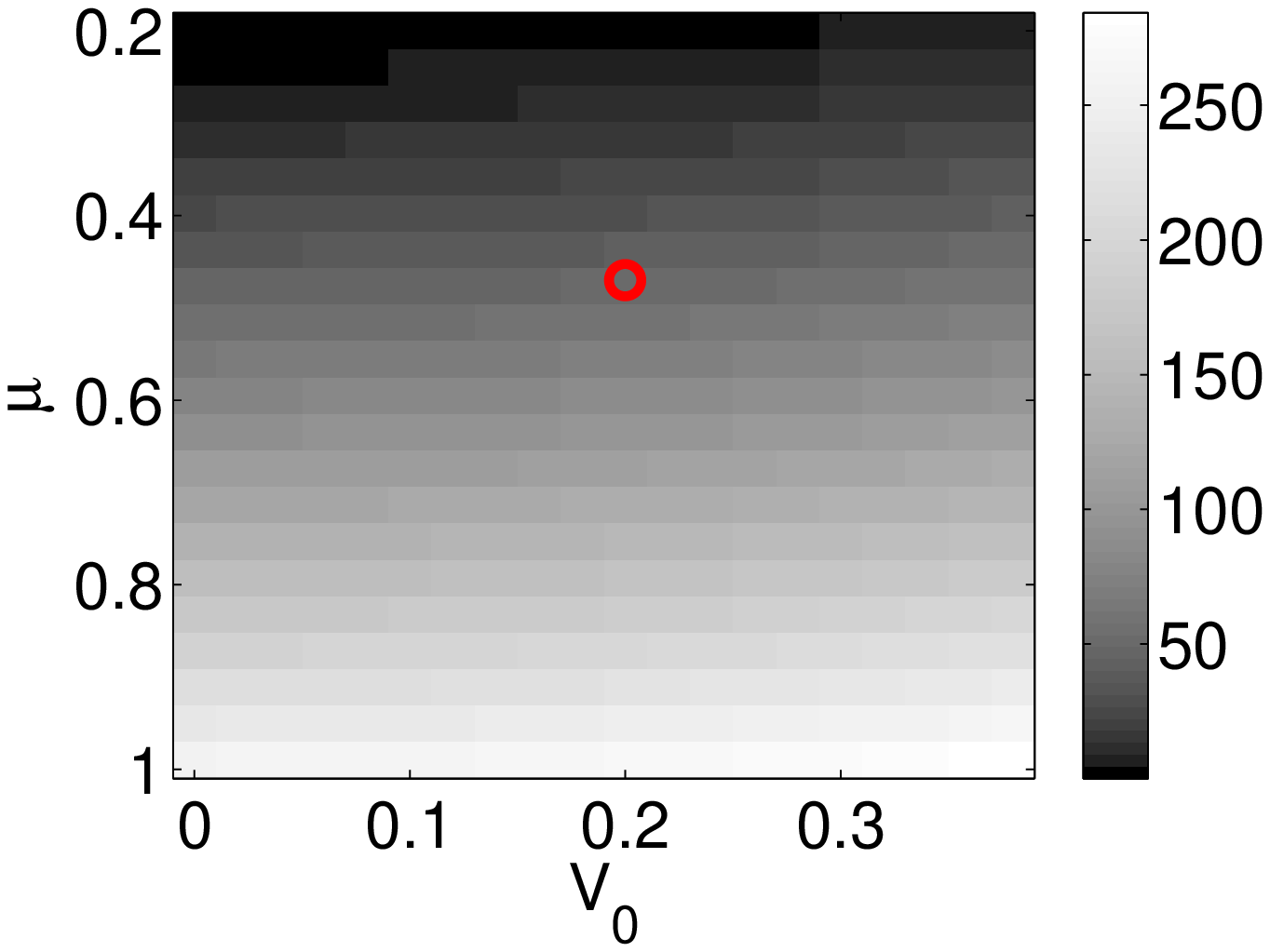}
\includegraphics[width=8cm]{\rootfig 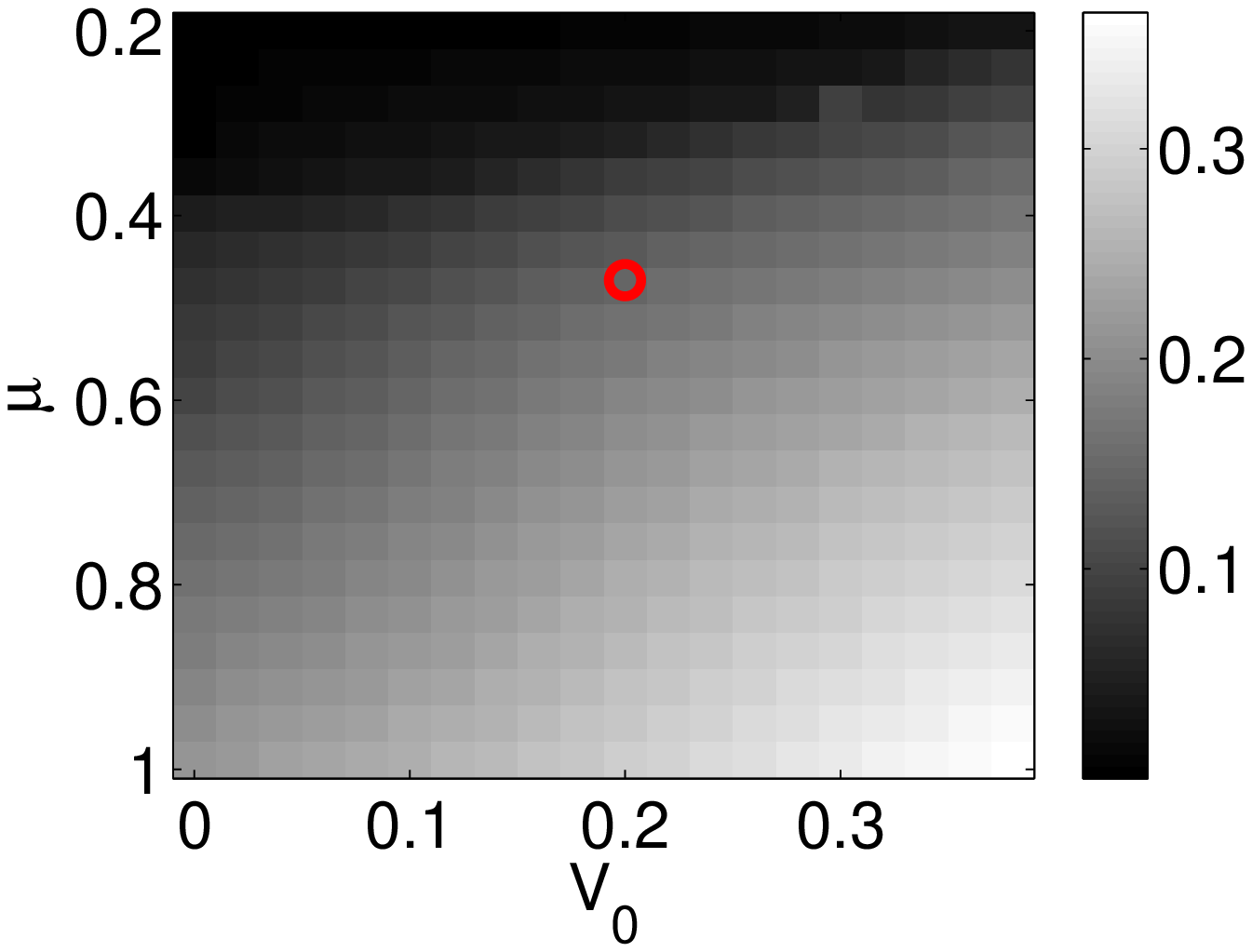}
\\
\includegraphics[width=8cm]{\rootfig 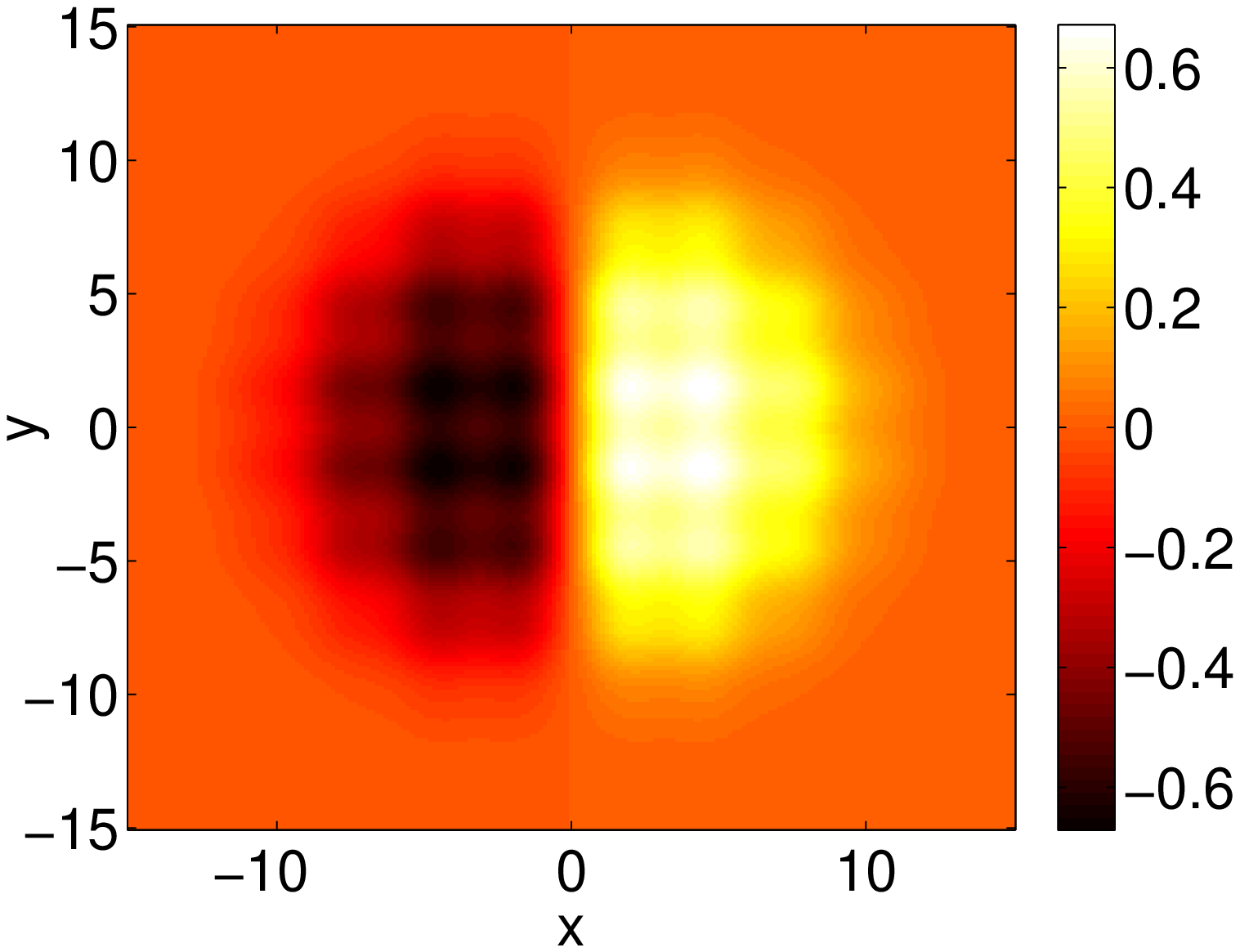}
~~\includegraphics[width=7.65cm,height=6.75cm]{\rootfig 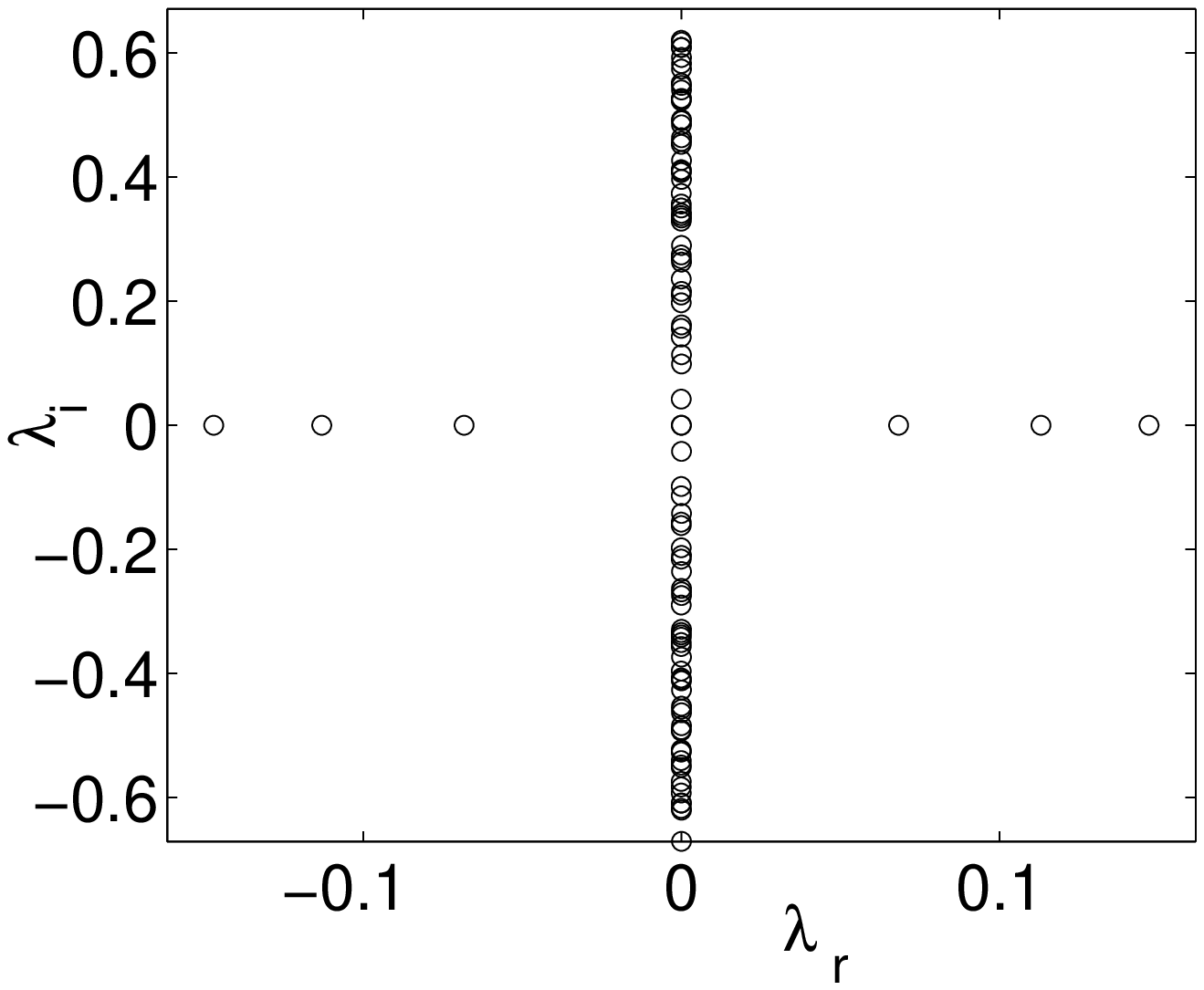}~~~
\caption{(Color Online) 
The state $|0,1\rangle$ in the case of repulsive interatomic interactions.
The bottom row illustrates
a sample profile (left) and the corresponding eigenvalue spectrum (right)
for $(V_0,\mu)=(0.2,0.47)$ displaying the strong instability arising from the three real
eigenvalue pairs.}
\label{vfig2}
\end{figure}

We now turn to excited states
with $m+n=1$.
Figure~\ref{vfig2} shows features similar to the previous one,
but now for the
state $|1,0\rangle$. This branch is found
to always be unstable
due to the appearance of up to three real eigenvalue pairs.
The top panels depict the dependence of the number of atoms $N(V_0,\mu)$ (left),
and instability growth rate $S(V_0,\mu)$ (right) on the lattice
depth $V_0$
and the chemical potential $\mu$. A sample profile and its
eigenvalue spectrum are given in the bottom panels, indicating
the presence, in this case, of three real eigenvalues pairs.

\begin{figure}[tbp]
\includegraphics[width=8cm]{\rootfig 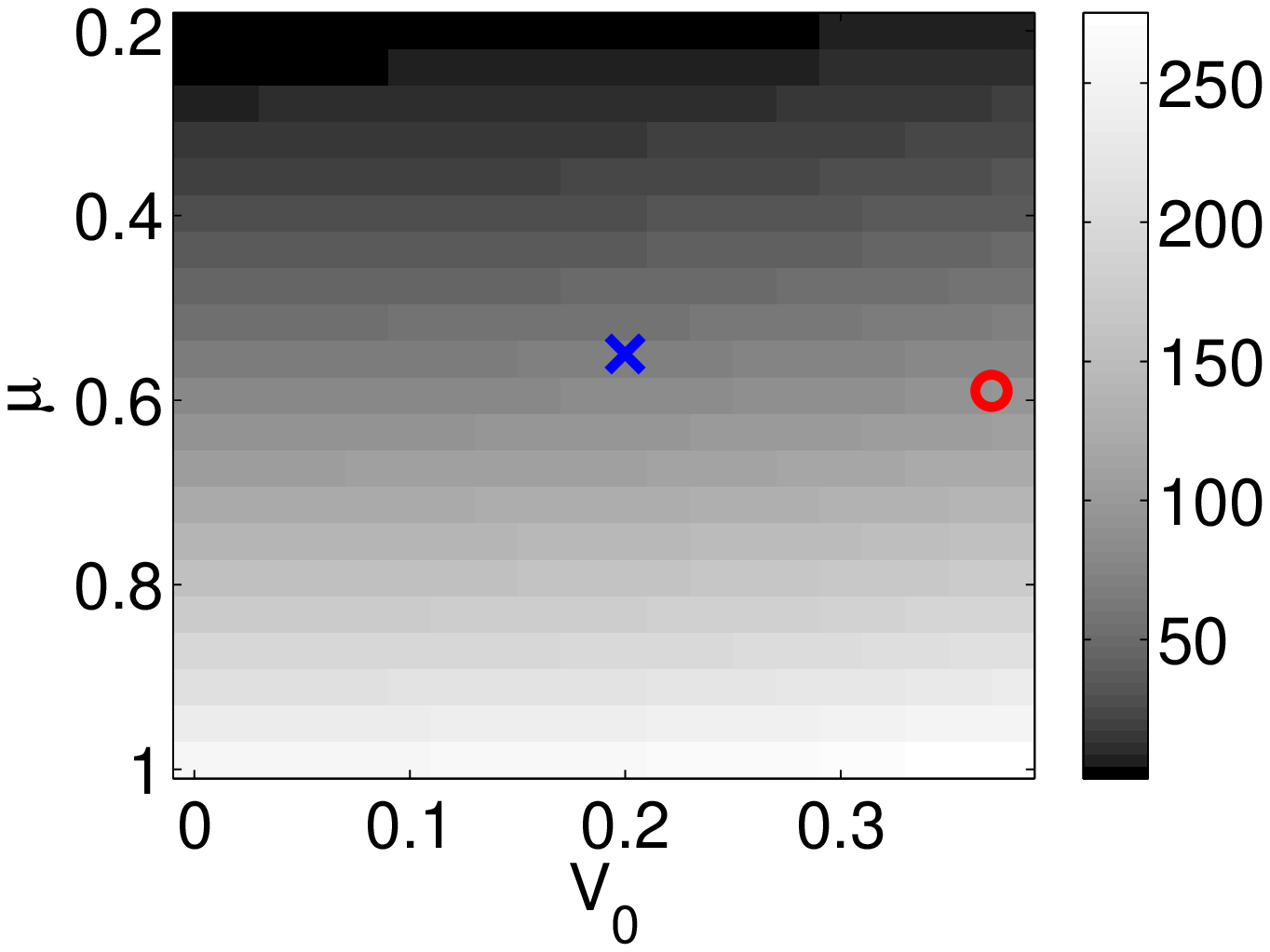}
\includegraphics[width=8cm]{\rootfig 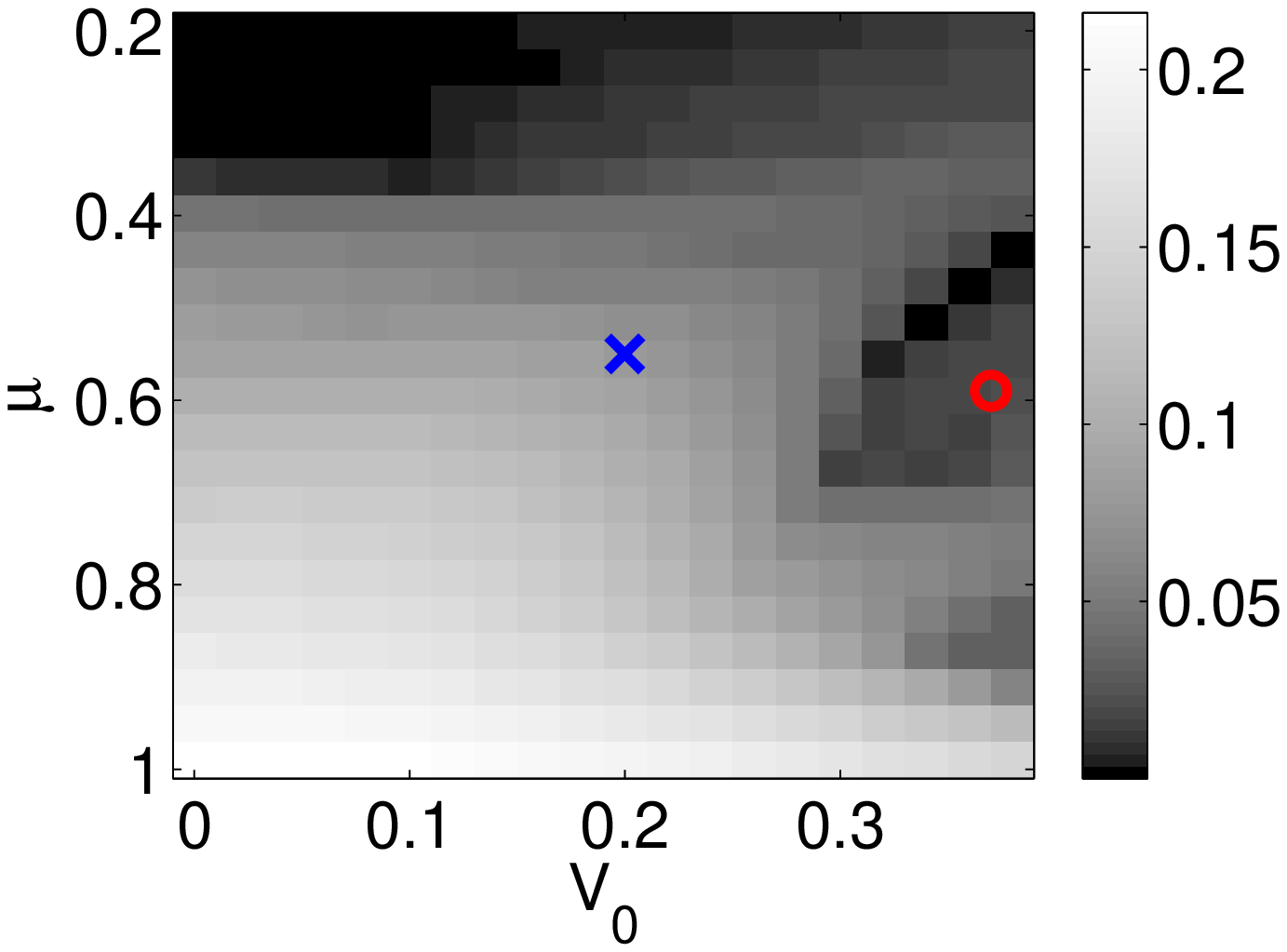}
\\
\includegraphics[width=8cm]{\rootfig 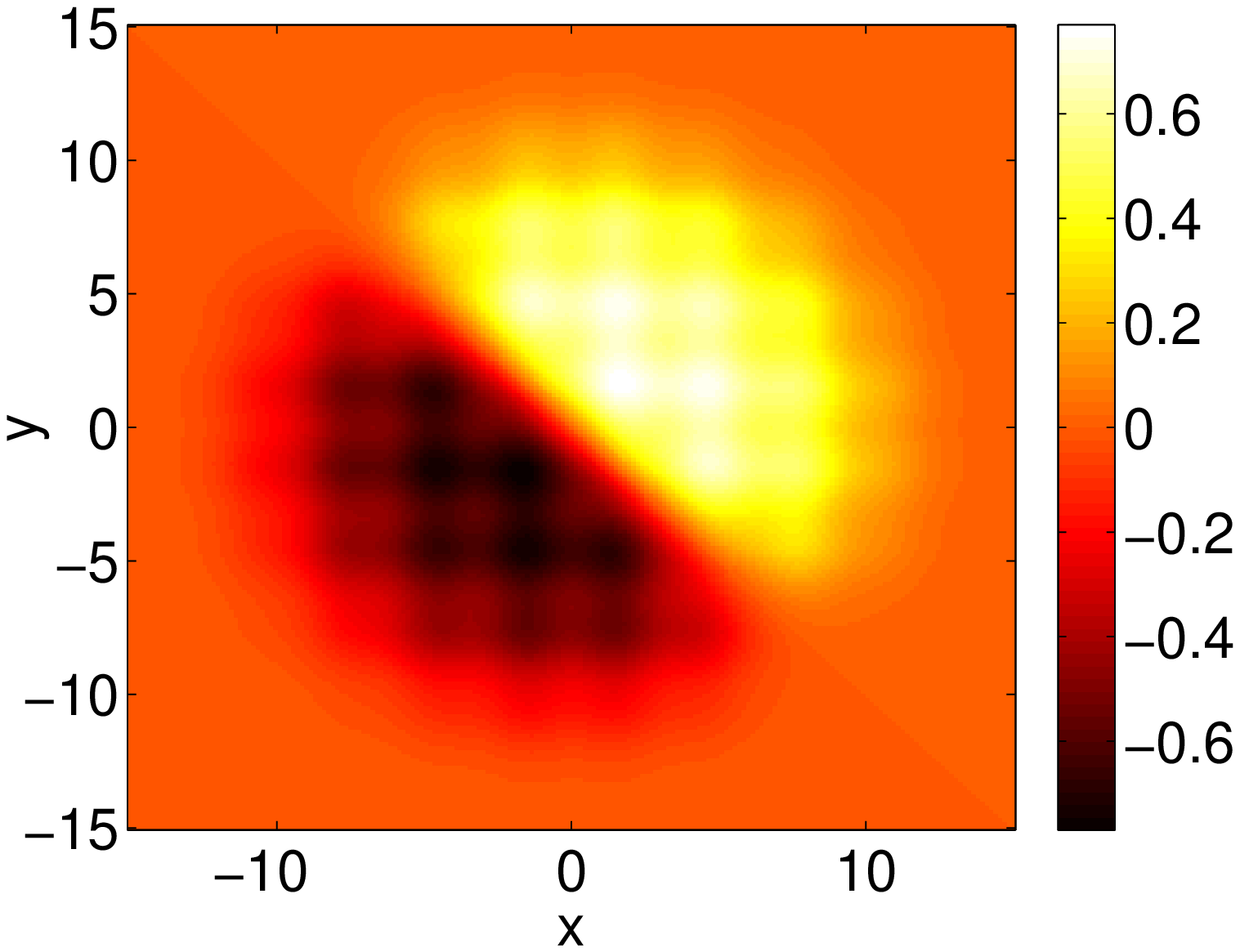}
~~\includegraphics[width=7.65cm,height=6.75cm]{\rootfig 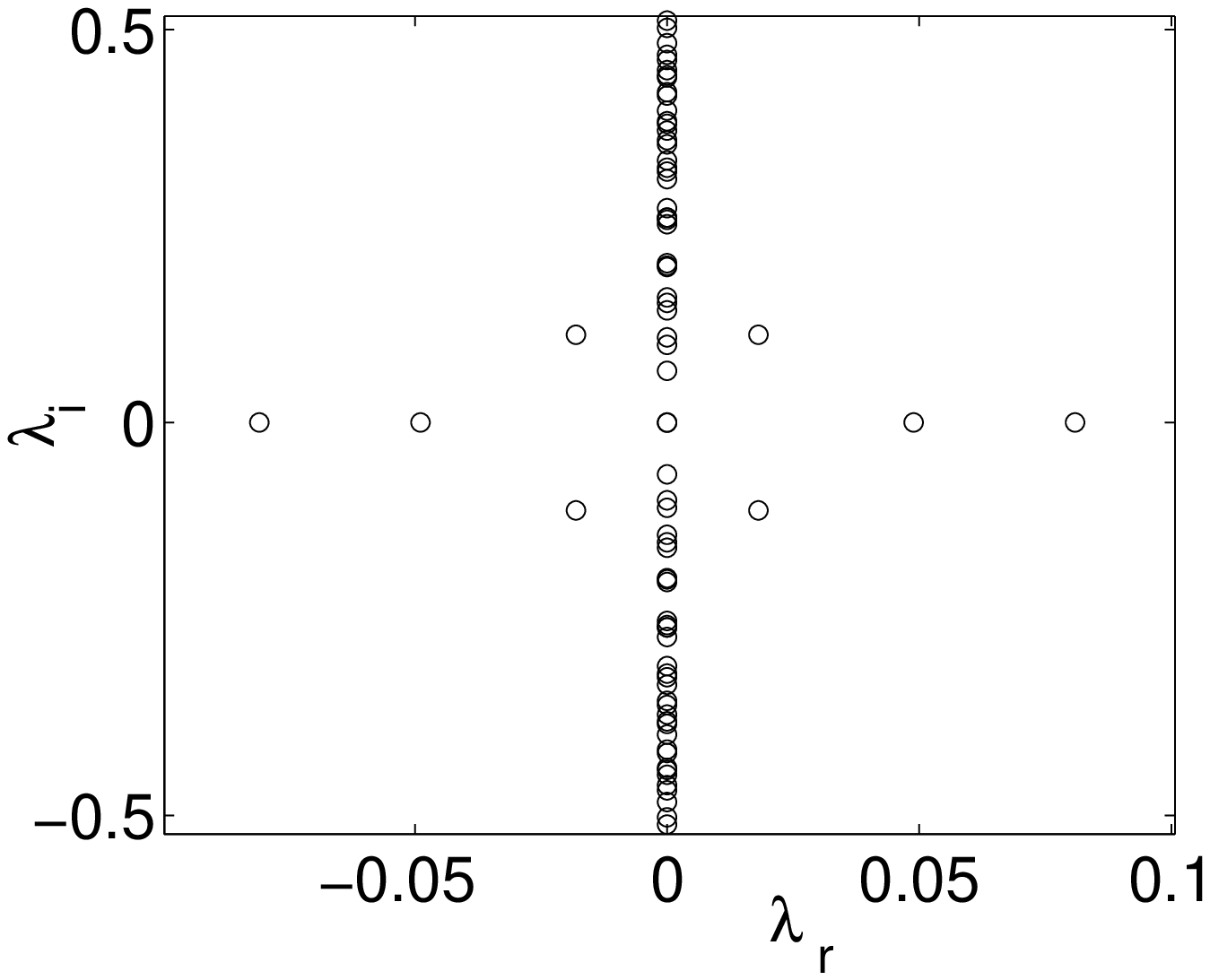}~~~
\\
\includegraphics[width=8cm]{\rootfig 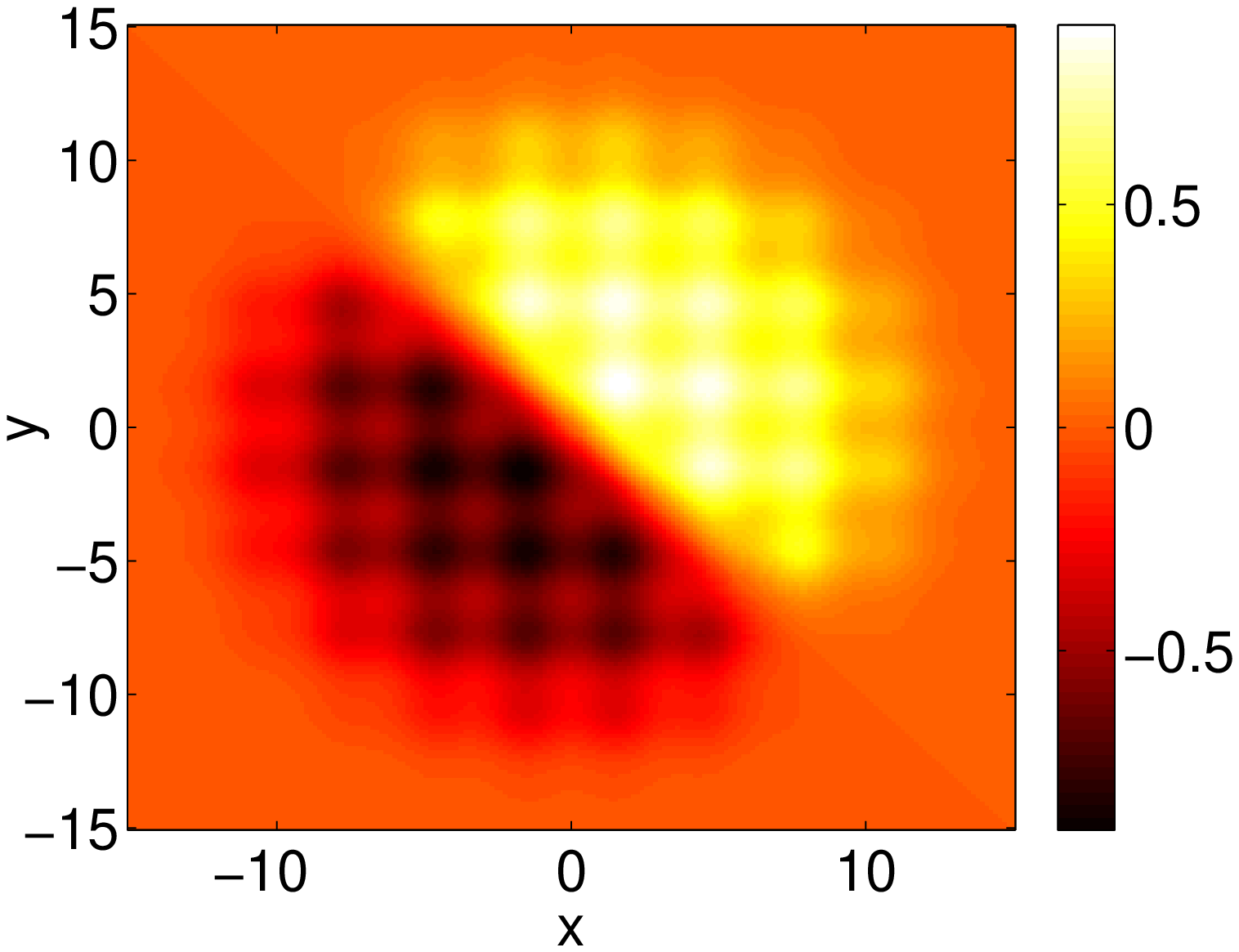}
~~\includegraphics[width=7.65cm,height=6.75cm]{\rootfig 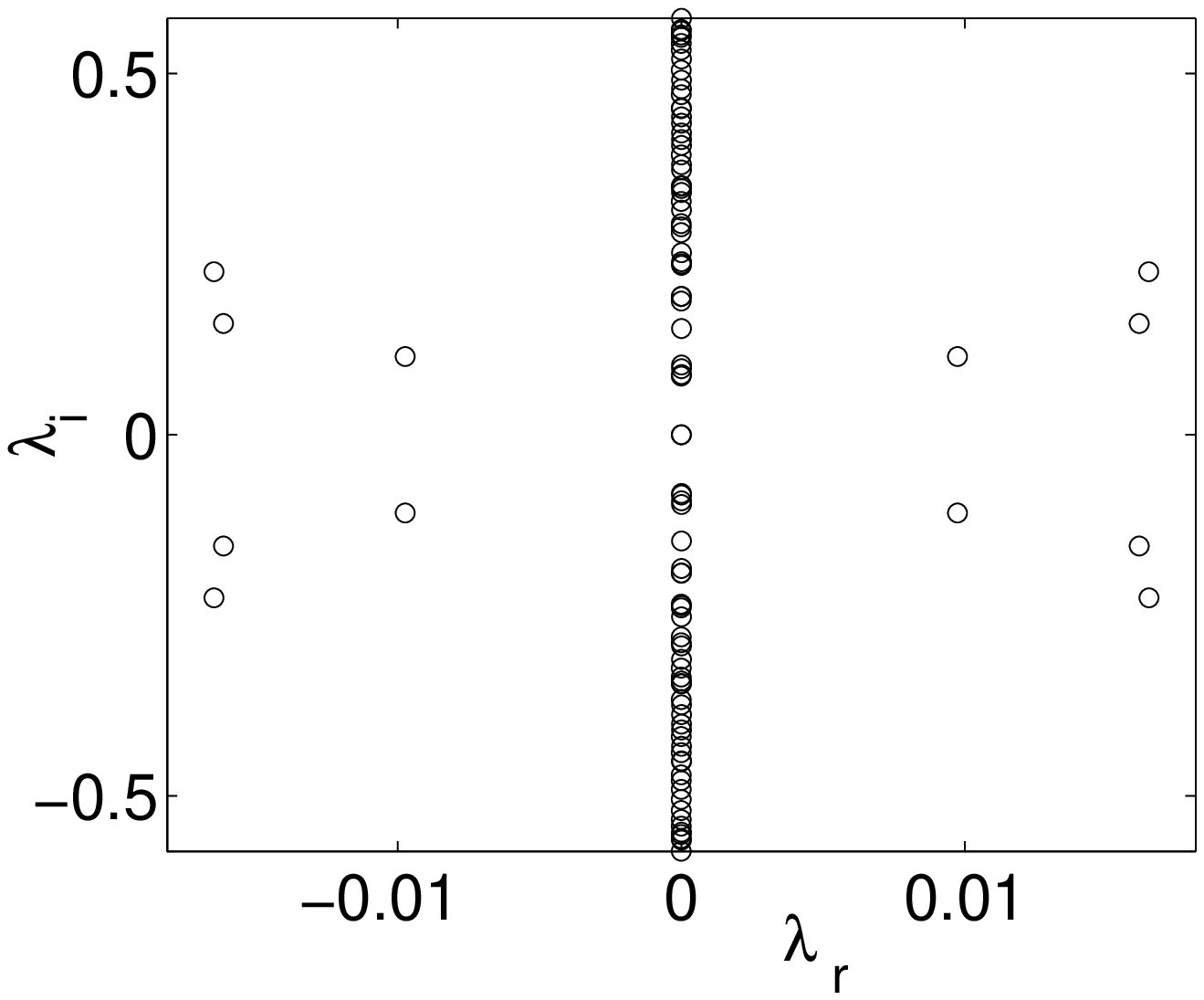}~~~
\caption{(Color Online) Same as the previous figures but
for the $|1,0\rangle+|0,1\rangle$ state in the case of repulsive interatomic interactions.
This state is always unstable due to at least an eigenvalue quartet and up
to two other real pairs. Note that there exists a region of weak instability, where only
small magnitude quartets are present.
The middle and bottom panels show the contour plots of this state
and its linearization spectrum for
$(V_0,\mu)=(0.2,0.55)$ (see blue cross in top panels) 
and $(0.37,0.59)$ (see red circle in top panels), respectively.}
\label{vfig4}
\end{figure}

The next state we consider 
is
the $|1,0\rangle+|0,1\rangle$ state which is presented in Fig.~\ref{vfig4}.
This state
always possesses a quartet of complex eigenvalues,
and up to two additional pairs of real eigenvalues, and is
thus unstable for all $\mu$.
It it
worth noticing, however, that the instability weakens to relatively
benign small magnitude complex quartets for intermediate
values of the chemical potential, roughly $\mu \in (0.4,0.9)$, and large lattice depths,
$V_0 >0.3$. This suggests that such a configuration should be
long-lived enough that it could be observable in
experiments with
repulsive condensates.

\begin{figure}[tbp]
\includegraphics[width=8cm]{\rootfig 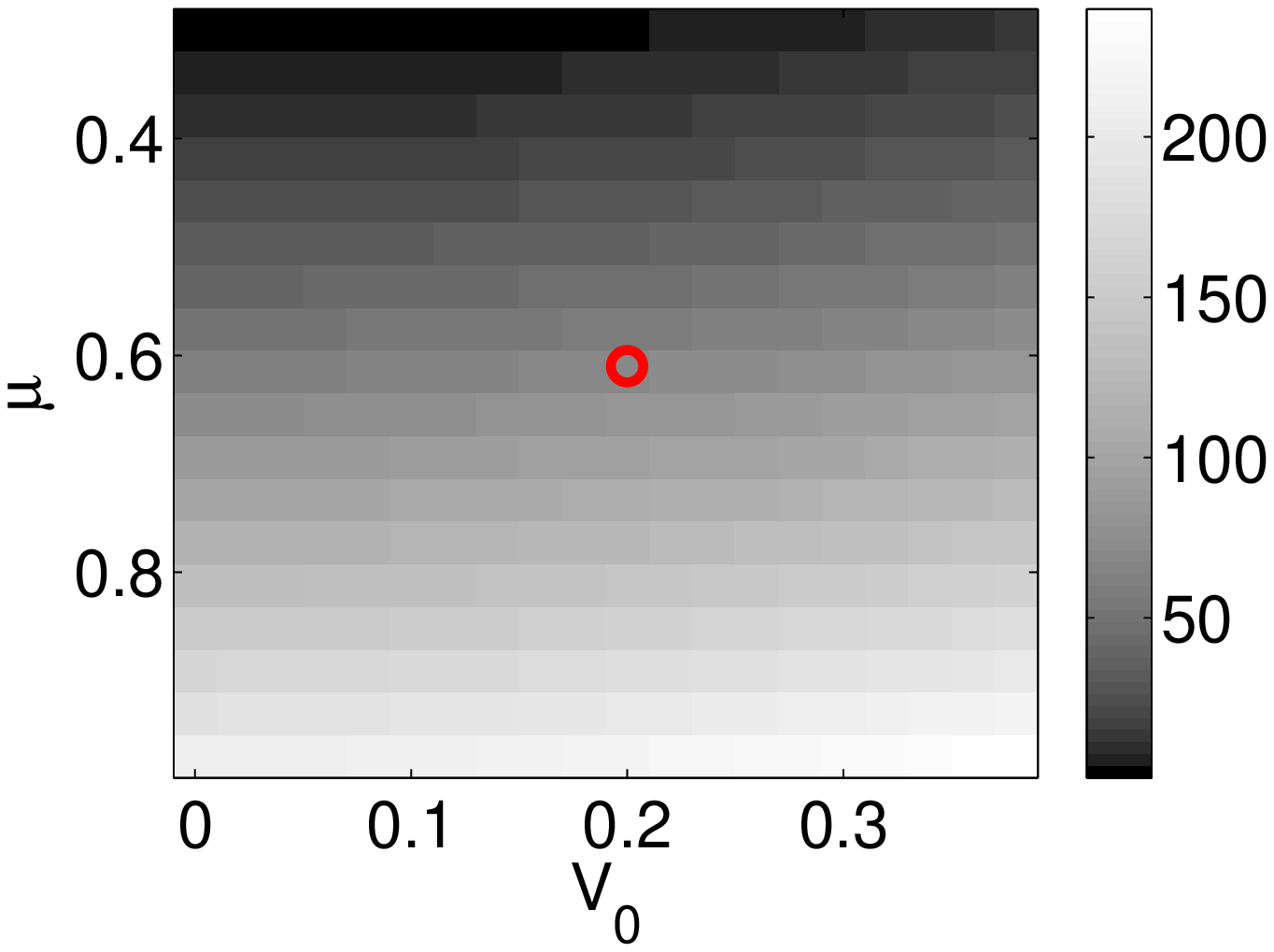}
\includegraphics[width=8cm]{\rootfig 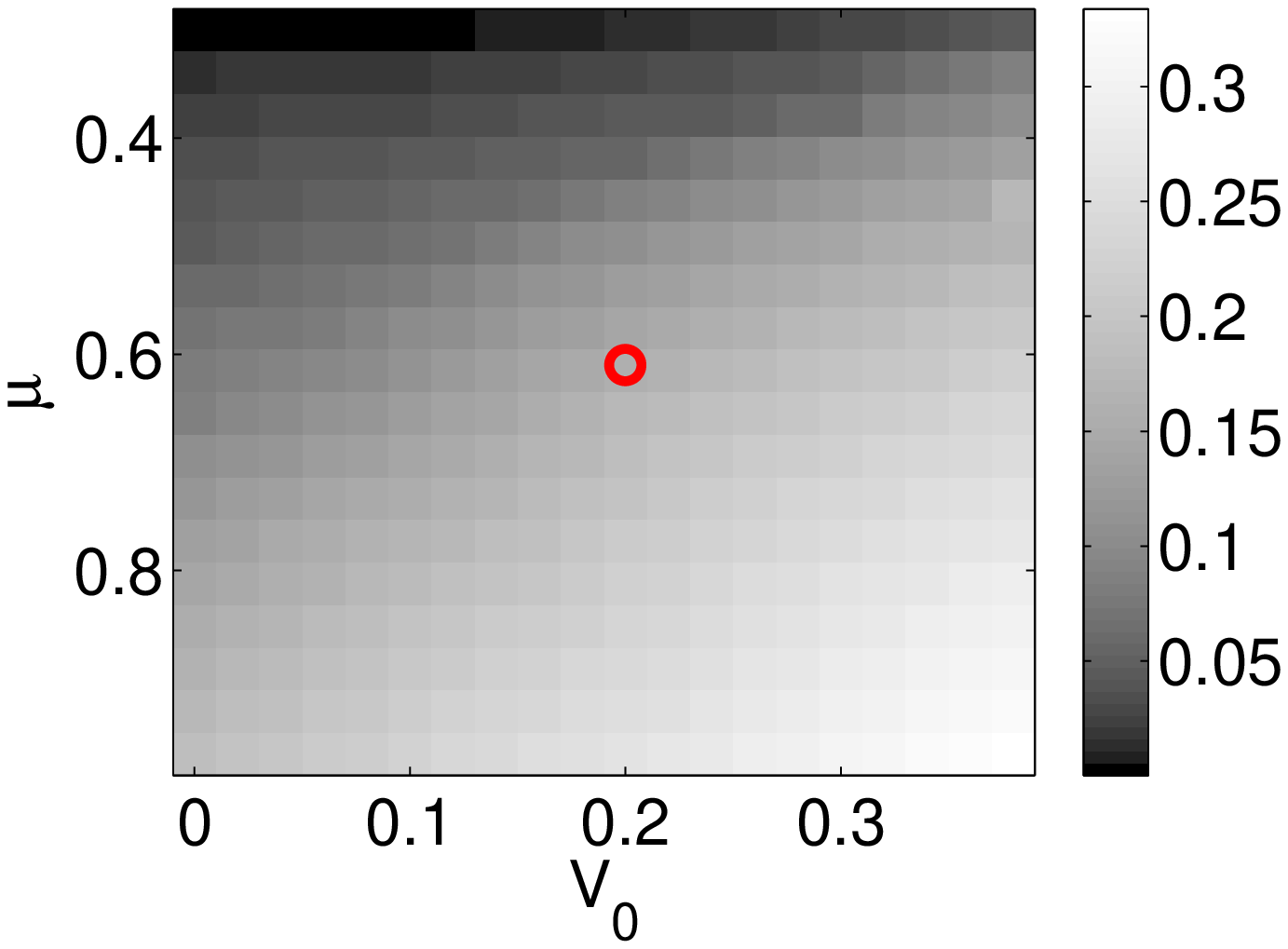}
\\
\includegraphics[width=8cm]{\rootfig 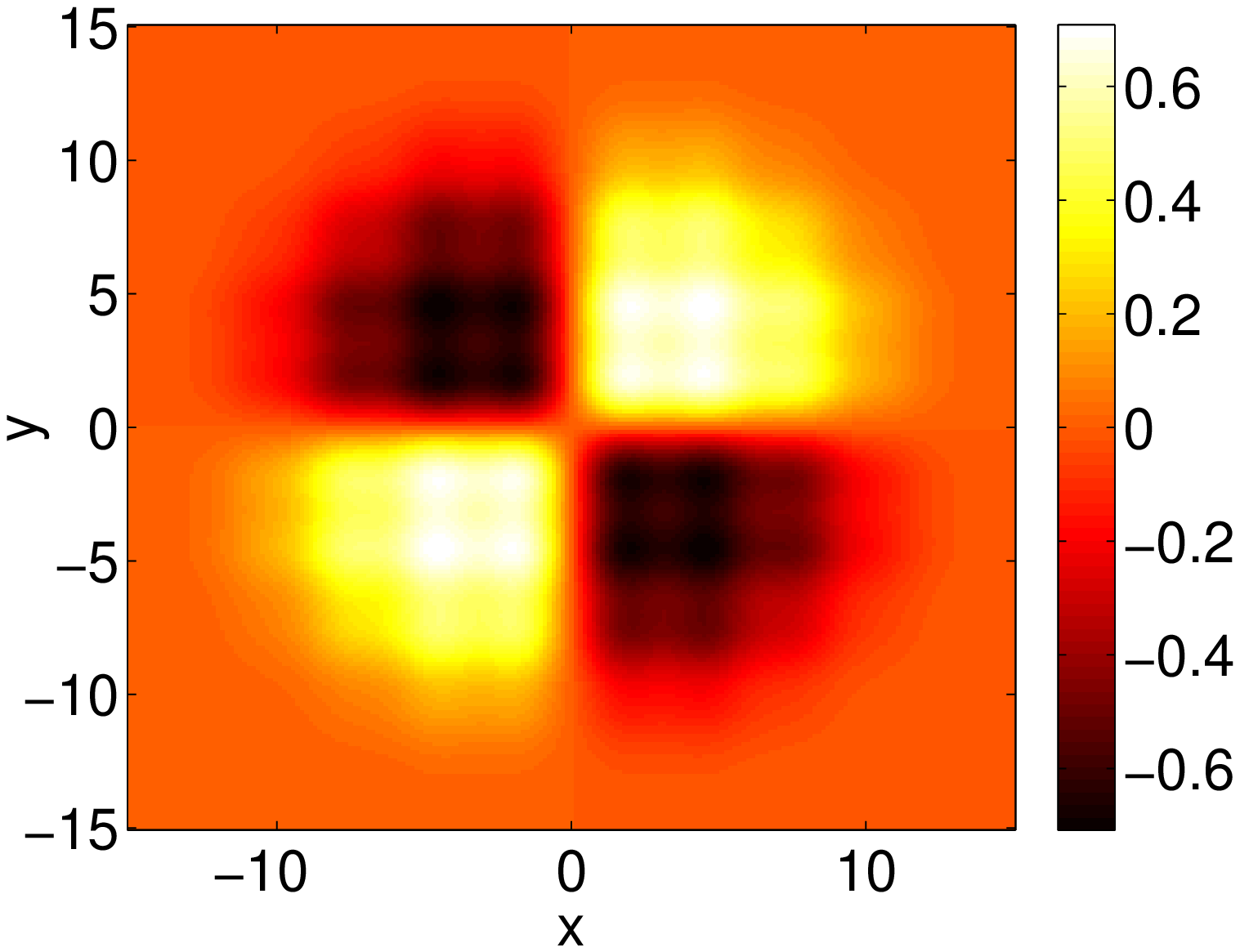}
~~\includegraphics[width=7.65cm,height=6.75cm]{\rootfig 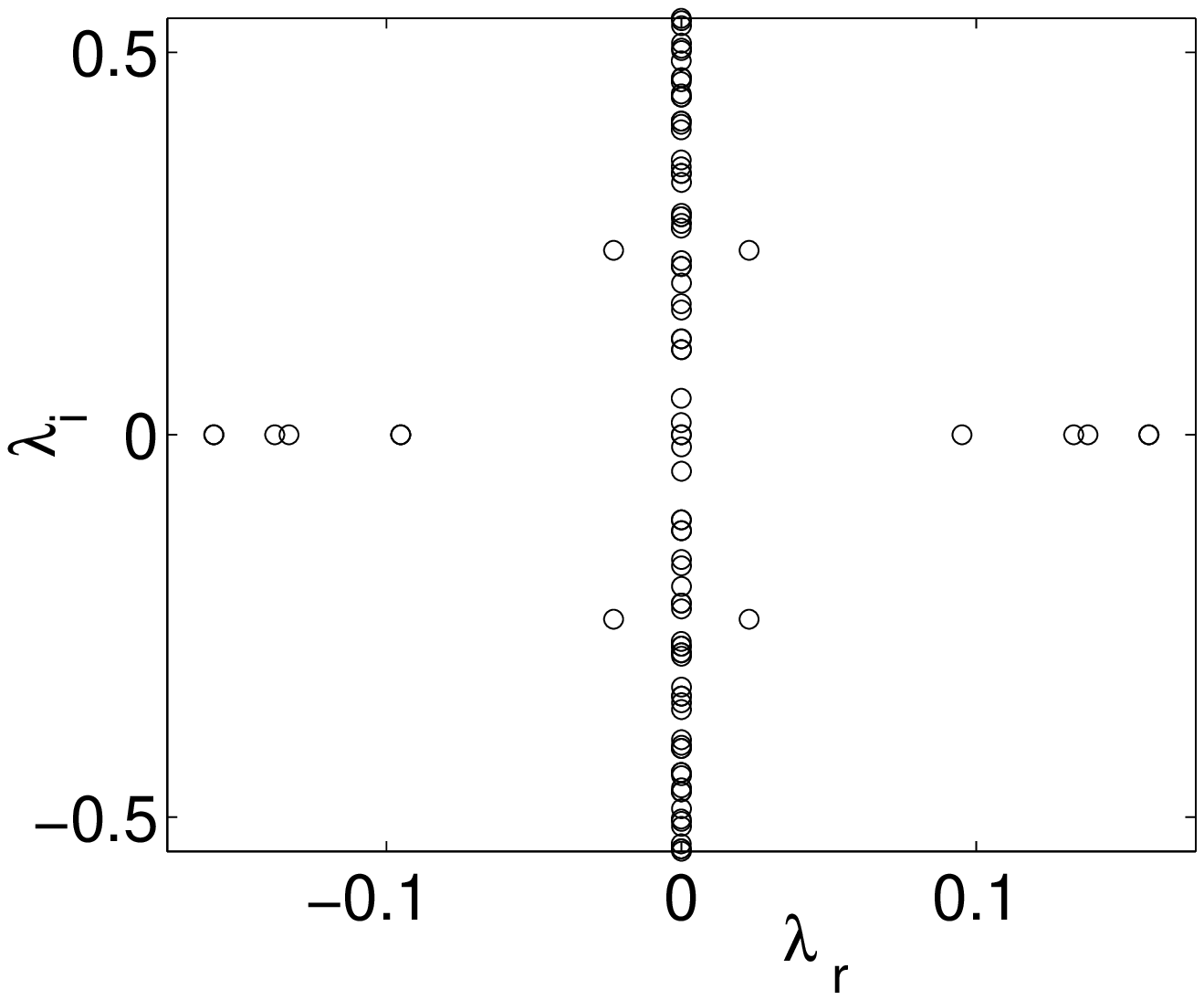}~~~
\caption{Same as in Fig.~\ref{vfig2}, but for the
state $|1,1\rangle$ in the case of repulsive interatomic interactions
for parameter values $(V_0,\mu)=(0.2,0.61)$. }
\label{vfig3}
\end{figure}

Next, we consider the
states with $n+m=2$ (again for repulsive interatomic interactions),
starting with the $|1,1\rangle$ branch in Fig.~\ref{vfig3}.
The branch is also always unstable, possessing a complex
quartet and then up to four additional real pairs for
larger values of $\mu$. The instability is shown in the
right subplots of Fig.~\ref{vfig3}, where the spectral
plane of the bottom right panel corresponds to the solution of the bottom left one,
for parameter values $(V_0,\mu)=(0.2,0.61)$. It is interesting to note
that such
states are reminiscent of the domain-walls presented in Ref.~\cite{malom} 
(here the domain-wall is imposed by the difference in phase), which however
were found as potentially stable structures
in multi-component condensates.


\begin{figure}[tbp]
\includegraphics[width=8cm]{\rootfig 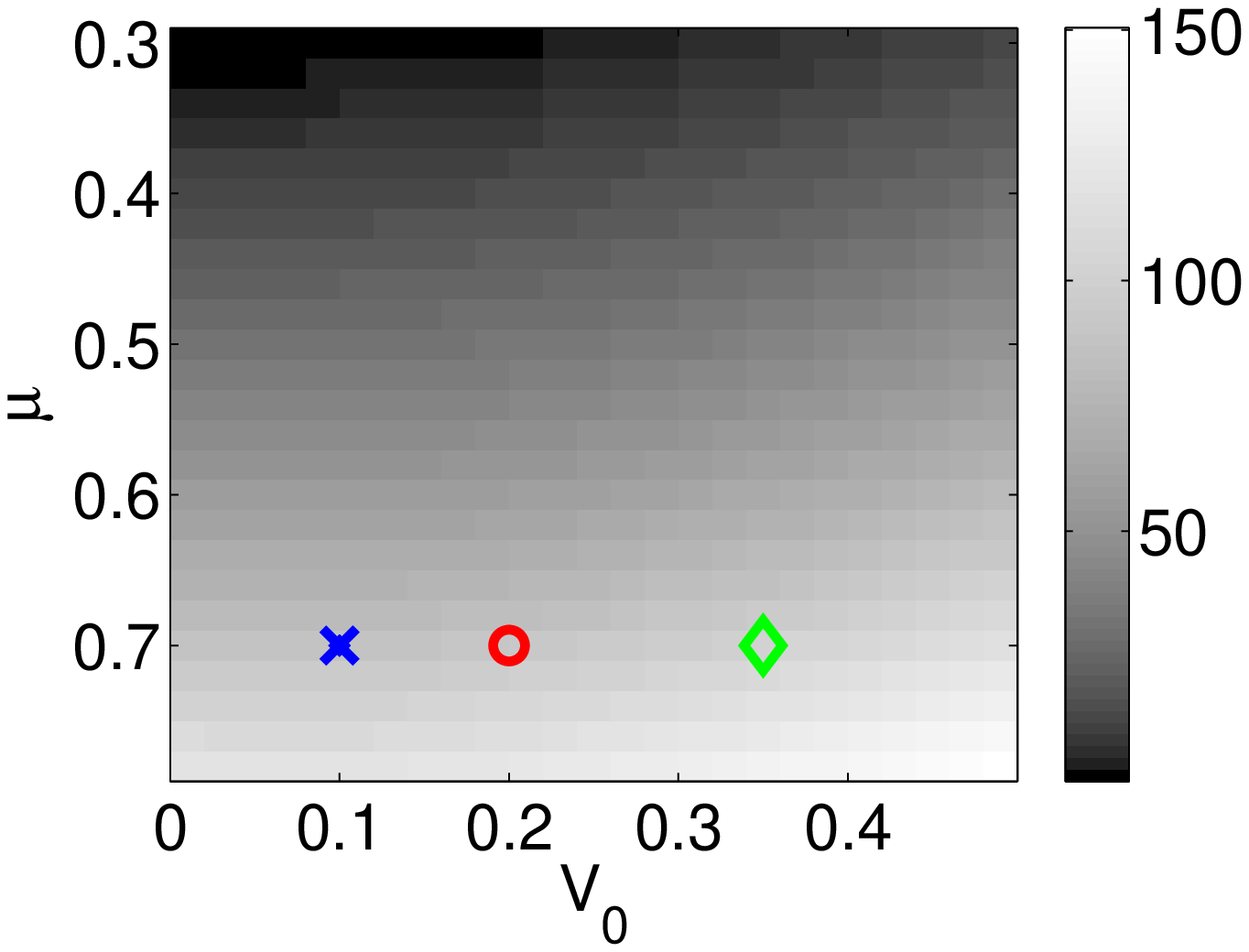}
\includegraphics[width=8cm]{\rootfig 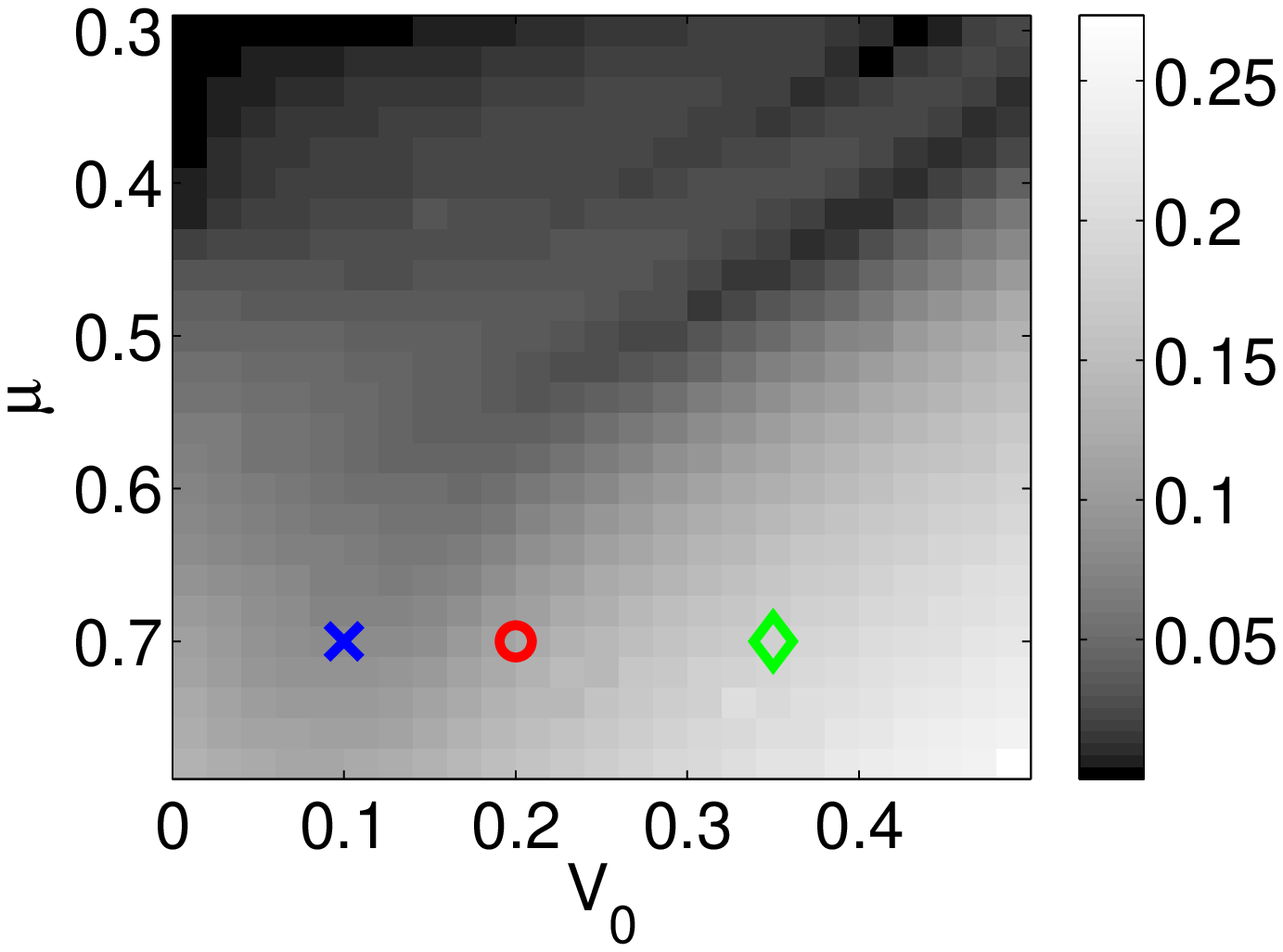}
\\
\includegraphics[width=8cm]{\rootfig 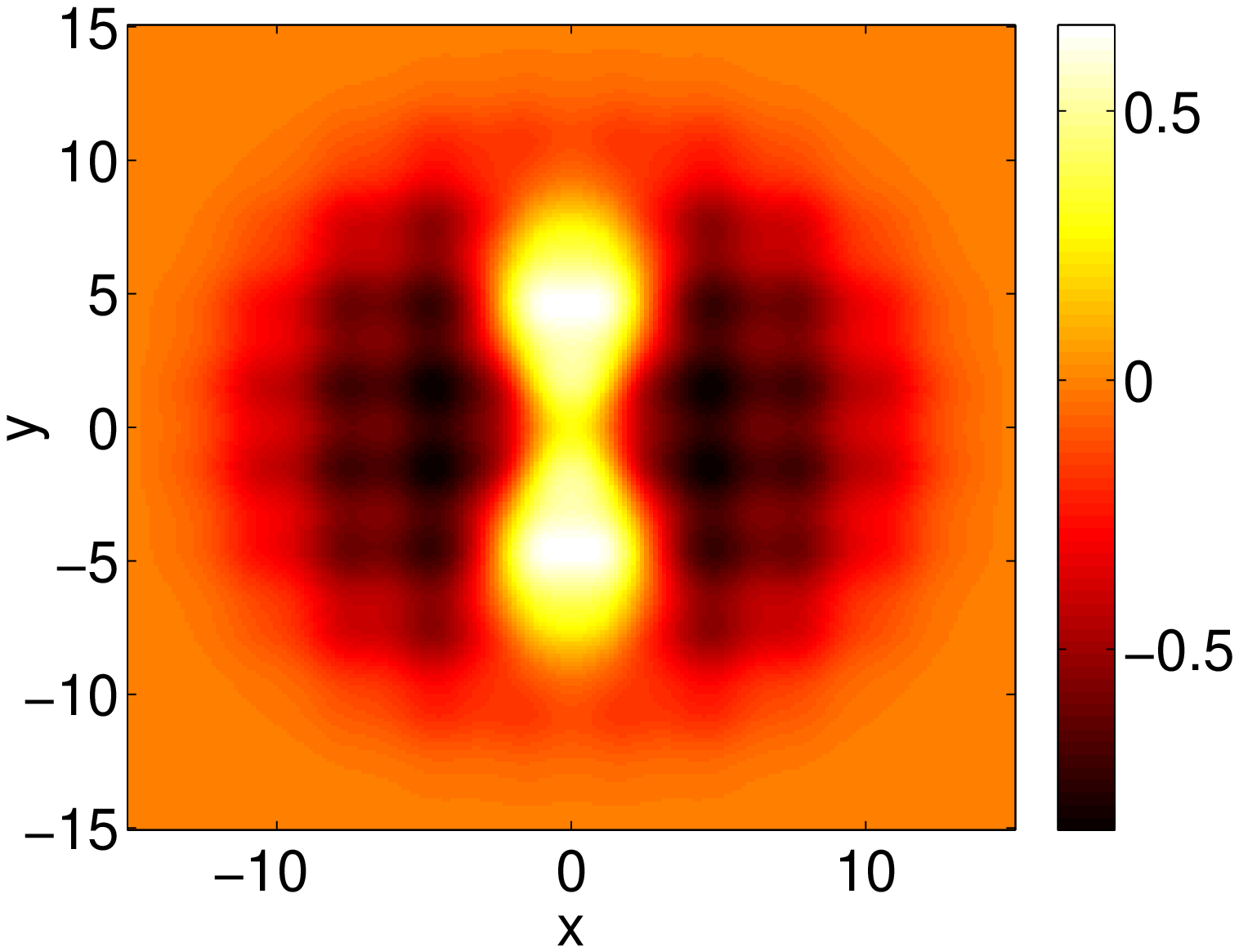}
~~\includegraphics[width=7.65cm,height=6.75cm]{\rootfig 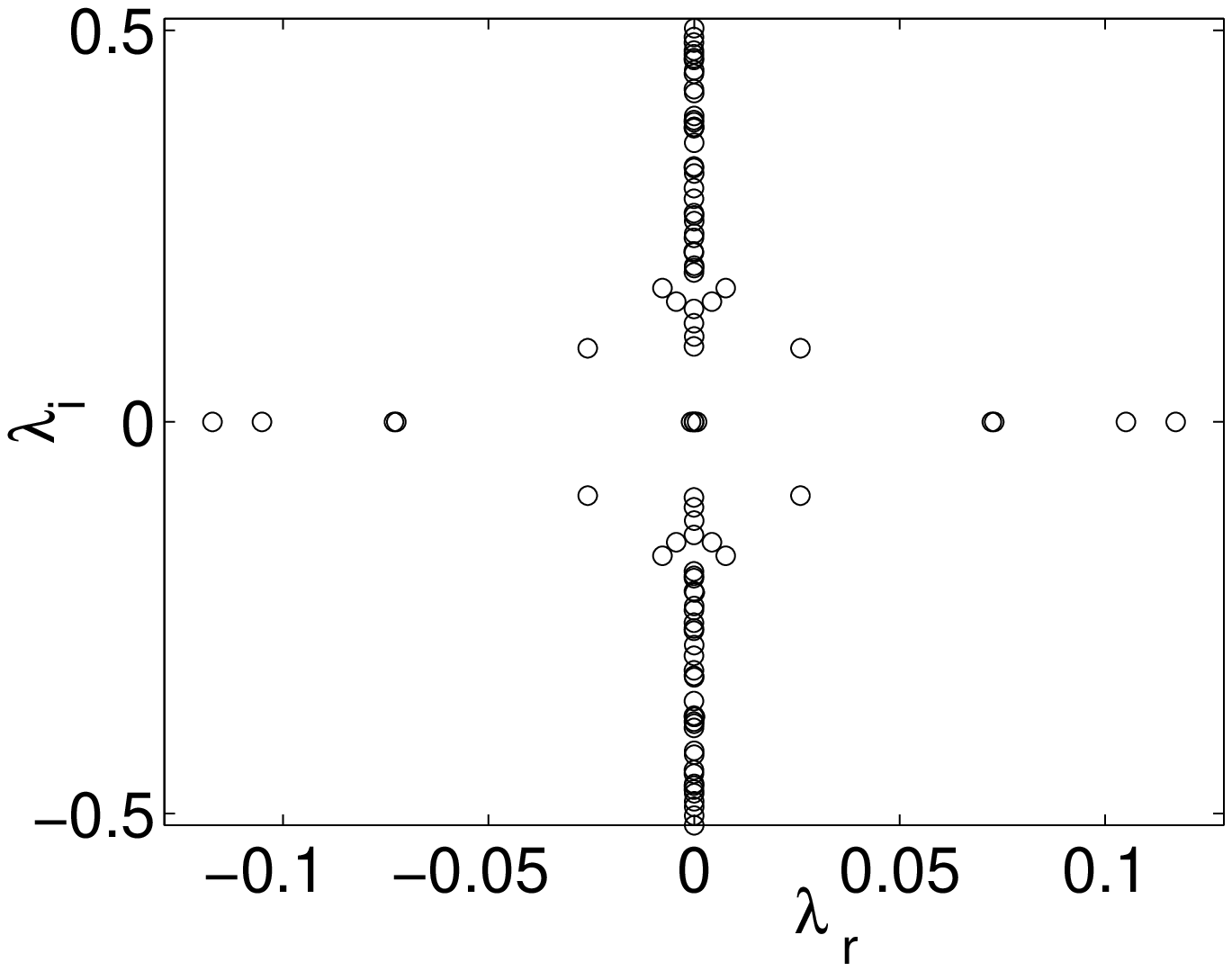}~~~
\\
\includegraphics[width=8cm]{\rootfig 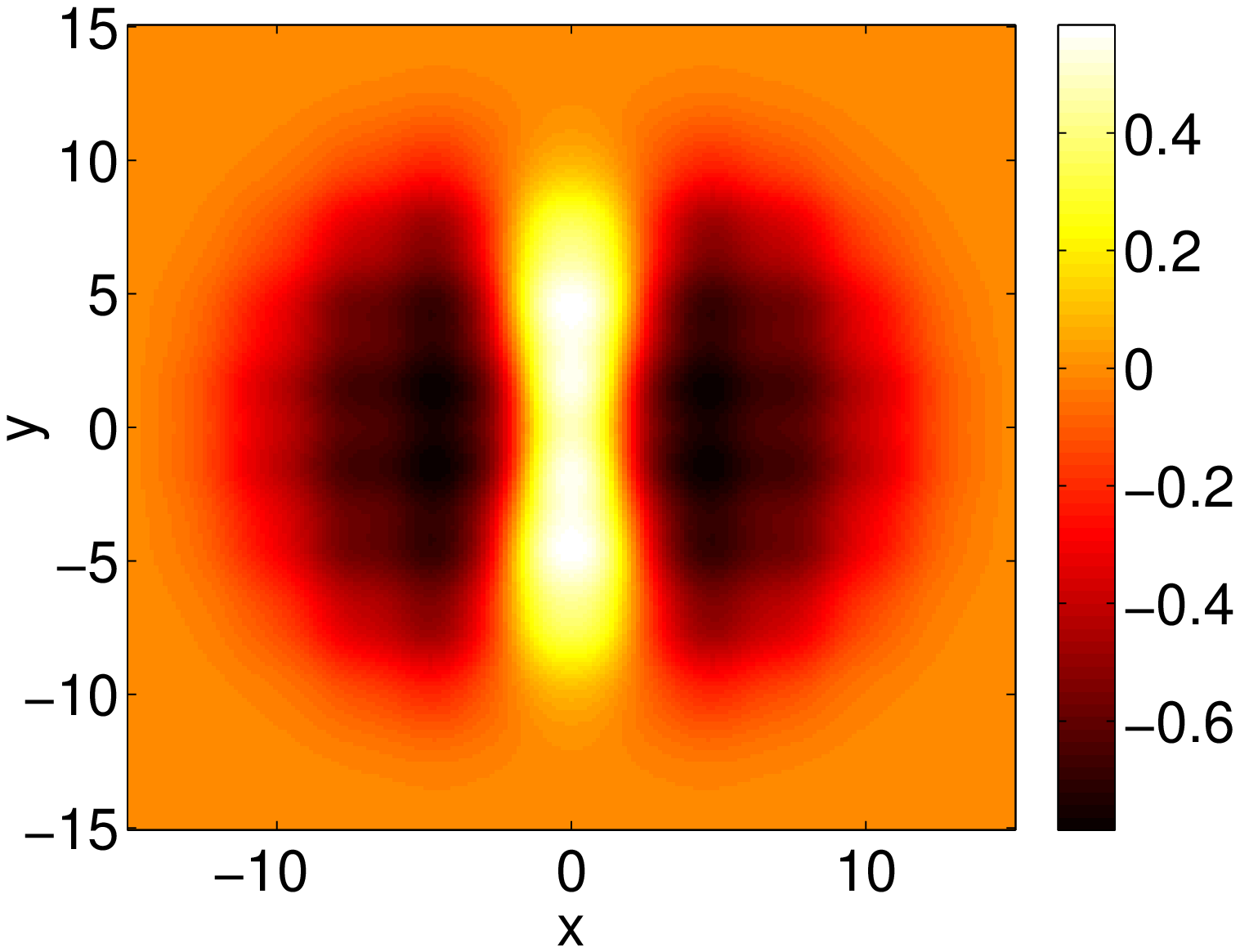}
\includegraphics[width=8cm]{\rootfig 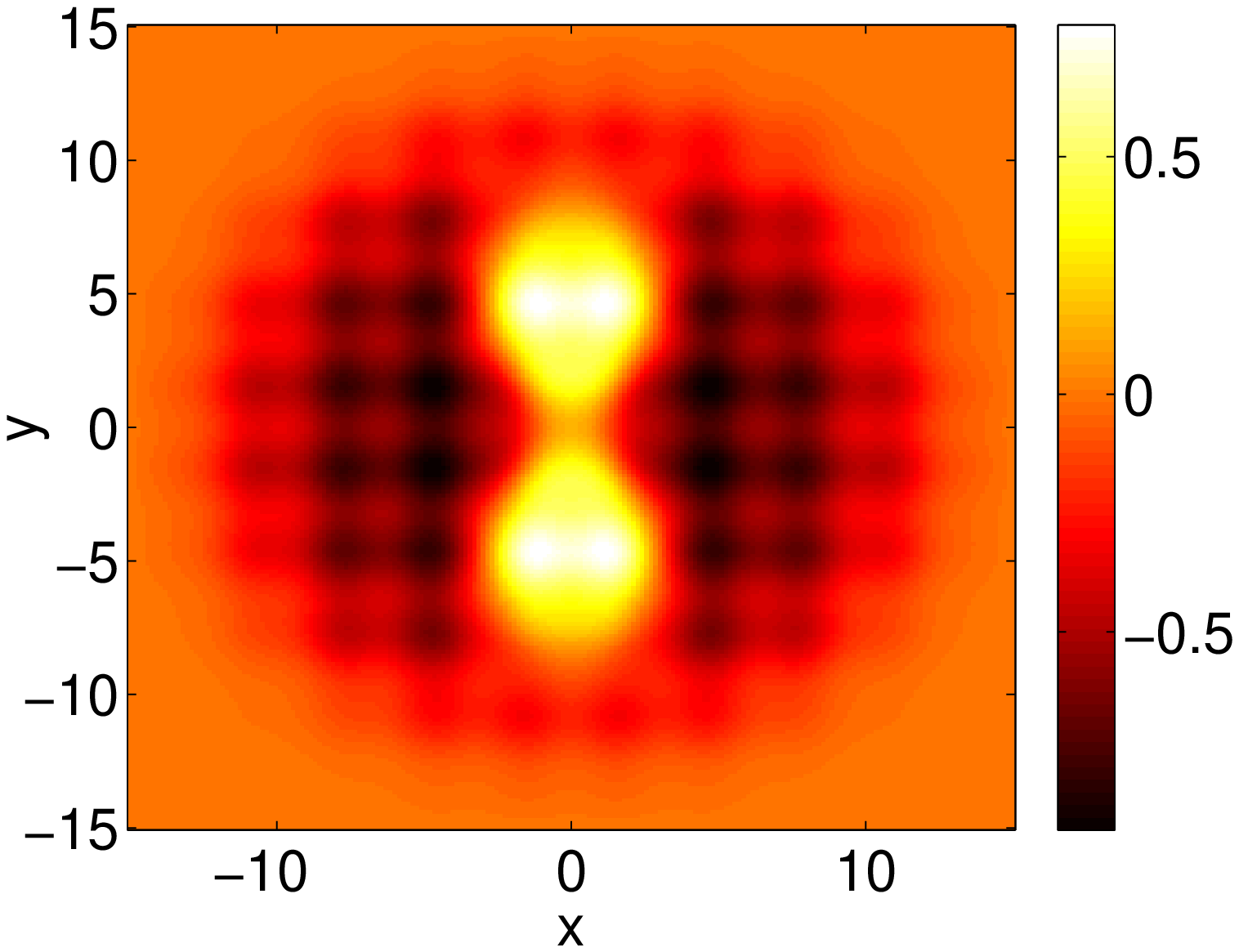}
\caption{(Color Online) Same as in Fig.~\ref{vfig2}, but for the
state $|2,0\rangle$ (in the case of repulsive interatomic interactions) 
with parameter values $(V_0,\mu)=(0.2,0.7)$ (see red circle in top panels).
The bottom row shows profiles for smaller, $V_0=0.1$ (left, see blue cross in top panels), 
and larger, $V_0=0.35$ (right, see green diamond in top panels), 
values of the optical lattice depth.}
\label{vfig20d}
\end{figure}

Next, the case of the $|2,0\rangle$ state
is shown in Fig.~\ref{vfig20d}. Here, there are up to three complex quartets
along with three real pairs of eigenvalues.
It is notable that for higher values of the lattice
depth, these states are
deformed as the lattice ``squeezes'' the central maximum separating
the two minima
(see middle left panel of the figure). The contour plot shown in the
bottom right panel suggests that further
increase of lattice
depth may lead to a new configuration
altogether when the two local maxima eventually pinch off of
each other.
This deformation is a direct consequence of the presence of the (repulsive) nonlinearity, which
results in drastically different configurations as compared to
the linear limit of the structure.

\subsubsection{Dynamics}

\begin{figure}[t]
\begin{center}
\begin{tabular}{cccc}
(a)&
(b)&
(c)&
(d)\\
\includegraphics[height=5.5cm]{\rootfig 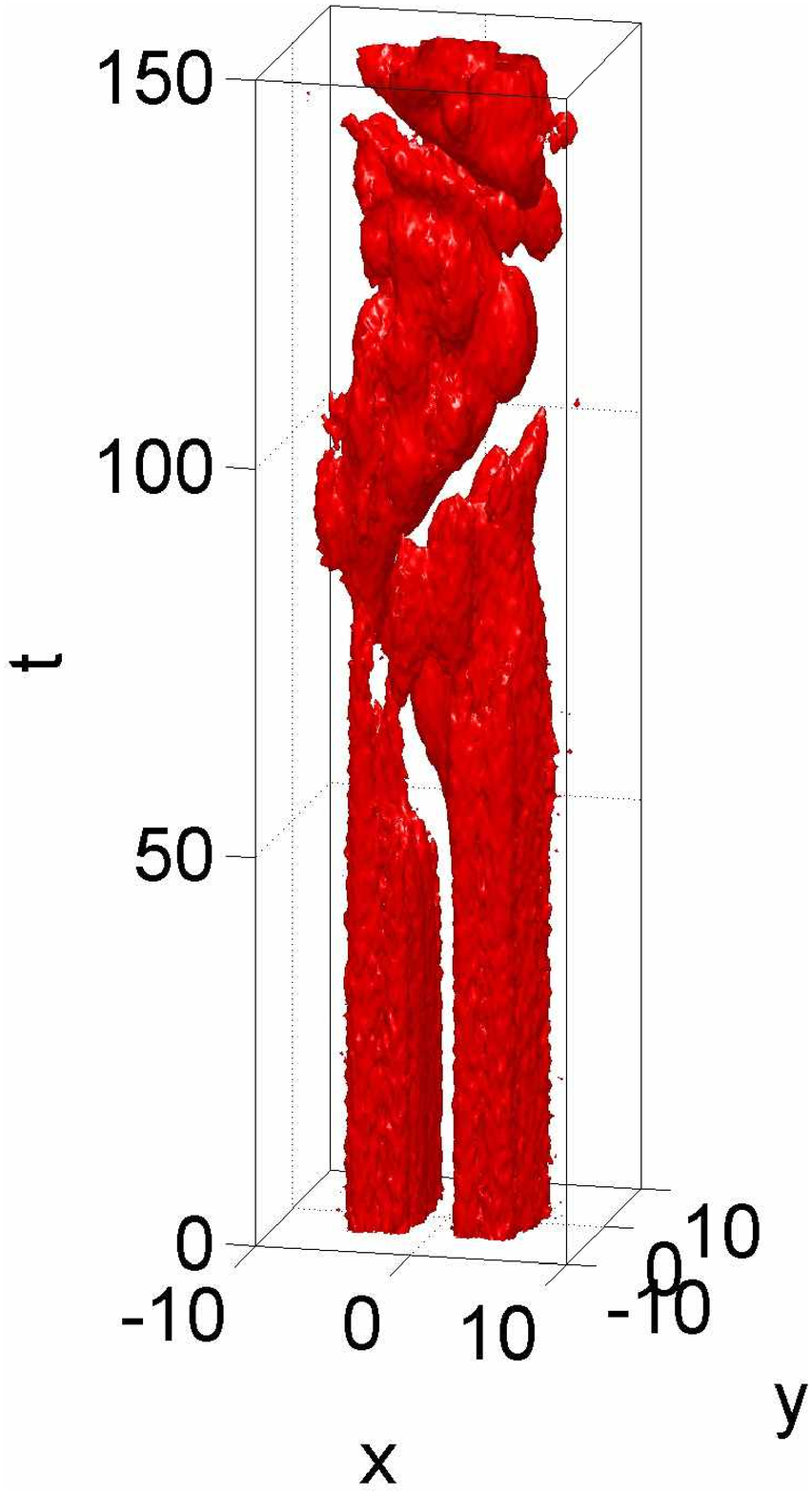}&
\includegraphics[height=5.5cm]{\rootfig 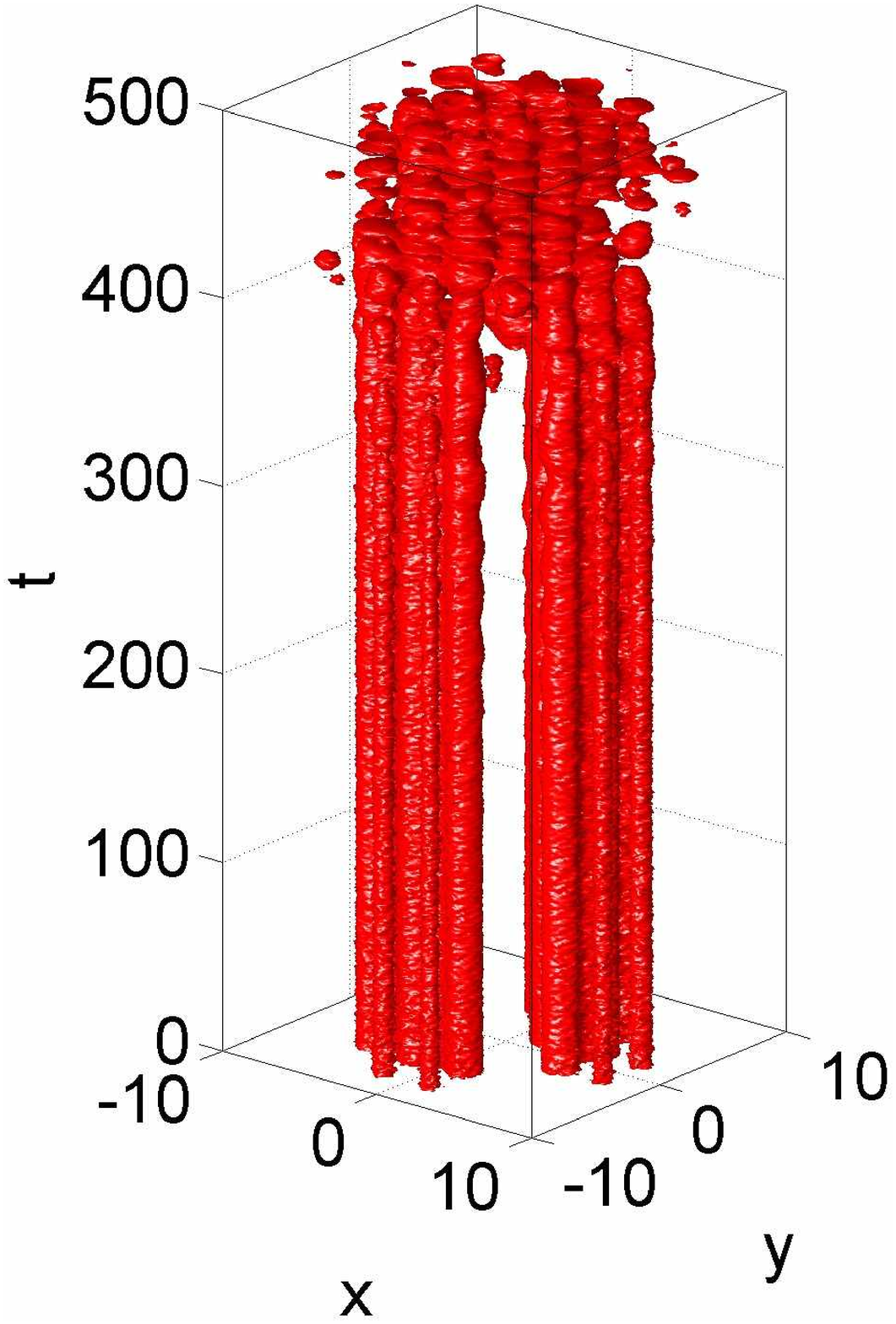}&
\includegraphics[height=5.5cm]{\rootfig 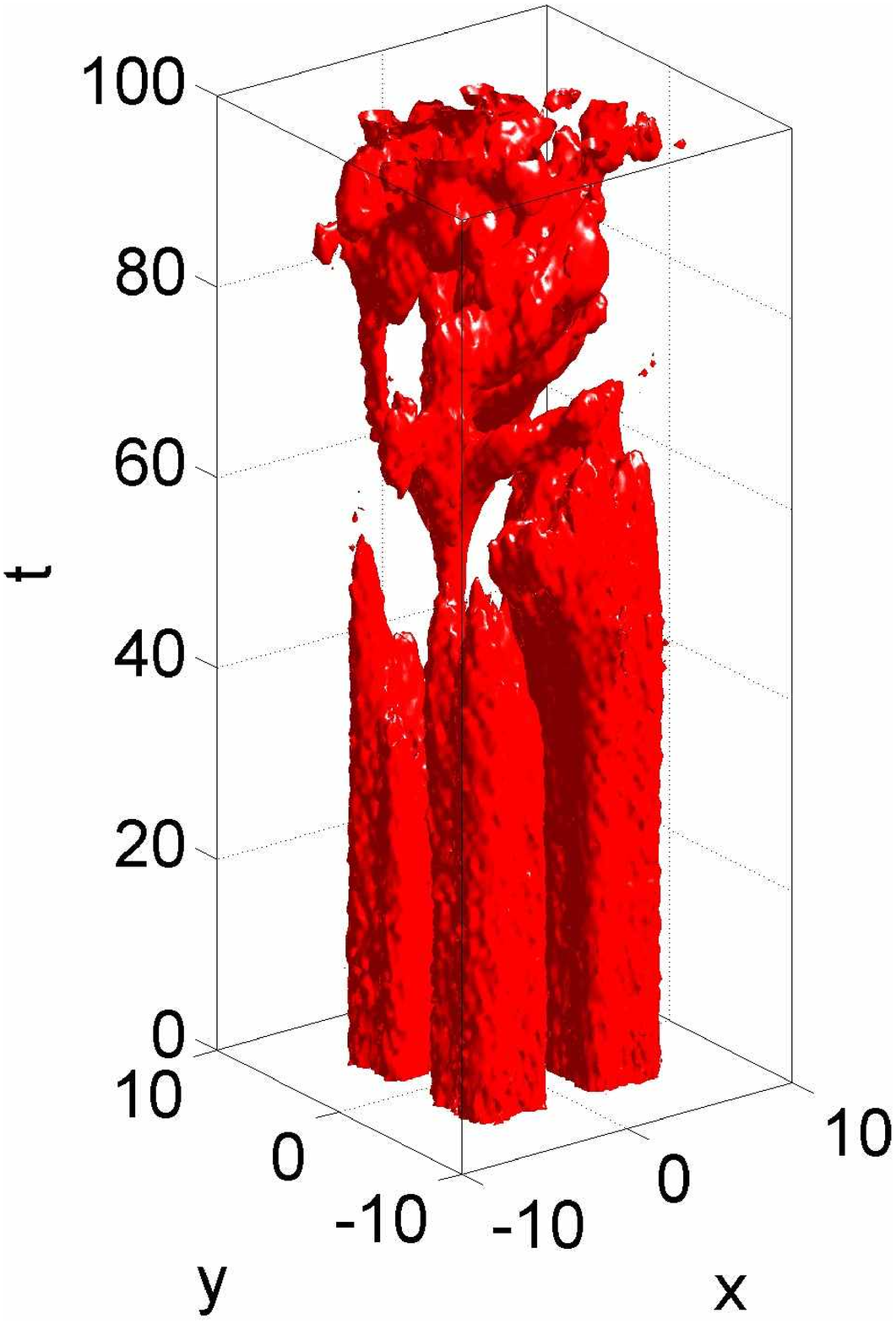}&
\includegraphics[height=5.5cm]{\rootfig 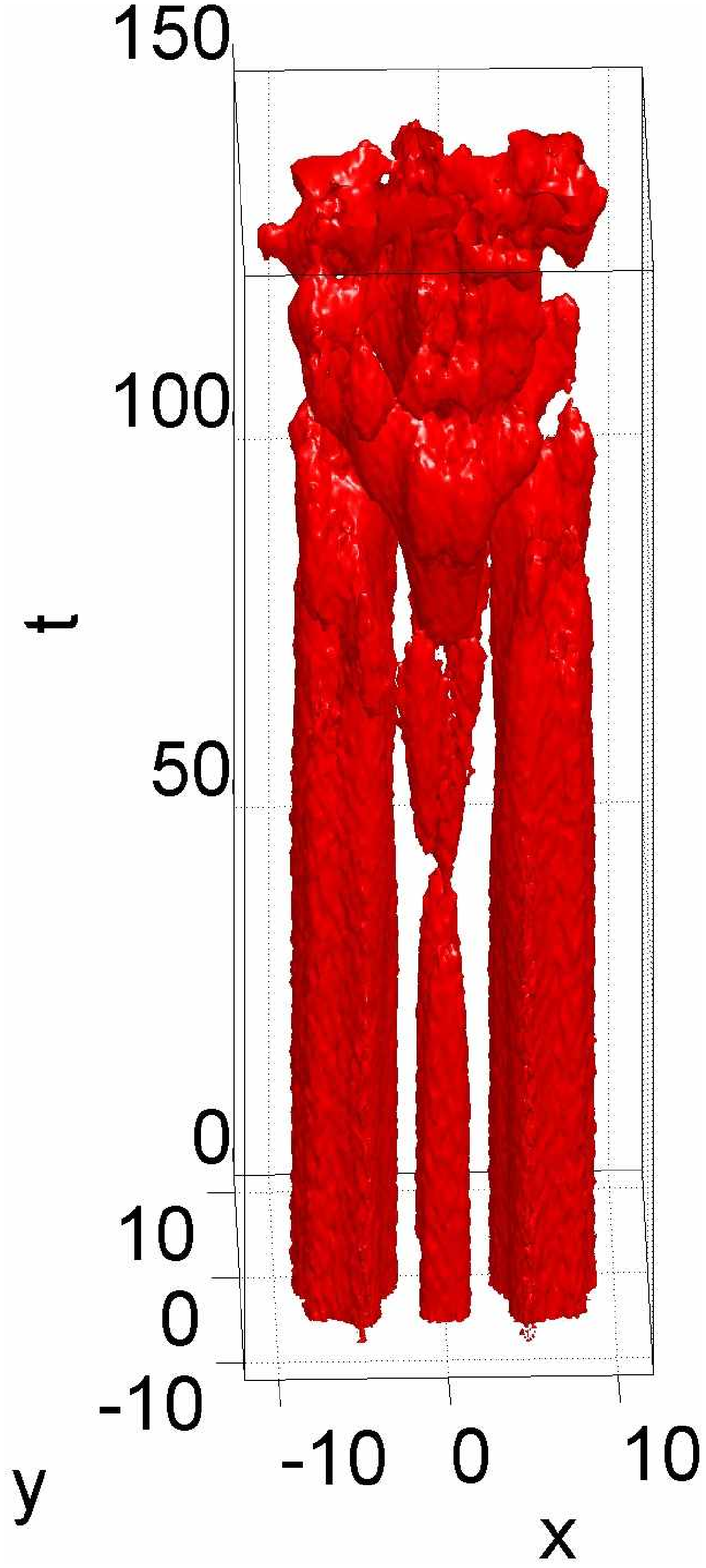}
\end{tabular}
\caption{(Color Online) 
The dynamics of the unstable states in the case of repulsive interatomic interactions.
Panels (a) and (b) show, respectively, the evolution of the states 
$|1,0\rangle$ and $|1,0\rangle+|0,1\rangle$.
It is clear that
the state $|1,0\rangle+|0,1\rangle$
is subject to a weaker oscillatory instability
for the parameter values
mentioned in the bottom row of Fig.~\ref{vfig4}
and, as a result, the original configuration persists for a long time.
Panels (c) and (d) show the dynamics
of the states with $m+n=2$, namely (c) $|1,1\rangle$, and (d) $|2,0\rangle$.
All these solutions ultimately degenerate into ground-state-like
configurations. The density isosurfaces are taken at $a=0.5$ with the exception of (d) at $a=0.4$.}
\label{df_dyno}
\end{center}
\end{figure}

We performed numerical simulations to investigate the evolution of
typical unstable states in the case of
repulsive interatomic interactions, using similar time-stepping
schemes as discussed above in the case of attractive interactions.
Apart from the ground state, all excited states presented
in the previous section were predicted to be unstable, and this is
confirmed in this section.
In the particular case of the state $|1,0\rangle+|0,1\rangle$ which was found
to be weakly unstable (see bottom row of Fig.~\ref{vfig4}),
the instability takes a considerable
time to manifest itself.
The evolution of this state
is depicted in panel (b) of
Fig.~\ref{df_dyno}; it is clearly seen that the state persists up to $t \approx 300$.
On the other hand, panel (a) shows the evolution of the state $|1,0\rangle$
which persists up to $t\approx 40$, while the bottom row of panels shows the
dynamics of the (c) $|1,1\rangle$ state
and of the (d) $|2,0\rangle$, both persisting also up to $t\approx 40$.
All of these excited states degenerate into ground-state-like configurations.
Notice that transient vortex-like structures seem to appear
during this process 
but they do not
persist in the eventual dynamics
and are hence not further discussed here.
%

\section{Conclusions and discussion}

In summary, we have studied the structure and the stability
of a pancake-shaped condensate (with either attractive or repulsive interatomic interactions)
confined in a potential with both a harmonic and an optical lattice component.
Starting from the non-interacting limit, and exploiting the smallness of the
harmonic trap strength, we have employed a multiscale perturbation method to find the
discrete energies and the corresponding eigenmodes of the pertinent 2D linear
Schr\"{o}dinger equation. Then, we used the results found in this
linear (non-interacting) limit
in order to identify states
persisting in the nonlinear (interacting)
regime as well. This investigation revealed that 
the most fundamental states (emanating from combinations of the ground state
and the first few excited states in the two orthogonal directions of the optical lattice)
can indeed be continued in the nonlinear regime. To demonstrate this continuation, we used
two-parameter diagrams involving the effective strength of
the nonlinearity (through the chemical potential) and the optical lattice depth.

Excited states were typically found to be unstable. The instability was found to result in
either wavefunction collapse or a robust single-lobed structure in
the case of attractive interactions; on the other hand, in the case
of repulsive interactions, the instability was always found to lead
to the ground state of the system.
Nevertheless, noteworthy exceptions of stable or very weakly unstable states were also revealed.
These include
the $|1,0\rangle+|0,1\rangle$ and the $|1,1\rangle$ states
in the case of attractive interatomic interactions. Moreover, in the case of
repulsive interactions, the same state, $|1,0\rangle+|0,1\rangle$,
was found (in
certain parameter regimes) to be only very weakly unstable. Direct numerical
simulations confirmed that the instability of this state is indeed weak, and it manifests itself at
large times, an order of magnitude larger than the ones pertaining to the manifestation of instabilities
of other excited states. Thus, it is clear that the obtained
results suggest that the state $|1,0\rangle+|0,1\rangle$ has a good chance
to be observed in experiments with either an attractive or a repulsive pancake condensate.
It is especially important to highlight that these states are
stabilized (or quasi-stabilized) only in the presence of a sufficiently
strong optical lattice; hence the latter potential plays a {\it critical}
role in determining the stability of the states presented herein.

As described in Section III B, the parameters used in our analysis have been chosen in order 
to facilitate convenience of the numerical computations, while also within range of experimentally 
achievable limits of atom number, chemical potential, harmonic oscillator frequencies, and optical lattice 
depth and periodicity. Furthermore, we believe that the stability and spatial structure of the states 
examined here can be examined experimentally. For example, we imagine utilizing a BEC held in a 
pancake-shaped harmonic trap, created by an optical field. By using an optical trap rather than a magnetic trap, 
the scattering length of a BEC may be adjusted using a Feshbach resonance. We envision that 
an optical lattice potential is ramped on and superimposed on a BEC with an interatomic scattering length 
tuned to be near zero. Once the lattice has reached the desired depth, the scattering length can be further 
adjusted with a magnetic field to be either positive or negative (the latter option would need to be within 
a region of stability that does not result in collapse of the BEC). Finally, phase imprinting 
techniques \cite{phimp}
can be used 
to generate the desired phase profile of the BEC. By optically examining the state of the BEC at various points 
in time after phase profile imprinting, the stability of the generated states can then be examined and compared 
with our numerical results and stability analysis. For example, with an optical lattice frequency of 
$\omega_L = 2\pi \times 120$ Hz (as in the example of Section III B), the time unit of our 
dynamical evolution plots is 1.3 ms. 
This implies that for the cases we have examined, signatures of instability would be typically visible on 
the experimentally feasible 10 to 100 ms timescale. We therefore believe that our predictions could be examined 
with current experimental techniques.

There are various directions along which one can extend the present
considerations. A natural one is to extend the analysis
to fully
3D condensates and examine the persistence
and stability of higher-dimensional variants of the presented states.
A perhaps more subtle
direction is to consider a different
basis of linear eigenfunctions in the 2D problem, namely instead of
the Hermite-Gauss basis used here, to focus on the Laguerre-Gauss basis
of the underlying linear problem with the parabolic potential. Under such a choice, it would be
interesting to
examine how solutions of that type, including one-node and multi-node
ring-like structures (see, e.g., Ref.~\cite{gregh} and references therein),
are deformed in the presence of the lattice and how their stability
is correspondingly affected. 
Finally, as discussed above, it appears that the setting
considered herein should be directly accessible to present experiments
with pancake-shaped BECs. In view of that, it would be particularly relevant
to examine which ones among the
structures presented in this work can survive
for evolution times that are of interest within the time scales of an experiment.
\\

\ack{P.G.K. and R.C.G. gratefully acknowledge the support of NSF-DMS-0505663,
and P.G.K. additionally acknowledges support from NSF-DMS-0619492,
NSF-CAREER and the Alexander von Humboldt Foundation.
B.P.A. acknowledges support from the
Army Research Office and NSF Grant No.~MPS-0354977.
The work of D.J.F. was partially supported by the Special Research Account of
the University of Athens.}
\\
\\


\begin{thebibliography}{99}

\bibitem{book1} C.J. Pethick and H. Smith,
{\it Bose-Einstein Condensation in Dilute Gases},
Cambridge University Press (Cambridge, 2001).

\bibitem{book2} L. Pitaevskii and S. Stringari,
{\it Bose-Einstein Condensation}, Oxford University Press
(Oxford, 2003).

\bibitem{rmp} F.\ Dalfovo,
S.\ Giorgini, L. P.\ Pitaevskii, and S.\ Stringari,
Rev.\ Mod.\ Phys.\ \textbf{71}, 463 (1999).

\bibitem{ourbook} 
P.G. Kevrekidis, D.J. Frantzeskakis, and R. Carretero-Gonz{\'a}lez (eds.),
{\it Emergent nonlinear phenomena in Bose-Einstein condensates. Theory and experiment}
(Springer-Verlag, Berlin, 2008).

\bibitem{bloch:02} M.\ Greiner, 
O.\ Mandel, T.\ Esslinger, T.W.\ H\"{a}nsch, and I.\ Bloch,
Nature (London) \newblock {\bf 415}, 39 (2002).

\bibitem{abdu} F.Kh. Abdullaev, A. Gammal,
A.M. Kamchatnov and L. Tomio,
Int. J. Mod. Phys. B {\bf 19}, 3415 (2005).

\bibitem{fetter} A.L.\ Fetter and A.A.\ Svidzinksy, \newblock J.\ Phys.:
Cond.\ Matter \textbf{13}, R135 (2001).

\bibitem{pgk1} P.G. Kevrekidis, R. Carretero-Gonz{\'a}lez,
D. J. Frantzeskakis and I. G. Kevrekidis, Mod. Phys. Lett. B
{\bf 18}, 1481 (2004).

\bibitem{pgk2} P.G. Kevrekidis and D.J. Frantzeskakis,
Mod. Phys. Lett. B {\bf 18}, 173 (2004).

\bibitem{konotop} V. Brazhnyi and V.V. Konotop,
Mod. Phys. Lett. B {\bf 18}, 627 (2004).

\bibitem{morsch} O. Morsch and M.K. Oberthaler, Rev. Mod. Phys. {\bf 78}, 179 (2006).

\bibitem{landau} L.D. Landau and E.M. Lifshitz, {\it Quantum Mechanics} (Pergamon Press, Oxford, 1987).

\bibitem{yuri1} Yu.S. Kivshar, T.J. Alexander, and S.K. Turitsyn, Phys. Lett. A {\bf 278}, 225 (2001).

\bibitem{kivshar} Yu.S. Kivshar and T.J. Alexander, in {\it Proceeding of the APCTP-Nankai
Symposium on Yang-Baxter Systems, Nonlinear Models and Their Applications}, edited by Q-Han Park {\it et al.} (World Scientific, Singapore, 1999).

\bibitem{konotop1}
P.G. Kevrekidis, V.V. Konotop, A. Rodrigues and D.J. Frantzeskakis
J. Phys. B: At. Mol. Opt. Phys. {\bf 38}, 1173 (2005).

\bibitem{mtol1} T. Kapitula and P.G. Kevrekidis, Chaos {\bf 15}, 37114 (2005);
T. Kapitula and P.G. Kevrekidis, Nonlinearity {\bf 18}, 2491 (2005).

\bibitem{gregh} G. Herring, L.D. Carr, R. Carretero-Gonz{\'a}lez, P.G. Kevrekidis,
and D.J. Frantzeskakis, 
Phys. Rev. A {\bf 77} 023625 (2008).

\bibitem{yuri2} E.A. Ostrovskaya, M.K. Oberthaler, and Yu.S. Kivshar in
{\it Emergent nonlinear phenomena in Bose-Einstein condensates. Theory and experiment}, 
P.G. Kevrekidis, D.J. Frantzeskakis, and R. Carretero-Gonz{\'a}lez (eds.),
(Springer-Verlag, Berlin, 2008).


\bibitem{gt} G. Theocharis, D.J. Frantzeskakis, P.G. Kevrekidis, R. Carretero-Gonz{\'a}lez,
and B.A. Malomed, Math. Comput. Simul. {\bf 69}, 537 (2005); Phys. Rev. E {\bf 71}, 017602 (2005).

\bibitem{bs} P.G. Kevrekidis, D.J. Frantzeskakis, R. Carretero-Gonz{\'a}lez, B.A. Malomed,
G. Herring, and A.R. Bishop, Phys. Rev. A {\bf 71}, 023614 (2005).

\bibitem{vortex} P.G. Kevrekidis, R. Carretero-Gonz{\'a}lez,
G. Theocharis, D.J. Frantzeskakis and B.A. Malomed
J. Phys. B {\bf 70}, 3647 (2003); K.J.H. Law, L. Qiao, P.G. Kevrekidis 
and I.G. Kevrekidis, arXiv:0803.3251. 

\bibitem{2dbec} A. G\"{o}rlitz, J.M. Vogels, A.E. Leanhardt, C. Raman, T.L. Gustavson,
J.R. Abo-Shaeer, A.P. Chikkatur, S. Gupta, S. Inouye, T. Rosenband, and W. Ketterle,
Phys. Rev. Lett. {\bf 87}, 130402 (2001);
D. Rychtarik, B. Engeser, H.-C. N\"{a}gerl, and R. Grimm, Phys. Rev. Lett. {\bf 92}, 173003 (2004).

\bibitem{gpe1d} V. M.\ P\'{e}rez-Garc\'{\i}a, H.\ Michinel, and H.\ Herrero,
Phys.\ Rev.\ A \textbf{57}, 3837 (1998).

\bibitem{becol}
J.H. Denschlag, J.E. Simsarian, H. Haffner, C. McKenzie,
A. Browaeys, D. Cho, K. Helmerson, S.L. Rolston, and W.D. Phillips,
\newblock J. Phys. B: At. Mol. Opt. Phys. {\bf 35}, 3095
(2002).


\bibitem{sulem} C. Sulem and P.L. Sulem,
\newblock {\it The Nonlinear Schr{\"o}dinger Equation},
Springer-Verlag (New York, 1999).

\bibitem{baizakov} B.B. Baizakov, B.A. Malomed and M. Salerno, Europhys. Lett. {\bf 63}, 642 (2003).


\bibitem{Mihalache05}
D. Mihalache, D. Mazilu, F. Lederer, B.A. Malomed, L.-C. Crasovan,
Y.V. Kartashov, and L. Torner.
Phys. Rev. A {\bf 72}, 021601(R) (2005).

\bibitem{malom} B.A. Malomed, H.E. Nistazakis, D.J. Frantzeskakis, and P.G. Kevrekidis,
Phys. Rev. A {\bf 70}, 043616 (2004).



\bibitem{phimp} S. Burger, K. Bongs, S. Dettmer, W. Ertmer, K. Sengstock, A. Sanpera,
G.V. Shlyapnikov, and M. Lewenstein, Phys.\ Rev.\ Lett.\ {\bf 83}, 5198 (1999); 
J. Denschlag, J.E. Simsarian, D.L. Feder, C.W. Clark, 
L.A. Collins, J. Cubizolles, L. Deng, E.W. Hagley, K. Helmerson, 
W.P. Reinhardt, S.L. Rolston, B.I. Schneider, and W.D. Phillips, 
Science {\bf 287}, 97 (2000).




\end{thebibliography}
\end{document}